\documentclass[osd, manuscript]{copernicus}
\usepackage{float}
\usepackage{graphicx}
\usepackage{subcaption}

\begin{document}

\nolinenumbers

\title{Statistical Analysis of the Phosphate Data of the World Ocean Database 2013}
\Author[1]{Joscha}{Reimer}

\affil[1]{Kiel University,
Department of Computer Science,
Algorithmic Optimal Control - CO$_2$ Uptake of the Ocean,
24098 Kiel, Germany}

\runningtitle{TEXT}
\runningauthor{TEXT}

\correspondence{Joscha Reimer (joscha.reimer.edu@web.de)}

\received{}
\pubdiscuss{}

\maketitle

\begin{abstract}
    The phosphate data of the World Ocean Database 2013 are extensively statistically analyzed by splitting the measurement results into a long scale, i.e., climatological, and a short scale part. Means, medians, absolute and relative standard deviations, interquartile ranges, quartile coefficients of dispersion, correlations and covariances are estimated and analyzed. The underlying probability distributions are investigated using visual inspection as well as statistical tests. All presented methods are applicable to other data as long as they satisfy the postulated assumptions.
\end{abstract}

\introduction

Phosphate is a limiting nutrient for phytoplankton and therefore of central importance for understanding marine ecosystems (cf. \cite[4]{Bigg2003}). In the context of marine biogeochemical and ecosystem modeling, phosphate data play a dominant role as they are available in a high number and by their good spatial and temporal coverage (cf. \cite{Kriest2010}).

A basic source for freely available and quality-controlled oceanic measurement data is the World Ocean Database, a project established by the Intergovernmental Oceanographic Commission of UNESCO. The most recent release is the World Ocean Database 2013 which is described in \cite{Boyer2013} and \cite{Johnson2013}. Therein, millions of measurement data for phosphate concentrations in the ocean are available. 

The World Ocean Atlas, a project by the National Oceanographic Data Center in the U.S., provides analyzed and aggregated data from the World Ocean Database. The most recent release of their analysis of phosphate data is introduced in \cite{Garcia2014} as part of the World Ocean Atlas 2013 version 2. Means, standard deviations and standard errors of the means are provided there on one and five degree spatial grids at selected depth levels, down to 500 ${\rm m}$ as monthly and seasonal averages and down to 5500 ${\rm m}$ as annual averages. The provided data are climatological data (cf. \cite[1.2.1]{Storch1999}), i.e., data from different years have been used to calculate these properties of an average year. This has been done only for (space-time) grid boxes with enough data available. For all other grid boxes, the means (and only these) have been interpolated.

In this paper, we extended the analysis and aggregation provided in the World Ocean Atlas in several respects. The most important one is the split of the measurement results into a long scale part, i.e., climatological part, and a short scale part, i.e., noise part from a climatological perspective. This splitting allows to separately analyze and quantify the noise part in measurements and a noise free climatological part. Moreover, the noise part is useful in assessing the spatial and temporal resolution used in the statistical analysis.

For each grid box, the climatological part is primarily described by its climatological mean, i.e., the average concentration in an average year, and the climatological variability, i.e., the usual deviation between the average concentration in an actual year and the climatological mean. The noise part is primarily described for each grid box by the usual deviation between the average concentration in an actual year and the result of its measurement, and thus includes the variation of the concentration inside a grid box as well as measurement errors.

Compared to the World Ocean Atlas, we used a slightly different spatial and temporal resolution. We took a monthly temporal resolution within each year in all depth layers down to the sea floor, allowing us to capture time-dependent changes in deeper layers, which can be especially important for fitting time-dependent models to these data. To counteract the sparseness of the measurement data, especially in deeper layers, the vertical resolution was decreased. Details are given in Appendix \ref{subsec:lsms}.

Additionally to monthly means, we provide medians, absolute and relative standard deviations as well as interquartile ranges and quartile coefficients of dispersion, all of these for the entire ocean.  Finally, we quantified the statistical dependencies using correlations and covariances and investigated the underlying probability distributions using visual inspection as well as statistical tests.

We feel confident that the results of this detailed statistical analysis may improve the understanding of the marine phosphate concentration. They may also be used to increase the accuracy of model fitting procedures (cf. \cite[4]{Walter1997} or \cite[2.1.4]{Seber2003}) of marine biogeochemical models, and the information gain through new measurements (cf. \cite[6]{Walter1997} or \cite[5]{Pronzato2013}). Our approach is also applicable to other marine concentrations satisfying the assumptions made.

Our method for the statistical analysis is presented in Section \ref{sec: method}, the results obtained for the phosphate concentration are presented in Section \ref{sec: result} and in Section \ref{sec: conclusion}, we draw our conclusion.

 \section{Methods used in the Statistical Analysis} \label{sec: method}

To lay the foundation of the statistical analysis the statistical model is introduced and statistical assumptions postulated. Methods to estimate the associated expected values and to quantify the associated variabilities are presented afterwards. Thereafter, methods to quantify the statistical dependencies including covariances and correlations are introduced. The section closes with an investigation of the underlying probability distributions.

\subsection{Statistical Model} \label{subsec: statistical model}

We conducted our statistical analysis on a space-time grid described in detail in Appendix \ref{subsec:lsms} and define:
\begin{itemize}
\item $\mathcal{X}_s$ as the set of all spatial grid boxes, identified e.g. by their center, 
\item $\mathcal{X}_t$ as the set of all time intervals within one year (which are the same for all years), 
\item $\mathcal{X}_a$ as the set of years,
\end{itemize}
for which a statistical analysis should be carried out. We then identify:
\begin{itemize}
\item $\mathcal{X} := \mathcal{X}_s \times \mathcal{X}_t \times \mathcal{X}_a \times \mathbb{N}$ as the set of all possible measurement point ,
\end{itemize}
where each measurement point $(s, t, a, n) \in \mathcal{X}$ in the grid box $(s,t,a)$ has an unique index $n$
and define:
\begin{itemize}
\item $X \subseteq \mathcal{X}$ as the set of all points where measurement data are actually available and
\item $y(s, t, a, n)$ as the result of the measurement at $(s, t, a, n) \in X$.
\end{itemize} 
For the analysis presented below we use the sets:
\begin{itemize}
\item $A(s, t) := \{a ~|~ (s, t, a, n) \in X\}$ with all years where measurements are available at $(s, t) \in \mathcal{X}_s \times \mathcal{X}_t$,
\item $N(s, t, a) := \{n ~|~ (s, t, a, n) \in X\}$  with all indices of available measurements at $(s, t, a) \in \mathcal{X}_s \times \mathcal{X}_t \times \mathcal{X}_a$.
\end{itemize}

\subsubsection*{Random Fields and Statistical Assumptions}

In the statistical analysis, we use the following main variables, which  are considered as random fields:
\begin{itemize}
\item
the measurement results $\eta$, defined on $\mathcal{X}$,
\item the true concentration $\delta$, without noise, averaged in each grid box, defined on $\mathcal{X}_s \times \mathcal{X}_t \times \mathcal{X}_a$,
\item the noise $\epsilon$, defined on $\mathcal{X}$, including the variability due to the discretization introduced by the space-time grid as well as by imperfect measuring instruments and methods.
\end{itemize}
Let $(s, t, a, n), (\hat{s}, \hat{t}, \hat{a}, \hat{n}) \in \mathcal{X}$ represent arbitrary measurement points for the rest of this section. It is assumed that the noise is additive and unbiased:
\begin{gather}
    \label{equ: random fields}
    \eta(s, t, a, n) = \delta(s, t, a) + \epsilon(s, t, a, n),\\
    \label{equ: assumption: unbiased noise}
    \mathbb{E}(\epsilon(s, t, a, n)) = 0. 
\end{gather}
Moreover, there is no interaction assumed between the true concentration and the noise as well as the noise at different points:
\begin{gather}
    \label{equ: assumption: concentration and noise independent}
    \delta(s, t, a) \text{ and } 
    \epsilon(\hat{s}, \hat{t}, \hat{a}, \hat{n}) \text{ are independent,}\\
    \label{equ: assumption: noise pairwise independent}
    \epsilon(s, t, a, n) \text{ and } \epsilon(\hat{s}, \hat{t}, \hat{a}, \hat{n}) \text{ are independent if } (s, t, a, n) \neq (\hat{s}, \hat{t}, \hat{a}, \hat{n}).
\end{gather}
The following additional assumptions were made due to the sparseness of the available data:
The noise is assumed to have equal distributions within the same grid box:
\begin{equation} \label{equ: assumption: epsilon equal distributed}
    \epsilon(s, t, a, n) \overset {d}{=} \epsilon(s, t, a, \hat{n}).
\end{equation}
The true concentration is assumed to have the same distribution at two points where only the year differs:
\begin{equation} \label{equ: assumption: delta equal distributed}
    \delta(s, t, a) \overset {d}{=} \delta(s, t, \hat{a}).
\end{equation}
The covariance of the true concentration is assumed to be invariant with respect to annual shifts:
\begin{equation} \label{equ: assumption: annual periodicity of covariance}
	\operatorname*{cov}(\delta(s, t, a), \delta(\hat{s}, \hat{t}, \hat{a}))
	=
	\operatorname*{cov}(\delta(s, t, a + z), \delta(\hat{s}, \hat{t}, \hat{a} + z))
    \text{ with } a + z, \hat{a} + z \in \mathcal{X}_a.
\end{equation}
The annual periodicity of the climatological process that is described is a justification for the last two assumptions.

\subsection{Climatological Means} \label{subsec: expected values}

The measurement results $\eta$ and the true concentration $\delta$ have the same expected values (cf. \cite[2.6.5]{Storch1999}) due to the additivity \eqref{equ: random fields} and unbiasedness \eqref{equ: assumption: unbiased noise} of the noise:
\begin{equation*}
    \mathbb{E}(\eta(s, t, a, n))
    =
   	\mathbb{E}(\delta(s, t, a)) + \mathbb{E}(\epsilon(s, t, a, n))
    =
    \mathbb{E}(\delta(s, t, a)).
\end{equation*}
Due to the assumed annual periodicity \eqref{equ: assumption: delta equal distributed} of the true concentration, its expected values do not depend on the year and thus can be defined as $\mu: \mathcal{X}_s \times \mathcal{X}_t \to {\mathbb{R}}$ with:
\begin{equation*}
	\mu(s, t) := \mathbb{E}(\delta(s, t, a))
	=
    \mathbb{E}(\eta(s, t, a, n))
\end{equation*}
which corresponds to the climatological mean concentration in the grid box $(s, t)$.

The climatological mean $\mu(s, t)$ could be estimated using the average of all measurement results available in the grid box $(s, t)$. However, this could result in a very inaccurate estimate if the number of available measurements varies from year to year. An estimate would then tend to the average true concentration in years with the most measurements available which can be significantly differ from the climatological mean.

As a remedy, we first estimated the average true concentration in the grid box without noise for each year using the average of the measurement results within the same year:
\begin{equation*}
	c(s, t, a)
    :=
    \frac{1}{|N(s, t, a)|} \sum_{n  \in N(s, t, a)} y(s, t, a, n)
    \qquad
    \text{ if }
    |N(s, t, a)| \geq 1
    .
\end{equation*}
The climatological mean in a grid box was then estimated by the average, i.e., the sample mean (cf. \cite[4.3.1]{Storch1999}), of the estimated true concentrations in the grid box for different years:
\begin{equation*}
	\mu(s, t)
    \approx
    m(s, t)
    :=
    \frac{1}{|A(s, t)|} \sum\limits_{a \in A(s, t)} c(s, t, a)
    \qquad
    \text{ if }
    |A(s, t)| \geq 1
    .
\end{equation*}

Alternatively, the median (cf. \cite[2.6.4]{Storch1999}) instead of the (arithmetic) mean could be used in the previously described calculations for which the true concentration would be estimated as:
\begin{equation*}
	\hat{c}(s, t, a)
    :=
    \operatorname*{median}_{n  \in N(s, t, a)} y(s, t, a, n)
    \qquad
    \text{ if }
    |N(s, t, a)| \geq 1
\end{equation*}
and the corresponding climatological mean as:
\begin{equation*}
    \mu(s, t)
    \approx
    \hat{m}(s, t)
    :=
    \operatorname*{median}_{a \in A(s, t)}~ \hat{c}(s, t, a)
    \qquad
    \text{ if }
    |A(s, t)| \geq 1
    .
\end{equation*}

In general, the median provides a more accurate estimate in case of outliers, otherwise the mean should be preferred (cf. \cite[4]{Linacre1992}). If the number of measurements in several years is low, both estimates for $\mu$ might lack accuracy. Thus, it is reasonable to choose the required number of years with measurements $|A(s, t)|$ sufficiently high. We decided to require measurements for at least two years, preventing one extraordinary year to cause a poor estimate.

At the grid boxes lacking enough measurements, values for the climatological mean were interpolated according to Appendix \ref{subsec:interpolation}. Without sufficient data to achieve a  meaningful interpolation, the average of the estimates could be used instead.

\subsection{Variabilities} \label{subsec:dispersion}

Due to the assumed independence of $\delta$ and $\epsilon$, see Equation \eqref{equ: assumption: concentration and noise independent}, the variances ($\operatorname*{var}$) (cf. \cite[2.6.7]{Storch1999}) of the measurement results $\eta$ are given by the sum of the variances of the true concentration $\delta$ and the measurement noise $\epsilon$:
\begin{equation*}
\begin{split}
	\operatorname*{var}&(\eta(s, t, a, n))
&=
	\operatorname*{var}(\delta(s, t, a)) + \operatorname*{var}(\epsilon(s, t, a, n))
\text{.}
\end{split}
\end{equation*}

The variances of the true concentration, describing the climatological variabilities, do not depend on the year due to the assumed annual periodicity \eqref{equ: assumption: delta equal distributed}. Hence, the standard deviation ($\operatorname*{sd}$) (cf. \cite[2.6.7]{Storch1999}) of the true concentration at a grid box was estimated by the sample standard deviation (cf. \cite[4.3.2]{Storch1999}) of all estimated true concentrations in this grid box and different years:
\begin{equation*}
	\operatorname*{sd}(\delta(s, t, a))
    \approx
\sqrt{\frac{1}{|A(s, t)| - 1} \sum\limits_{\hat{a} \in A(s, t)} (c(s, t, \hat{a}) - m(s, t))^2}
    \qquad
    \text{ if }
    |A(s, t)| \geq 2
    .
\end{equation*}

Under the assumption of equal distributions of the noises in a grid box \eqref{equ: assumption: epsilon equal distributed}, the variance of the noises, describing the short scale variability, do not depend on the number of measurements. Hence, the standard deviation of the noise for a specific grid box and a specific year was estimated by the sample standard deviation of all measurement results in this grid box and year:
\begin{equation*}
	\operatorname*{sd}(\epsilon(s, t, a, n))
	\approx
    \sqrt{\frac{1}{|N(s, t, a)| - 1} \sum_{\hat{n} \in N(s, t, a)} (y(s, t, a, \hat{n}) - c(s, t, a))^2}
    \qquad
    \text{ if }
    |N(s, t, a)| \geq 2
    .
\end{equation*}

The sample interquartile range as an approximation of the interquartile range ($\operatorname*{iqr}$) can be used as well to quantify the variability:
\begin{equation*}
	\operatorname*{iqr}(\delta(s, t, a))
    \approx
    \operatorname*{q\text{\tiny{75}\%}}_{\hat{a} \in A(s, t)} c(s, t, \hat{a}) - 
    \operatorname*{q\text{\tiny{25}\%}}_{\hat{a} \in A(s, t)} c(s, t, \hat{a})
    \qquad
    \text{ if }
    |A(s, t)| \geq 1
    ,
\end{equation*}
and
\begin{equation*}
	\operatorname*{iqr}(\epsilon(s, t, a, n))
	\approx
    \operatorname*{q\text{\tiny{75}\%}}_{\hat{n} \in N(s, t, a)} y(s, t, a, \hat{n}) - 
    \operatorname*{q\text{\tiny{25}\%}}_{\hat{n} \in N(s, t, a)} y(s, t, a, \hat{n})
    \qquad
    \text{ if }
    |N(s, t, a)| \geq 1
    .
\end{equation*}
Here, $\operatorname*{q\text{\tiny{25}\%}}$ and $\operatorname*{q\text{\tiny{75}\%}}$ denote the first and third quartile, respectively. The value of the quartile was linearly interpolated between two available values if necessary.

When interpreting the variability of a random variable, the variability relative to the expected value is usually more helpful than just the variability. Therefore, we calculate the relative standard deviations, i.e., the standard deviations divided by the means, and the quartile coefficients of dispersion, i.e., the interquartile ranges relative to the medians. Relative variabilities of the noise $\epsilon$ are meaningless since its expected values were assumed to be zero.

In the presence of outliers, interquartile ranges and quartile coefficients of dispersion perform better, otherwise standard deviations and relative standard deviations should be preferred.

Estimates of the variabilities in all variants are more accurate if a high number of measurements can be used. We computed these estimates only where at least three values were available and interpolated otherwise as described in Appendix \ref{subsec:interpolation}. The average of the estimates could be used in absence of sufficient data for interpolation.

\subsection{Statistical Dependencies} \label{subsection: statistical dependencies}

From the additivity of the noise \eqref{equ: random fields}, the assumptions \eqref{equ: assumption: concentration and noise independent}, \eqref{equ: assumption: noise pairwise independent} and the bilinearity of the covariance, we immediately deduce:
\begin{equation*}
\left.
\begin{array}{l}
    \operatorname*{cov}(\delta(s, t, a), \epsilon(\hat{s}, \hat{t}, \hat{a}, \hat{n})) = 
    \operatorname*{cov}(\epsilon(s, t, a, n), \epsilon(\hat{s}, \hat{t}, \hat{a}, \hat{n})) = 
    \operatorname*{cov}(\eta(s, t, a, n), \epsilon(\hat{s}, \hat{t}, \hat{a}, \hat{n})) = 0,\\
    \operatorname*{cov}(\eta(s, t, a, n), \eta(\hat{s}, \hat{t}, \hat{a}, \hat{n})) = 
    \operatorname*{cov}(\eta(s, t, a, n), \delta(\hat{s}, \hat{t}, \hat{a}, \hat{n})) =
    \operatorname*{cov}(\delta(s, t, a), \delta(\hat{s}, \hat{t}, \hat{a}))
\end{array}
\right\}
\text{ if } (s, t, a, n) \neq (\hat{s}, \hat{t}, \hat{a}, \hat{n}).
\end{equation*}
Thus, only the covariances of the true concentration  $\delta$ had to be estimated. The corresponding correlations were calculated using the estimated standard deviations.

\subsubsection*{Pointwise Covariances and Correlations}

The covariances of $\delta$ were assumed to be invariant with respect to annual shifts \eqref{equ: assumption: annual periodicity of covariance} and its covariance between two specific grid boxes and years was estimated by the sample covariance (cf. \cite[5.2.7]{Storch1999}) of all pairs of estimated true concentrations in the same grid boxes and with the same difference in the years:
\begin{equation*}
\begin{gathered}
	\operatorname*{cov}(\delta(s, t, a), \delta(\hat{s}, \hat{t}, \hat{a}))
	\approx
	\frac{1}{|B|-1}
	\sum_{(b, \hat{b}) \in B}
	(c(s, t, b) - m) (c(\hat{s}, \hat{t}, \hat{b}) - \hat{m}),
    \qquad
    \text{ if }
    |B| \geq 2
    \\
   	\text{ ~with~ }	
   	B
   	:=
   	\{
   	(a + z, \hat{a} + z)
   	~|~
   	a + z, \hat{a} + z \in A(s, t)
   	\},~
   	m = \frac{1}{|B|} \sum_{(b, \hat{b}) \in B} c(s, t, b)
    \text{ ~and~ }
   	\hat{m} = \frac{1}{|B|} \sum_{(b, \hat{b}) \in B} c(\hat{s}, \hat{t}, \hat{b})
       .
\end{gathered}
\end{equation*}
A higher $|B|$ results, as usual, in more accurate estimate.
The estimate of $\operatorname*{cov}(\delta(s, t, a), \delta(s, t, a))$ is equal to the estimate of $\operatorname*{var}(\delta(s, t, a))$ yielding a consistent estimate.

\subsubsection*{Covariance and Correlation Matrices}

The pointwise covariance estimates were processed into a covariance matrix (cf. \cite[2.8.7]{Storch1999}). For very large dimension, the covariance matrix can not be stored as a dense matrix due to limited memory. Hence, we used a sparse matrix and assumed that the covariance is zero where no estimate was available, to that effect that, the  number of stored entries corresponds to the number of estimated pointwise covariances, which can be controlled by $|B|$, the number of concentration estimates required for a covariance estimate. If estimated pointwise covariances close to zero are not stored and are thus implicitly assumed to be zero, the number of entries to be stored can be reduced even further.

In order to obtain a consistent estimate of a covariance matrix, it is not sufficient to just combine the individual estimates into a matrix. The resulting matrix has to be positive semidefinite and usually, even a well-conditioned positive definite estimate is preferred.

Often, so called shrinking methods are applied for this purpose (cf. \cite{Ledoit2004,Chen2010,Schaefer2005}). These tend to pull the most extreme matrix entries towards more central values, achieved by a convex combination of the covariance matrix estimate and some suitable chosen target matrix. A disadvantage of these methods is the alternation of all matrix entries (i.e., pointwise covariances). In extreme cases, one poorly estimated covariance can lead to a large impact of all other well estimated covariances.

Due to this and other disadvantages of these methods, we used the approach described in \cite{Reimer2019} and implemented in \cite{matrix-decomposition-1.2} where single off-diagonal entries are moved closer to zero generating a well-conditioned positive definite matrix. The $LDL^\top$ decomposition of the resulting covariance matrix were calculated by this approach as a byproduct and can be used to solve corresponding linear equations quickly. Furthermore permutation methods are applied to reduce the number of entries that must be stored allowing to efficiently process even larger matrices.

If the estimated variances, i.e. the diagonal values of the covariance matrix estimated, are considered sufficiently accurate, the diagonal values can be left unchanged by the algorithm, or the correlation matrix instead of the covariance matrix can be approximated by the algorithm. We decided to approximate the correlation matrix.

To get a well-conditioned matrix, we forced the approximation algorithm to ensure that each entry in the diagonal matrix $D$ of the $LDL^\top$ decomposition is at least 0.01. This threshold directly affects the condition number and the approximation error and can be adjusted to prioritize one of these two.

\subsubsection*{Correlations Dependencies on Distances}

In many applications in natural sciences, the correlations of random fields depend solely on the distance between the associated points and this we checked for the true concentration $\delta$. 

If the estimated correlations can be described by a function depending only on the distance of the associated points, the estimated correlations for points with the same distance must be (approximately) the same. Hence, we grouped all estimated correlations for points with the same distance. For each group, we calculated the interquartile range which must be close to zero. 0.1 could be a good threshold here, or 0.05 as a more restrictive value.

Moreover, if the correlations could be described even by a continuous function, the correlations must be close to each other if the distances of their associated points are close to each other. This can be also checked by grouping and calculating interquartile ranges.

Both can also be checked to some extend graphically by plotting the grouped correlations or the calculated interquartile ranges. To ensure significance of theses checks, the number of involved correlations has to be sufficiently large.

\subsection{Probability Distributions} \label{subsec:probability distributions}

In addition to the estimation of statistical parameters like expectation value or standard deviation, the type of the underlying probability distribution (cf. \cite[2.6.3]{Storch1999}) is of interest too. Visual inspection and statistical tests were used to analyze from what probability distribution the data may originate (cf. \cite[4]{Linacre1992}). We focused on normal distribution (cf. \cite[2.7.3]{Storch1999}) and log-normal distribution (cf. \cite[2.7.6]{Storch1999}).

Histogram (cf. \cite[5.2.1]{Storch1999}) and kernel density estimation (cf. \cite{Scott2015}) were used to give an idea of the underlying probability distribution. Box plots are an alternative which indicate the expected value, the spread, the skewness and outliers. P-P (probability-probability) and Q-Q (quantile-quantile) plots are useful to study the data with respect to a particular probability distribution.

For a particular probability distribution, Usually several statistical tests are available to verify if data originate from a particular probability distribution. Tests applied in this thesis regarding normal distributions were the Shapiro-Wilk test introduced in \cite{Shapiro1965}, the Anderson-Darling test introduced in \cite{Anderson1952} and the D'Agostino-Pearson test introduced in \cite{DAgostino1971} and \cite{DAgostino1973}. To check if the data originate from log-normal distributions, first the natural logarithm and afterwards the tests for normal distributions were applied to the data.

To check the probability distribution at a particular point, by visual inspection or by statistical tests, all values originating from random variables with the same probability distribution were used:
\begin{itemize}
\item For the measurement results  $\eta$:
all values $ y(\hat{s}, \hat{t}, \hat{a}, \hat{n})$ with  $(\hat{s}, \hat{t}, \hat{a}, \hat{n}) \in N(s, t, a)$.
\item 
For the true concentration  $\delta$: all values $c(s, t, a)$ with $ a \in A(s, t)$.
\item 
For the noise  $\epsilon$: 
all values
   $\left(y(\hat{s}, \hat{t}, \hat{a}, \hat{n}) - c(s, t, a)\right)$
    with 
    $(\hat{s}, \hat{t}, \hat{a}, \hat{n}) \in N(s, t, a)$.
\end{itemize}

It should be noted that these statistical tests, as well as the graphical methods, cannot ensure whether data are really realizations from a normal distribution or not. They just can determine a reasonable certainty.

 \section{Results of the Statistical Analysis} \label{sec: result}

We now present the results of our statistical analysis of the phosphate concentration data provided by the World Ocean Database 2013 presented in \cite{Boyer2013} and \cite{Johnson2013}, obtained as described in Section \ref{sec: method}, using the software mentioned in Appendix \ref{subsec:software} and a one degree resolution with 33 depth layers and a monthly time resolution as described in Appendix \ref{subsec:lsms}. They are not interpreted in a marine context.

The spatial and temporal distribution of the data is described in Subsection \ref{subsec: po4: number of measurements}, the climatological means in Subsection \ref{subsec: po4: expected values}, the long and short scale variabilities in Subsection \ref{subsec: po4: dispersion}, the covariances and correlations in Subsection \ref{subsec: po4: statistical dependencies}, and the investigation of the probability distributions in Subsection \ref{subsec: po4: probability distribution}.

\subsection{Spatial and Temporal Distribution} \label{subsec: po4: number of measurements}

All data in the World Ocean Database 2013 have been quality checked (cf. \cite[Section 3]{Johnson2013}). We used all phosphate data that passed quality control, which were more than 4.1 million measurements in total.

\begin{figure}[h]
    \centering
    \begin{subfigure}[b]{0.495\linewidth}
    	\includegraphics[width=1.0\linewidth]{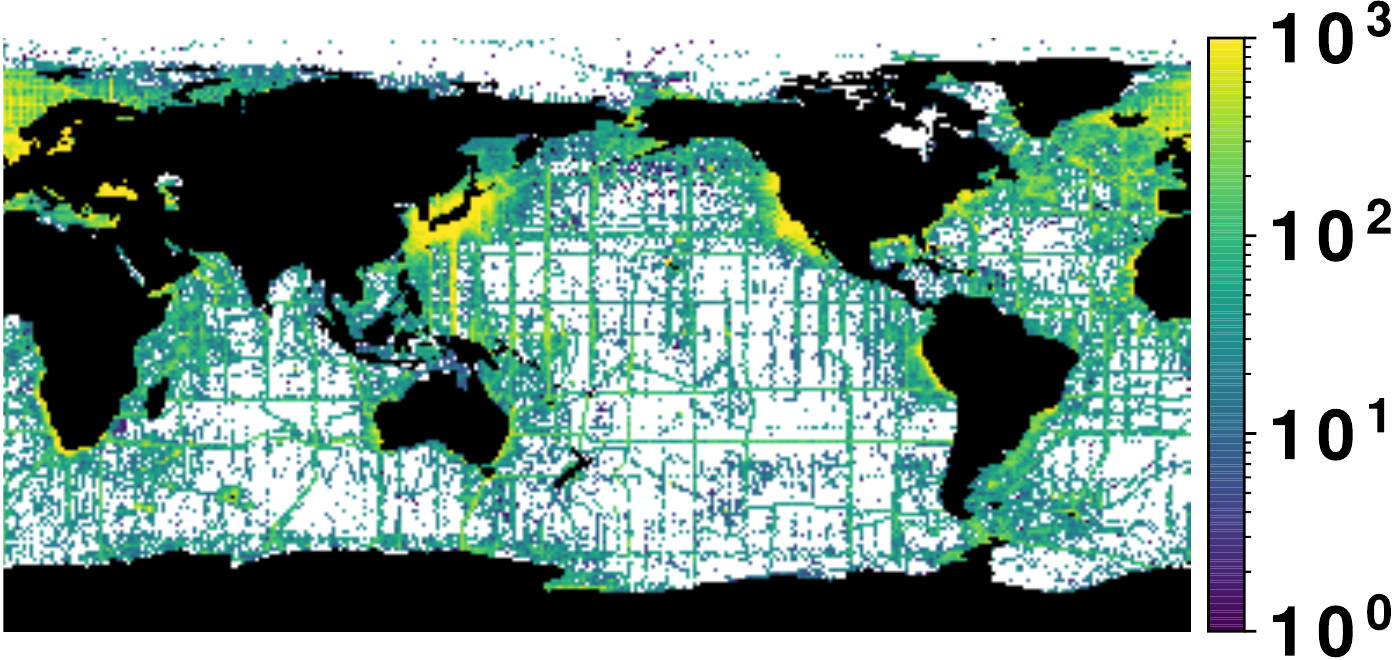}
    	\caption{in all depths (1$^{\circ}$ grid)} 
    	\label{fig:distribution:po4:space_all}
    \end{subfigure}
	\hfill
    \begin{subfigure}[b]{0.495\linewidth}
        \includegraphics[width=1.0\linewidth]{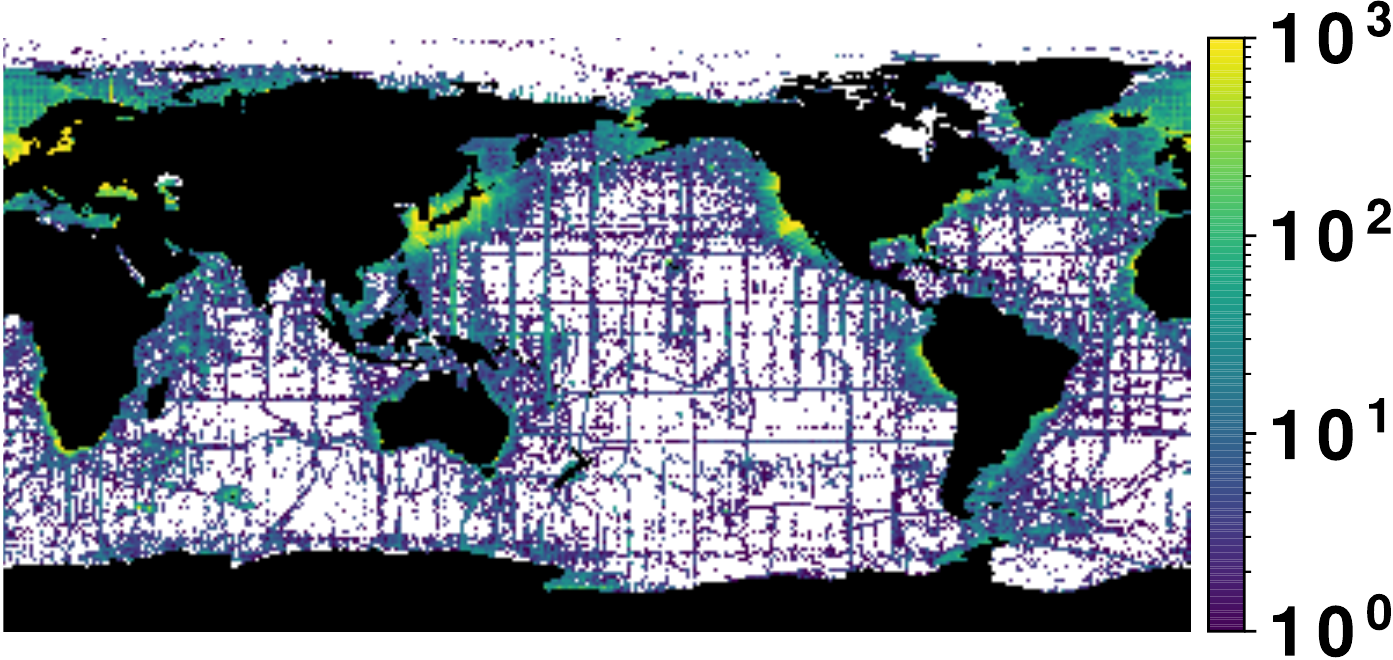}
        \caption{between 0 and 25 ${\rm m}$ depth (1$^{\circ}$ grid)} 
        \label{fig:distribution:po4:space_first}
    \end{subfigure}
	\begin{subfigure}[b]{0.32\linewidth}
    	\includegraphics[width=1.0\linewidth, height=0.615\linewidth]{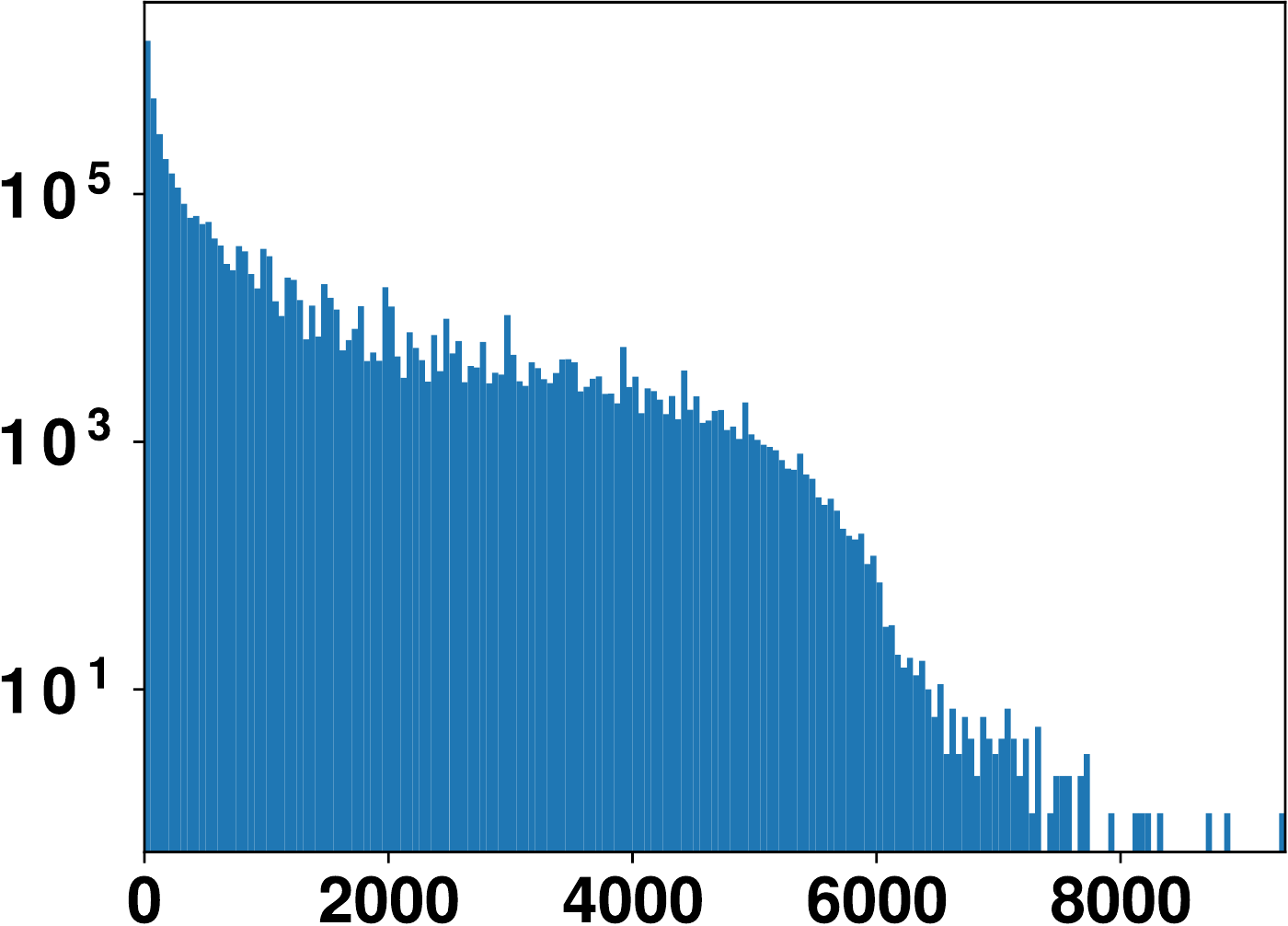}
\caption{per 50 m depth} 
    	\label{fig:distribution:po4:depth}
	\end{subfigure}   
	\hfill 
    \begin{subfigure}[b]{0.32\linewidth}
        \includegraphics[width=1.0\linewidth, height=0.65\linewidth]{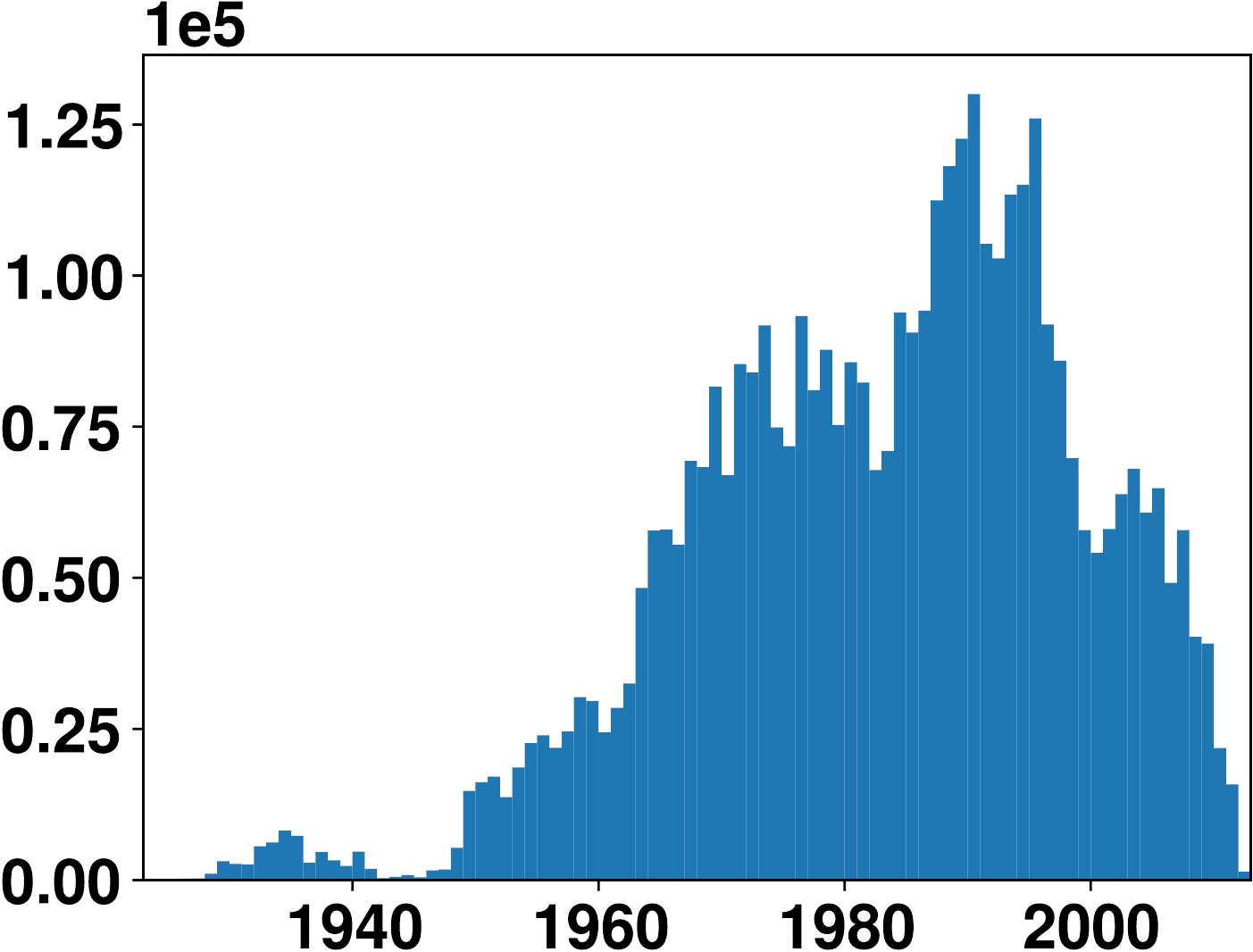}
        \caption{per year} 
        \label{fig:distribution:po4:year}
    \end{subfigure}
	\hfill
    \begin{subfigure}[b]{0.32\linewidth}
        \includegraphics[width=1.0\linewidth, height=0.65\linewidth]{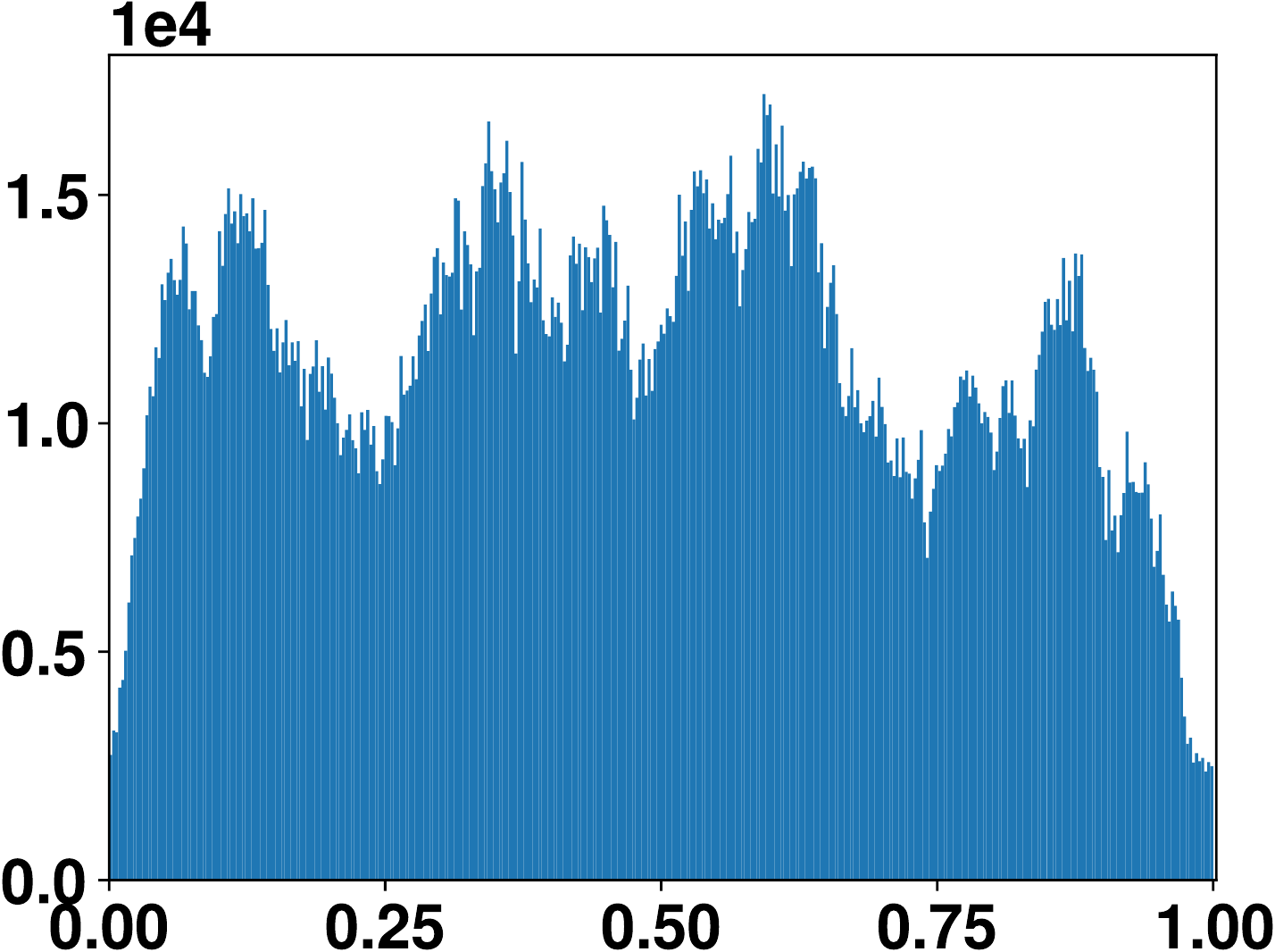}
        \caption{per day in the year} 
        \label{fig:distribution:po4:day}
    \end{subfigure}
    \vspace{-0.5em}
    \caption{Number of phosphate measurements with respect to space and time.} 
\end{figure}

Regarding the spatial distribution, the number of measurements decreases with the distance to the coast (Figure \ref{fig:distribution:po4:space_all}, \ref {fig:distribution:po4:space_first}) and with growing depth (Figure \ref{fig:distribution:po4:depth}). In time, the measurements range from 1923 to 2012, with the majority between 1963 and 2009 (Figure \ref{fig:distribution:po4:year}). Over most of the year, the measurements are uniformly distributed, significantly fewer measurements are only available in December and January (Figure \ref{fig:distribution:po4:day}).

\subsection{Climatological Means} \label{subsec: po4: expected values}

\begin{figure}[t!]
	\centering
	\begin{subfigure}[b]{0.495\linewidth}
		\includegraphics[width=1.0\linewidth]{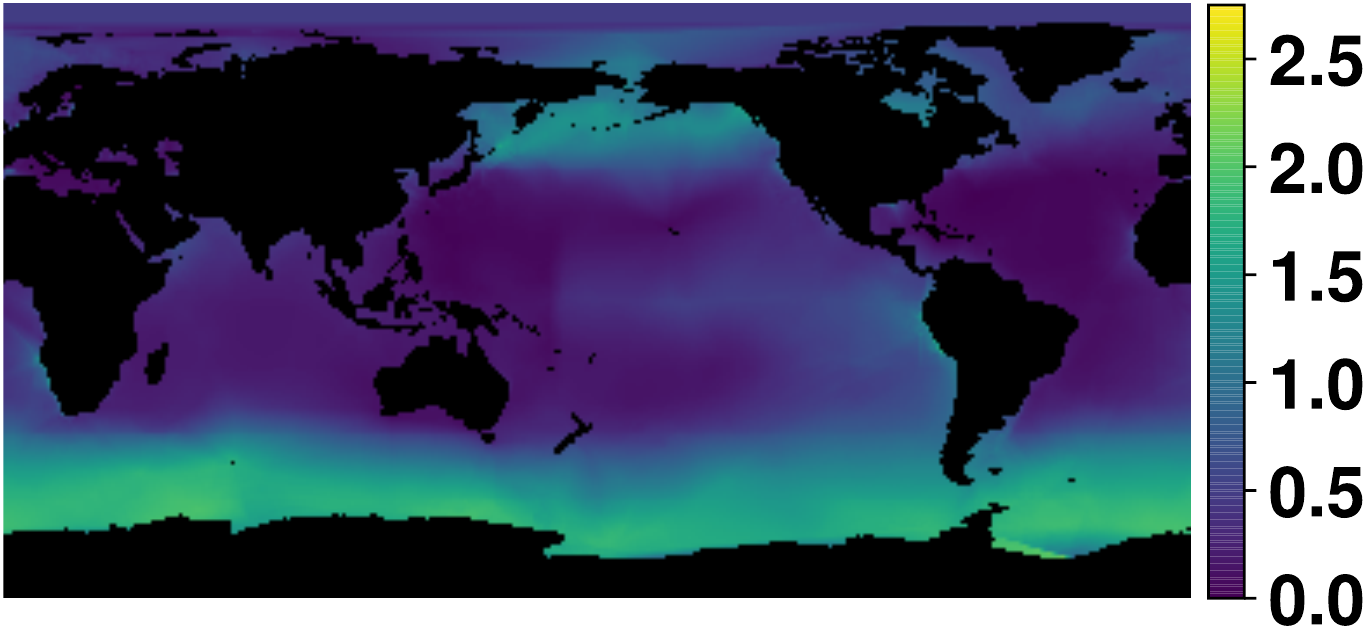}
\caption{water surface: averaged over time and 0 to 25 ${\rm m}$ depth}  
		\label{fig:mean:surface}
	\end{subfigure}
	\hfill
	\begin{subfigure}[b]{0.245\linewidth}
		\includegraphics[width=1.0\linewidth, height=0.9\linewidth]{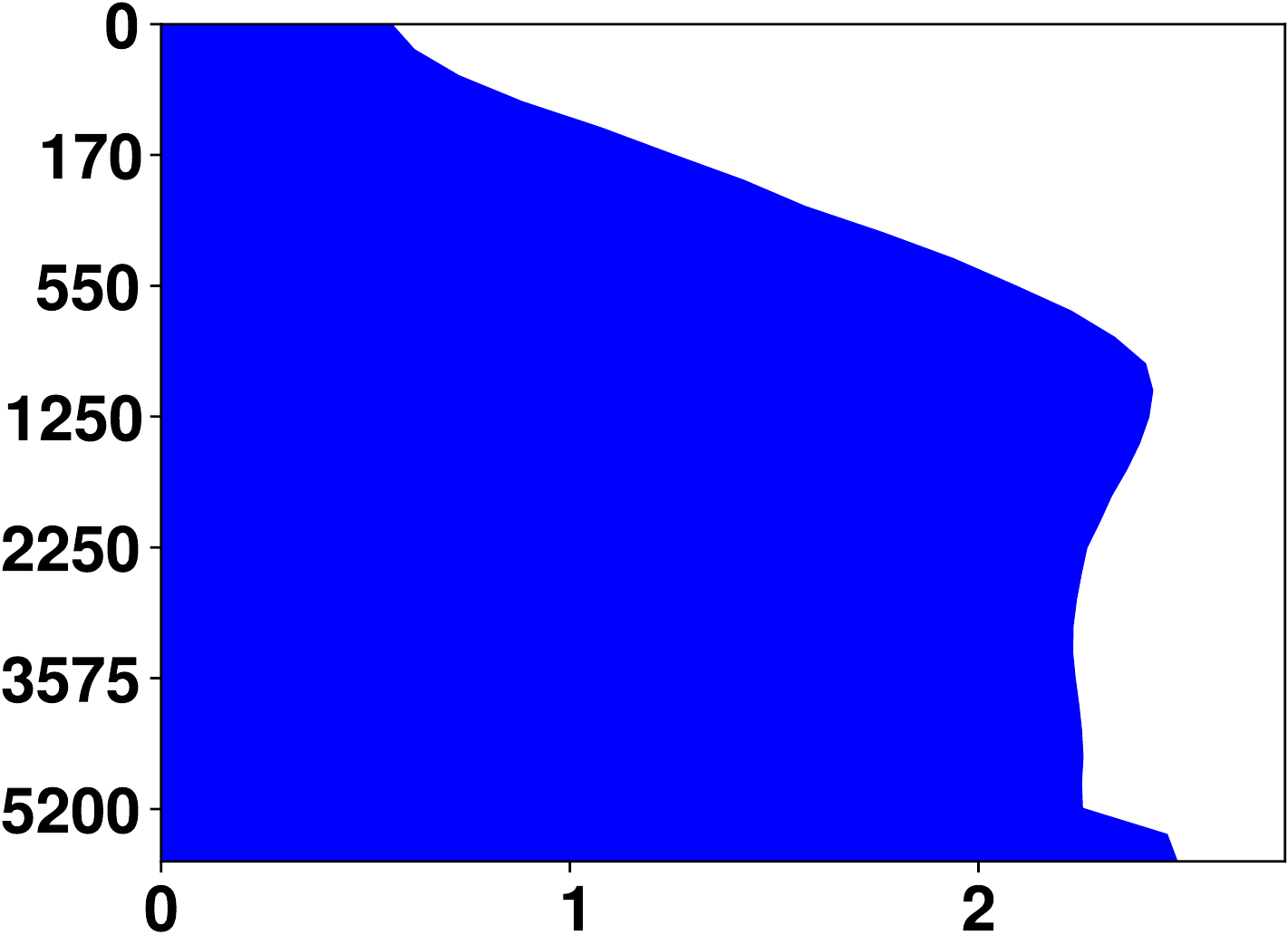}
\caption{averaged over all but depth} 
		\label{fig:mean:depth}
	\end{subfigure}
   	\hfill
   	\begin{subfigure}[b]{0.245\linewidth}
   		\includegraphics[width=1.0\linewidth, height=0.9\linewidth]{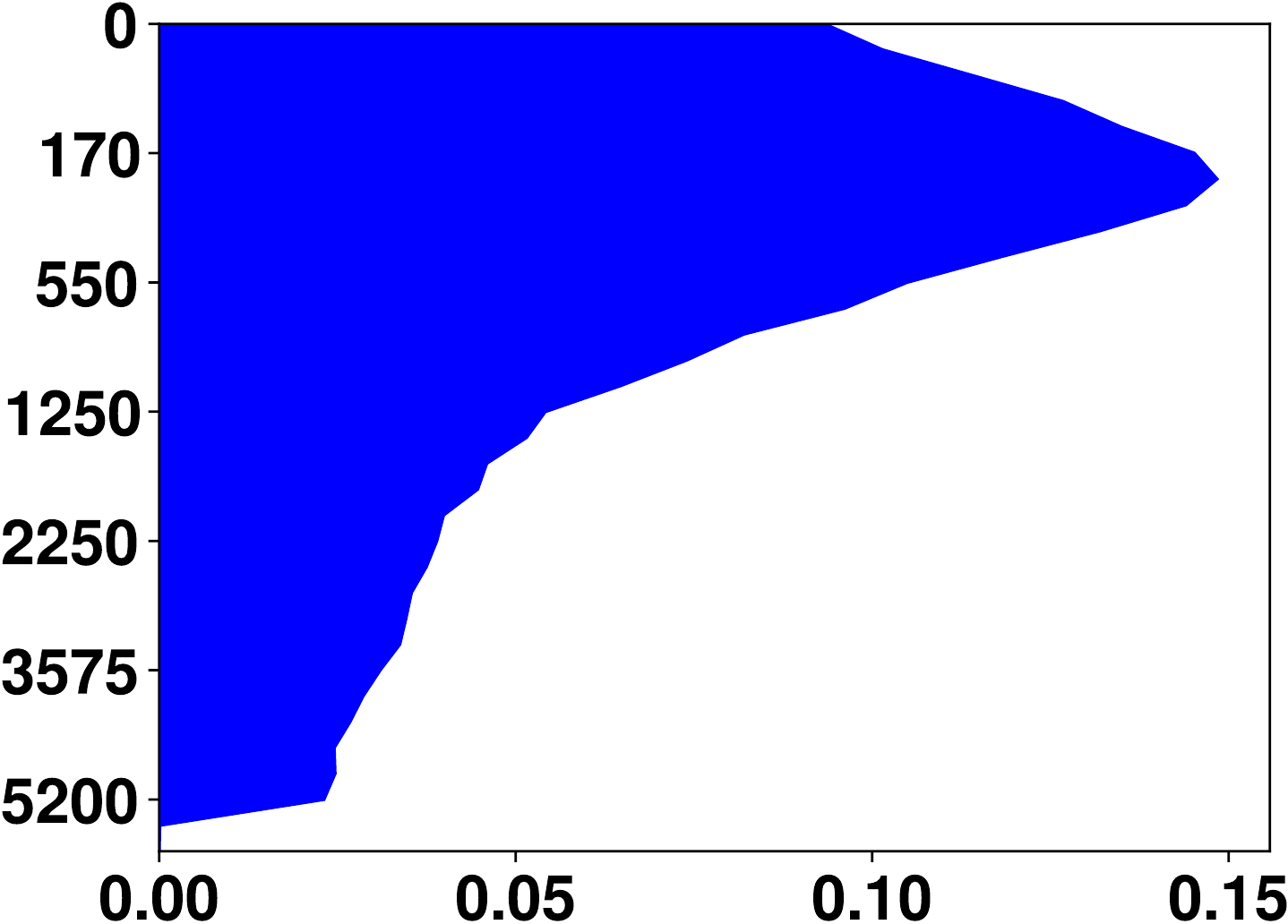}
\caption{average monthly change} 
   		\label{fig:mean:time_diff}
   	\end{subfigure}
              
   	\begin{subfigure}[b]{0.495\linewidth}
   		\includegraphics[width=1.0\linewidth]{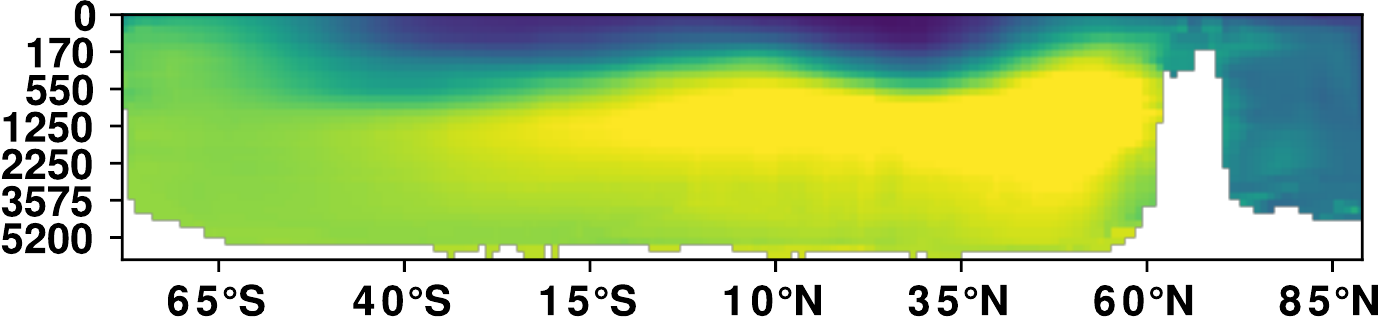}
\caption{Pacific Ocean: averaged over time and between 125$\degree$E and 70$\degree$W} 
   		\label{fig:mean:pacific}
   	\end{subfigure}
     \hfill    
   	\begin{subfigure}[b]{0.495\linewidth}
   		\includegraphics[width=1.0\linewidth]{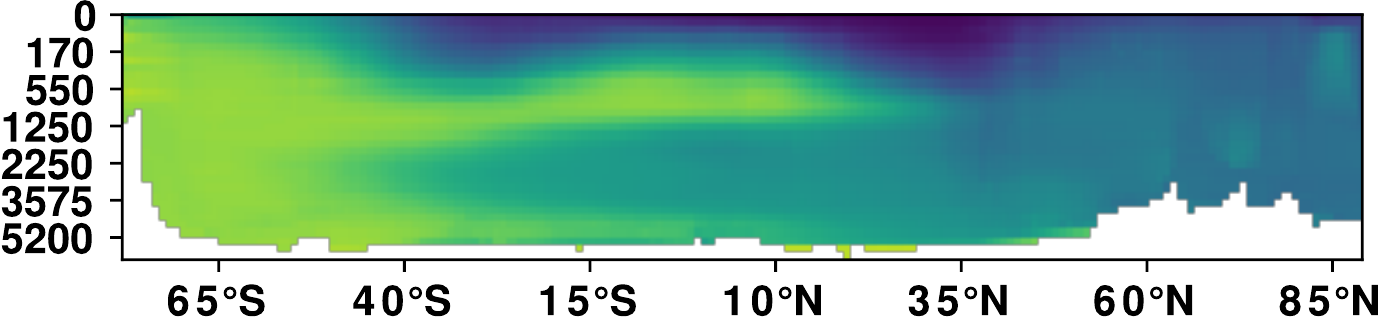}
\caption{Atlantic Ocean: averaged over time and between 70$\degree$W and 20$\degree$E} 
   		\label{fig:mean:atlantic}
   	\end{subfigure}
	\vspace{-0.5em}
	\caption{Climatological mean of phosphate in ${\rm mmol\,m}^{-3}$.}
	\label{fig:mean} 
\end{figure}

The climatological mean concentrations, i.e. the concentrations in an average year, were estimated once using the arithmetic mean and once using the median, (compare Subsection \ref{subsec: expected values}). The results using the arithmetic mean are described next and are plotted in Figure \ref{fig:mean}.

The average estimated climatological mean is 2.17 ${\rm mmol\,m}^{-3}$. The time averaged climatological means near the surface are shown in Figure \ref{fig:mean:surface}. Here, the highest values are at the Southern Ocean ranging from 1.5 to 2.2 ${\rm mmol\,m}^{-3}$. Other high values are at the north of the Pacific Ocean, ranging from 1.0 to 1.5 ${\rm mmol\,m}^{-3}$, and at the southeast of the Pacific Ocean, around 1.0 ${\rm mmol\,m}^{-3}$. Elsewhere, they are usually below 0.5 ${\rm mmol\,m}^{-3}$.

The climatological means averaged over all but the depth are presented in Figure \ref{fig:mean:depth}. They strictly increase from 0.6 to 2.5 ${\rm mmol\,m}^{-3}$ with growing depth up to 1200 ${\rm m}$. After that they remain constant at approximately 2.25 ${\rm mmol\,m}^{-3}$.

The average absolute change in the climatological means after one month is about 0.05 ${\rm mmol\,m}^{-3}$. The average absolute monthly changes depending on the depth are plotted in Figure \ref{fig:mean:time_diff}. They strictly increase from 0.09 to 0.15 ${\rm mmol\,m}^{-3}$ with growing depth up to around 250 ${\rm m}$. Afterwards they strictly decrease with growing depth. After a depth of 1500 ${\rm m}$ they are below 0.05 ${\rm mmol\,m}^{-3}$.

The climatological means in the Pacific Ocean depending on depth and latitude and averaged over longitude and time are shown in Figure \ref{fig:mean:pacific}. Near the surface they are usually between 0.2 ${\rm mmol\,m}^{-3}$ and 1 ${\rm mmol\,m}^{-3}$. Only south of 50$\degree$S they are between 1.1 ${\rm mmol\,m}^{-3}$ and 1.7 ${\rm mmol\,m}^{-3}$. After a few hundred meter depth they increase rapidly. After a depth of 500 ${\rm m}$, they are above 2 ${\rm mmol\,m}^{-3}$ and between 25$\degree$S and 60$\degree$N they are above 2.5 ${\rm mmol\,m}^{-3}$. Even deeper, they change only slightly.

The results for the Atlantic Ocean are shown in Figure \ref{fig:mean:atlantic}. They are similar to the one in the Pacific Ocean at the first few hundred meters depth as well as south of 50$\degree$S. However between 50$\degree$S and 30$\degree$N the average concentrations decrease from around 2.2 ${\rm mmol\,m}^{-3}$ to around 1.5 ${\rm mmol\,m}^{-3}$ at a depth of about 1250 ${\rm m}$ and then increase again slightly near the seafloor. North of 30$\degree$N, the values remain almost constant at 1.1 ${\rm mmol\,m}^{-3}$ at a depth greater than 1000 ${\rm m}$.

The estimated climatological means using the median instead of the arithmetic mean look quite similar. Their average absolute difference is 0.004 ${\rm mmol\,m}^{-3}$. It is 0.011 ${\rm mmol\,m}^{-3}$ near the surface and decreases with growing depth.

\subsection{Variabilities} \label{subsec: po4: dispersion}

The estimated variabilities of the measurement results $\eta$ as well as the short and the long scale variabilities, i.e., the variabilities of the noise $\epsilon$ and the true concentration $\delta$, are described next using standard deviations and relative standard deviations.

\subsubsection*{Standard Deviations}

\begin{figure}[t!]
	\centering
   	\begin{subfigure}[b]{0.495\linewidth}
   		\includegraphics[width=1.0\linewidth]{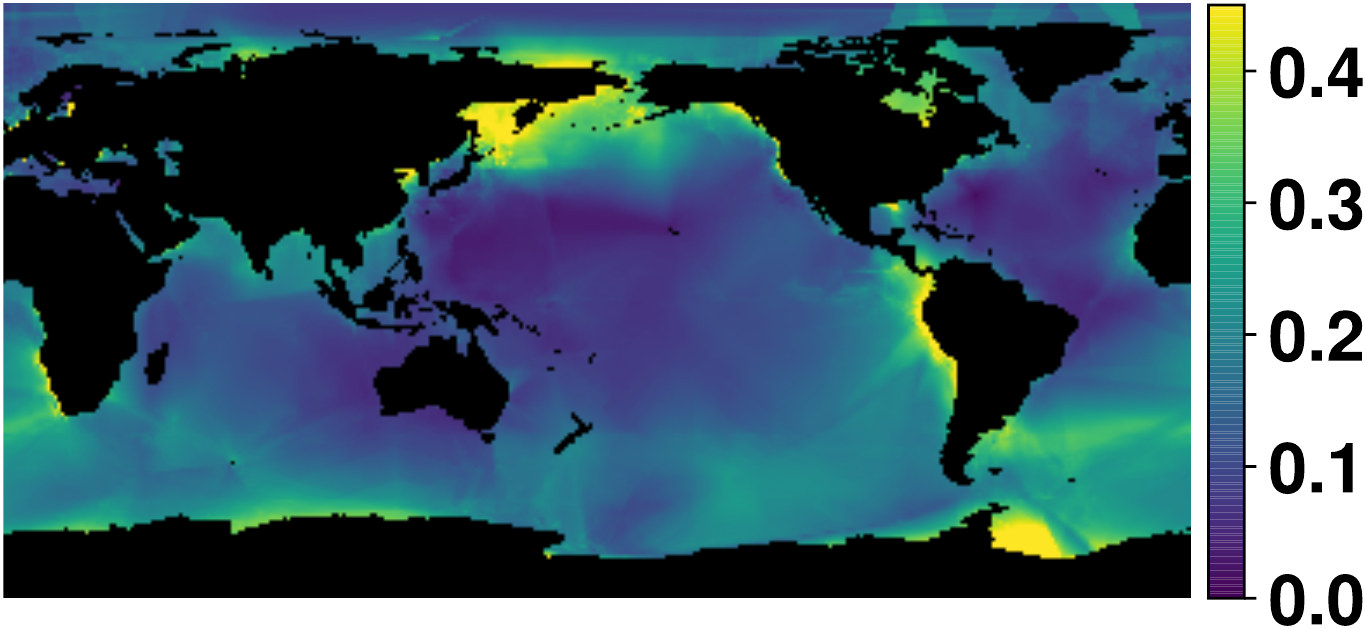}
  		\caption{water surface: averaged over time and 0 to 25 ${\rm m}$ depth}  
   		\label{fig:standard_deviation:surface}
   	\end{subfigure}
   	\hfill
   	\begin{subfigure}[b]{0.245\linewidth}
   		\includegraphics[width=1.0\linewidth, height=0.9\linewidth]{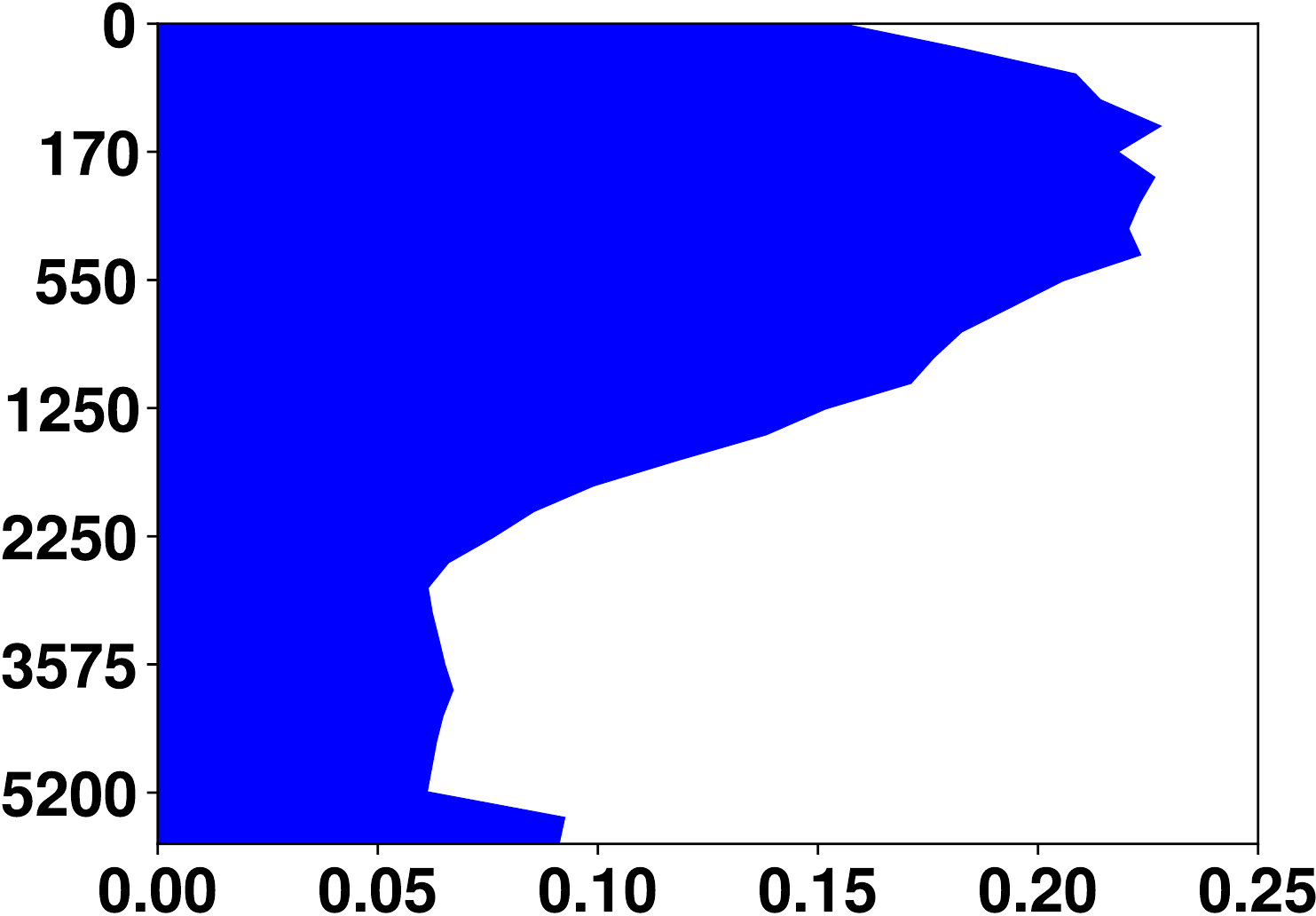}
   		\caption{averaged over all but depth}
   		\label{fig:standard_deviation:depth}
   	\end{subfigure}
   	\hfill
   	\begin{subfigure}[b]{0.245\linewidth}
   		\includegraphics[width=1.0\linewidth, height=0.9\linewidth]{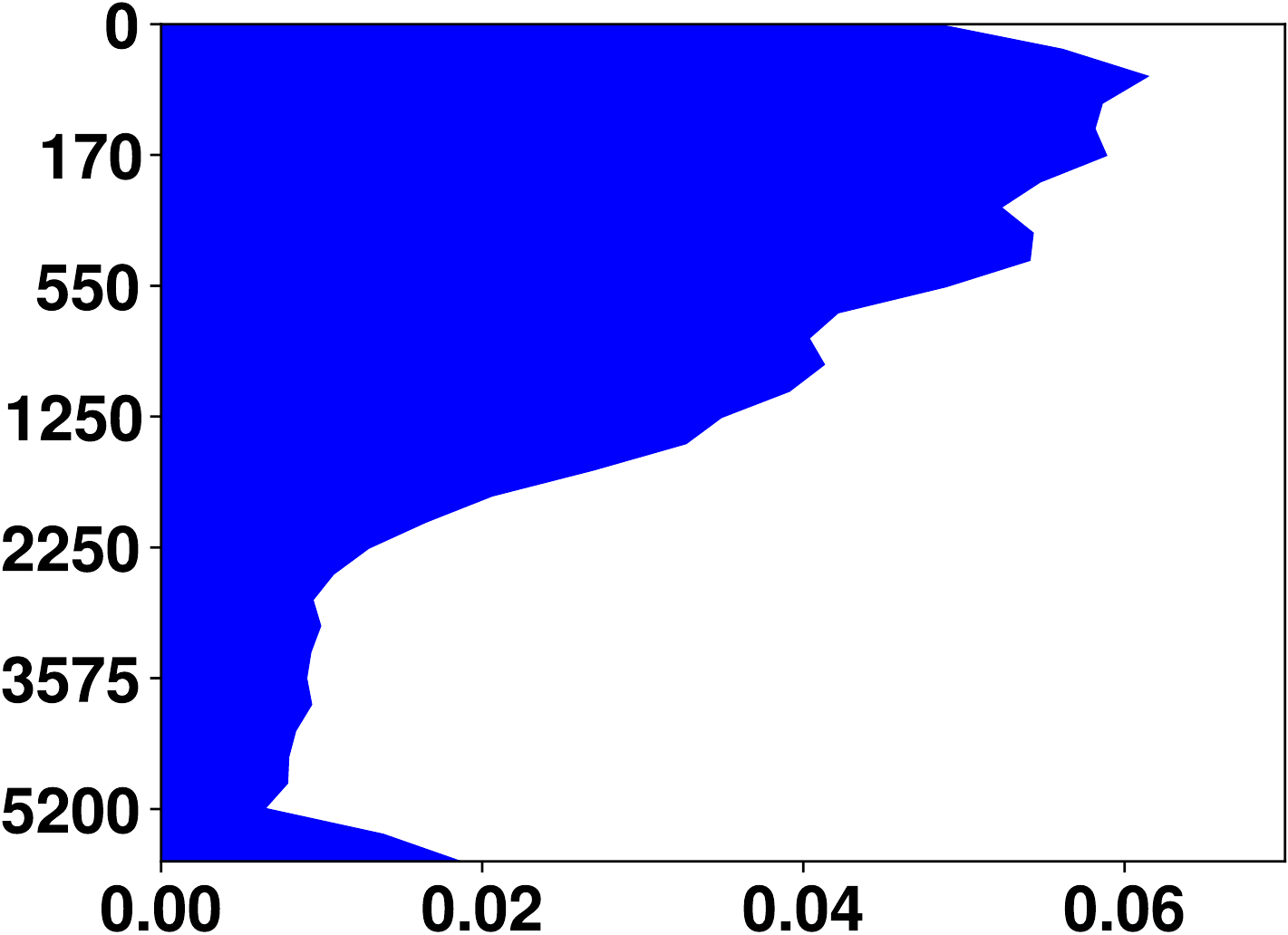}
   		\caption{average monthly change} 
   		\label{fig:standard_deviation:time_diff}
   	\end{subfigure}
  
   	\begin{subfigure}[b]{0.495\linewidth}
   		\includegraphics[width=1.0\linewidth]{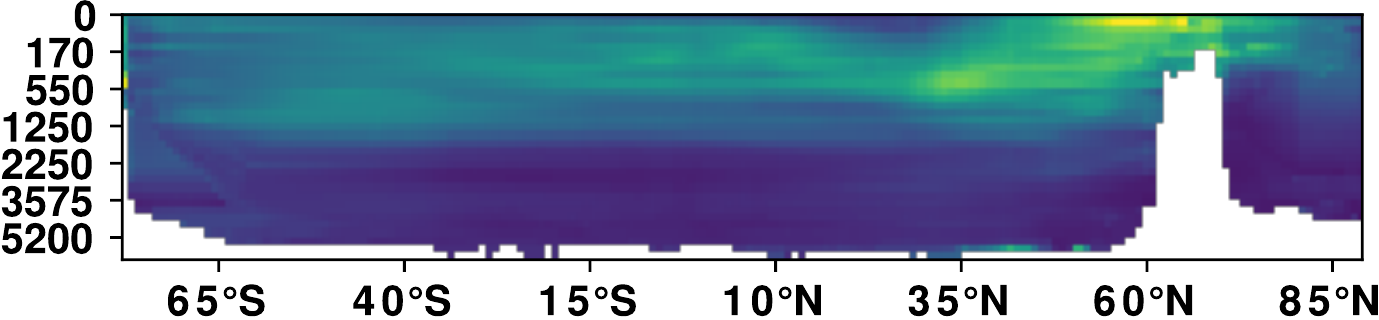}
   		\caption{Pacific Ocean: averaged over time and between 125$\degree$E and 70$\degree$W} 
   		\label{fig:standard_deviation:pacific}
   	\end{subfigure}
    \hfill    
   	\begin{subfigure}[b]{0.495\linewidth}
   		\includegraphics[width=1.0\linewidth]{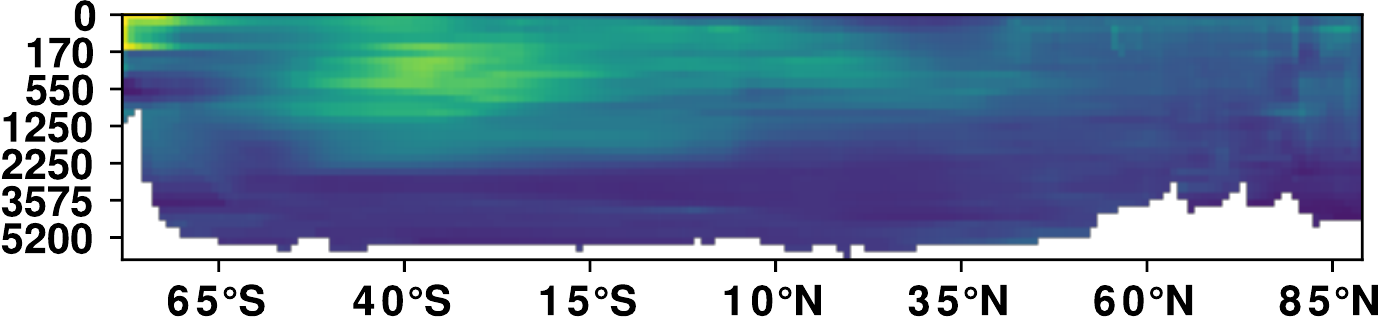}
   		\caption{Atlantic Ocean: averaged over time and between 70$\degree$W and 20$\degree$E}  
   		\label{fig:standard_deviation:atlantic}
   	\end{subfigure}
	\vspace{-0.5em}
	\caption{Standard deviation of phosphate measurements ($\eta$) in ${\rm mmol\,m}^{-3}$.} 
	\label{fig:standard_deviation}
\end{figure}

The estimated standard deviations of the measurement results $\eta$ are plotted in Figure \ref{fig:standard_deviation}. The average estimated standard deviation is 0.11 ${\rm mmol\,m}^{-3}$.

Its time averaged standard deviations near the surface are shown in Figure \ref{fig:standard_deviation:surface}. Here, the highest standard deviations are between 0.35 and 0.45 ${\rm mmol\,m}^{-3}$ around the eastern part of Russia, near the west and south of South America and the west coastal region of Southern Africa. Elsewhere near the coast or in the Southern Ocean, the standard deviations are between 0.15 and 0.25 ${\rm mmol\,m}^{-3}$. They are between 0.05 and 0.15 ${\rm mmol\,m}^{-3}$ in the remaining areas. 

The standard deviations averaged over all but the depth is plotted in Figure \ref{fig:standard_deviation:depth}. It increases from 0.15 ${\rm mmol\,m}^{-3}$ at the surface to 0.23 ${\rm mmol\,m}^{-3}$ at around 120 ${\rm m}$ depth. Afterwards it barely changes up to a depth of 500 ${\rm m}$. Then it strictly decreases to 0.06 ${\rm mmol\,m}^{-3}$ at a depth of 2500 ${\rm m}$ and hardly changes while going deeper.

The average absolute change in the standard deviation after one month is shown in Figure \ref{fig:standard_deviation:time_diff}. Up to a depth of 500 ${\rm m}$, the change is between 0.05 and 0.06 ${\rm mmol\,m}^{-3}$. Then it decreases to around 0.01 ${\rm mmol\,m}^{-3}$ at a depth of 2500 ${\rm m}$ whereupon it remains almost constant.

In Figure \ref{fig:standard_deviation:pacific} and \ref{fig:standard_deviation:atlantic}, the standard deviations in the Pacific Ocean and the Atlantic Ocean averaged over time and longitude are shown. They are between 0.25 and 0.35 ${\rm mmol\,m}^{-3}$ above 1000 ${\rm m}$ depth in the Pacific and the South Atlantic. Elsewhere they are lower than 0.25 ${\rm mmol\,m}^{-3}$ and deeper than 1500 ${\rm m}$ even lower than 0.15 ${\rm mmol\,m}^{-3}$. Peaks between 0.40 and 0.45 ${\rm mmol\,m}^{-3}$ are located in the far north of the Pacific as well as in in the south of the Atlantic and the Southern Ocean.

\begin{figure}[t!]
	\centering    
   	\begin{subfigure}[b]{0.495\linewidth}
   		\includegraphics[width=1.0\linewidth]{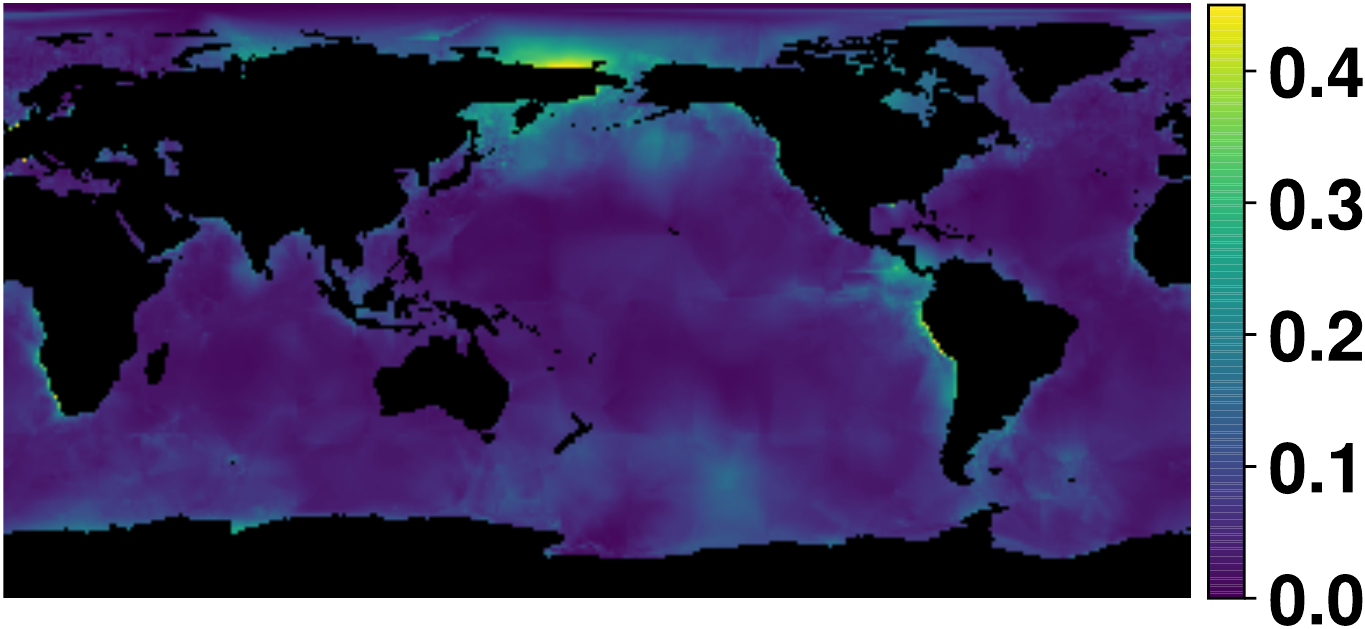}
   		\caption{water surface: averaged over time and 0 to 25 ${\rm m}$ depth}  
   		\label{fig:noise_standard_deviation:surface}
   	\end{subfigure}
   	\hfill
   	\begin{subfigure}[b]{0.245\linewidth}
   		\includegraphics[width=1.0\linewidth, height=0.9\linewidth]{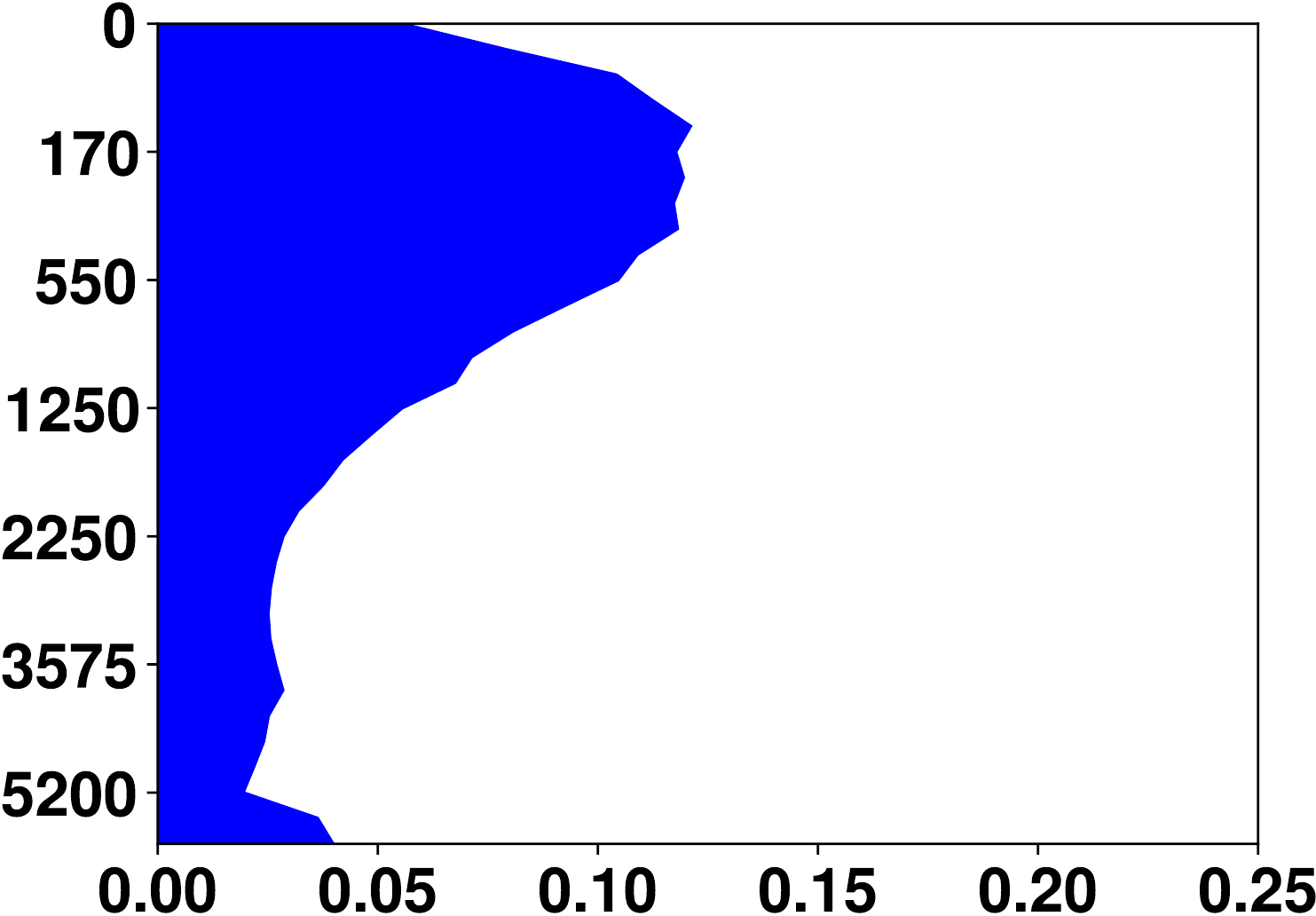}
   		\caption{averaged over all but depth}
   		\label{fig:noise_standard_deviation:depth}
   	\end{subfigure}
   	\hfill
   	\begin{subfigure}[b]{0.245\linewidth}
   		\includegraphics[width=1.0\linewidth, height=0.9\linewidth]{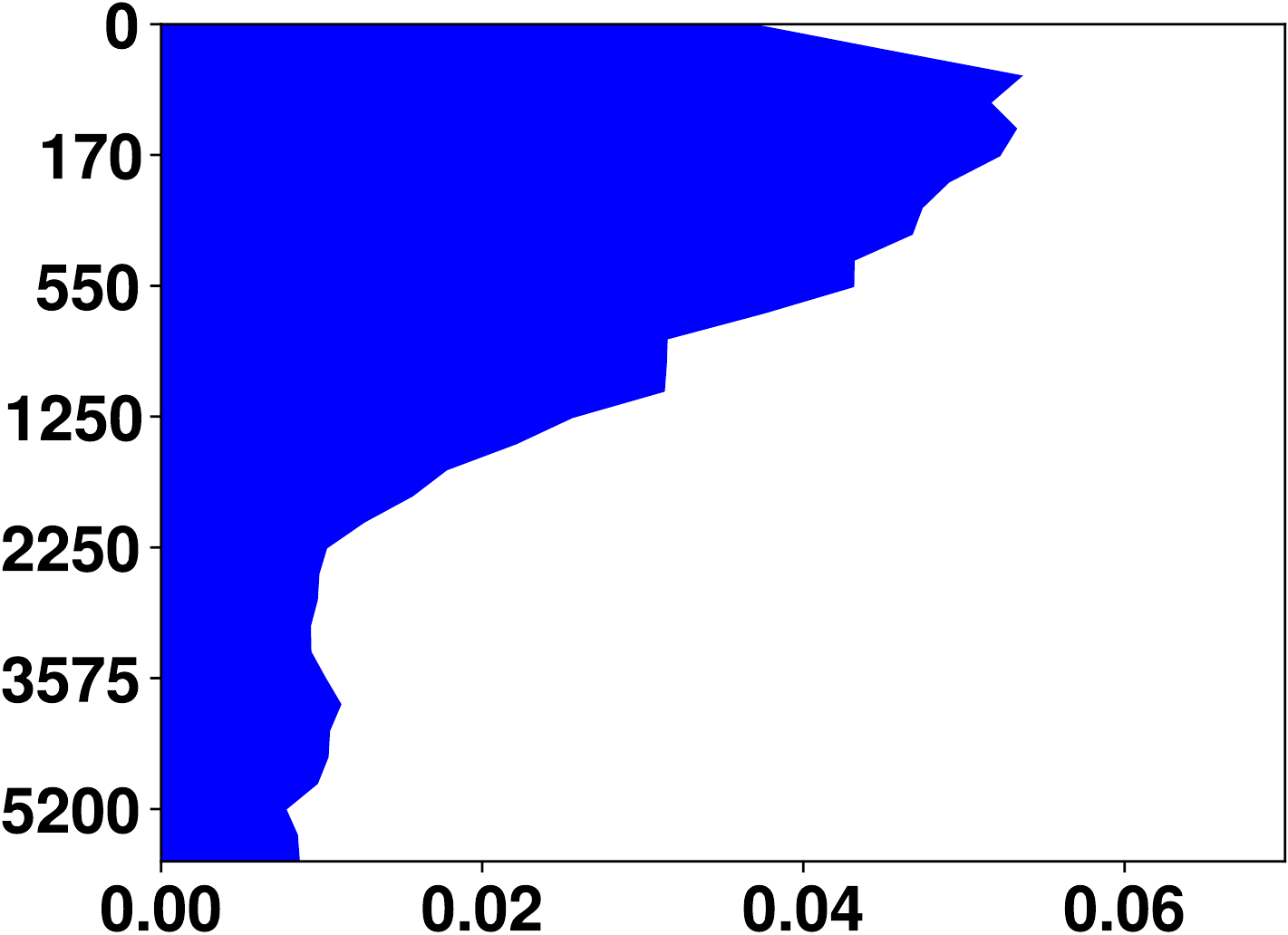}
   		\caption{average monthly change} 
   		\label{fig:noise_standard_deviation:time_diff}
   	\end{subfigure}

   	\begin{subfigure}[b]{0.495\linewidth}
   		\includegraphics[width=1.0\linewidth]{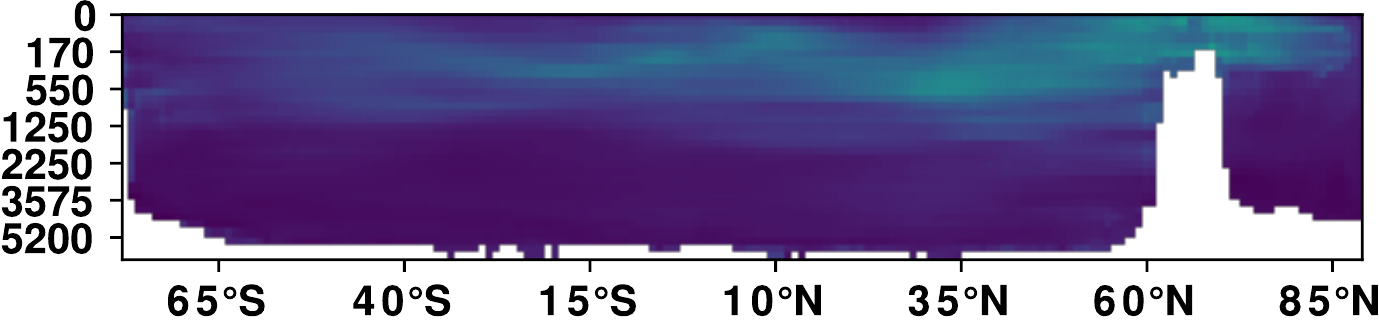}
   		\caption{Pacific Ocean: averaged over time and between 125$\degree$E and 70$\degree$W} 
   		\label{fig:noise_standard_deviation:pacific}
   	\end{subfigure}
    \hfill    
   	\begin{subfigure}[b]{0.495\linewidth}
   		\includegraphics[width=1.0\linewidth]{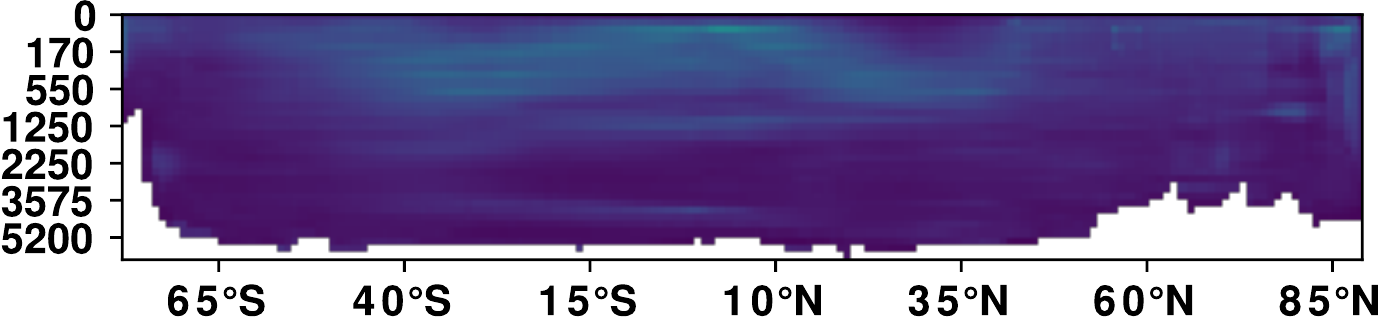}
   		\caption{Atlantic Ocean: averaged over time and between 70$\degree$W and 20$\degree$E}  
   		\label{fig:noise_standard_deviation:atlantic}
   	\end{subfigure}
	\vspace{-0.5em}
\caption{Short scale, i.e. noise, standard deviation of phosphate concentration ($\epsilon$) in ${\rm mmol\,m}^{-3}$.} 
	\label{fig:noise_standard_deviation}
\end{figure}

\begin{figure}[t!]
	\centering
   	\begin{subfigure}[b]{0.495\linewidth}
   		\includegraphics[width=1.0\linewidth]{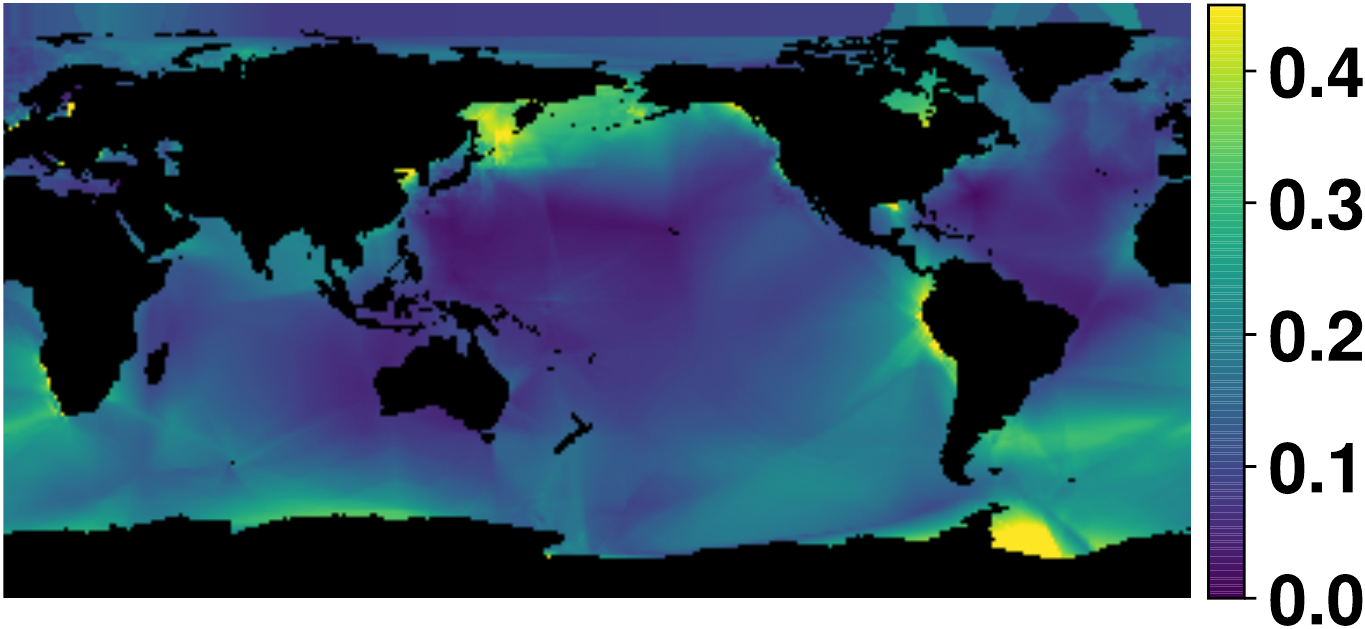}
   		\caption{water surface: averaged over time and 0 to 25 ${\rm m}$ depth} 
   		\label{fig:concentration_standard_deviation:surface}
   	\end{subfigure}
   	\hfill
   	\begin{subfigure}[b]{0.245\linewidth}
   		\includegraphics[width=1.0\linewidth, height=0.9\linewidth]{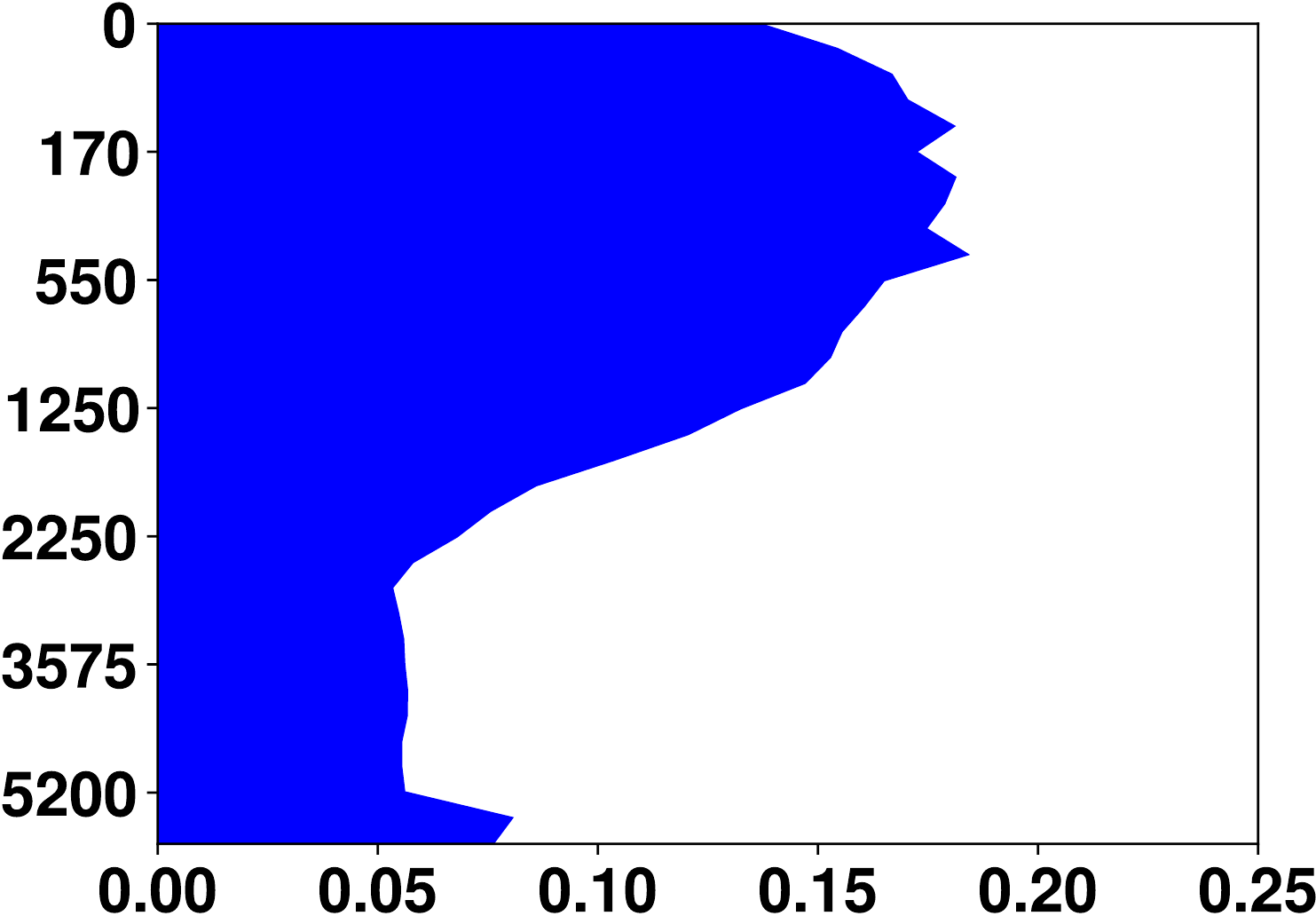}
   		\caption{averaged over all but depth}
   		\label{fig:concentration_standard_deviation:depth}
   	\end{subfigure}
   	\hfill
   	\begin{subfigure}[b]{0.245\linewidth}
   		\includegraphics[width=1.0\linewidth, height=0.9\linewidth]{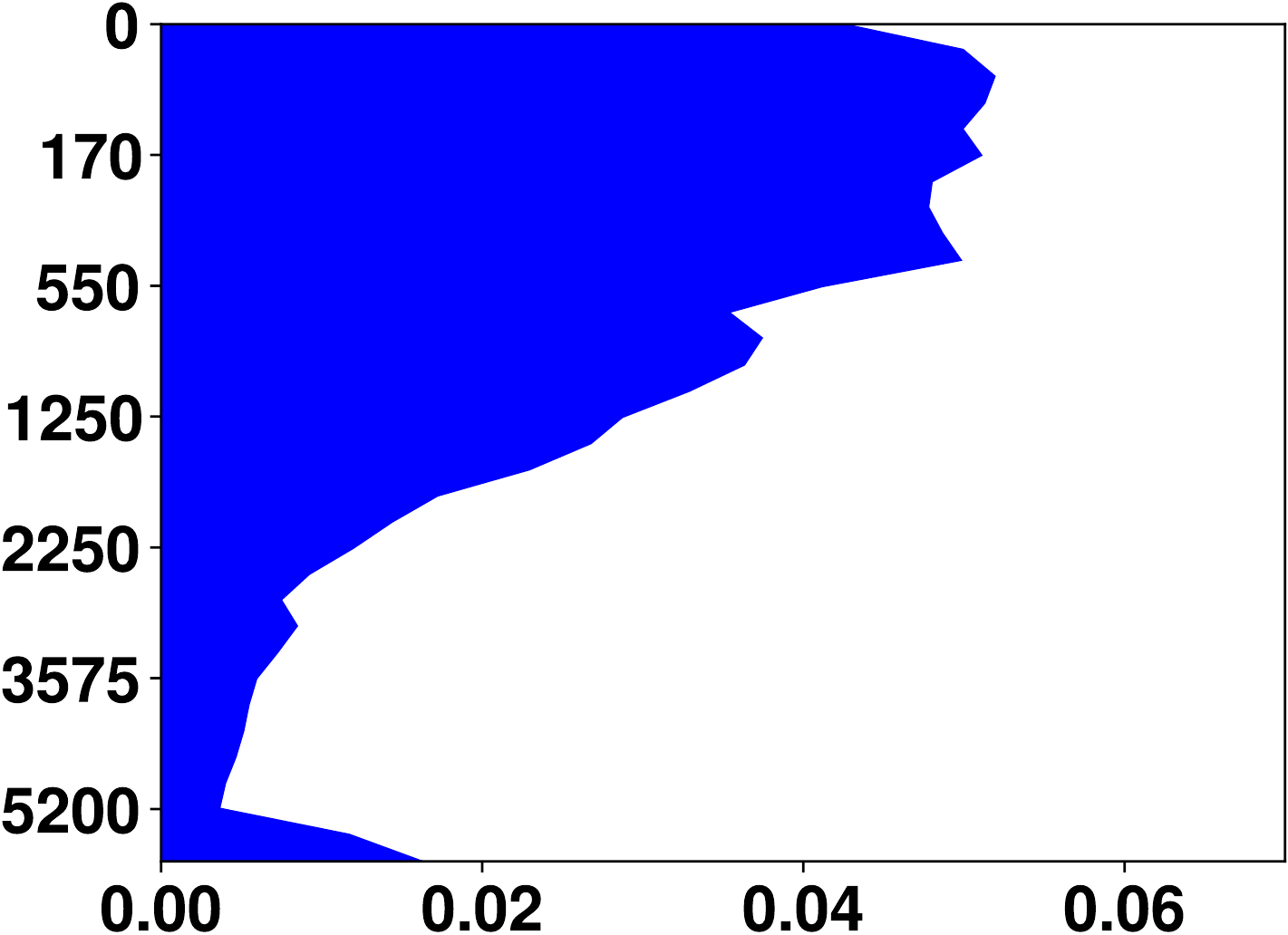}
   		\caption{average monthly change} 
   		\label{fig:concentration_standard_deviation:time_diff}
   	\end{subfigure}

   	\begin{subfigure}[b]{0.495\linewidth}
   		\includegraphics[width=1.0\linewidth]{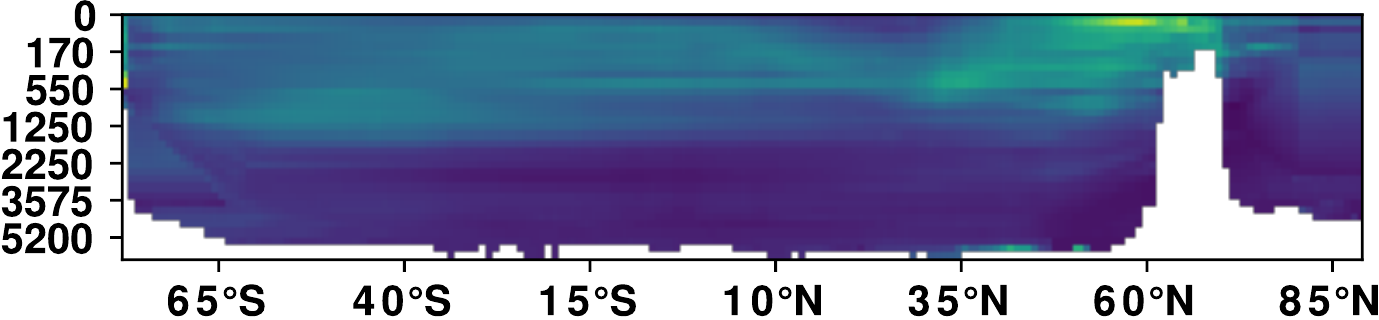}
   		\caption{Pacific Ocean: averaged over time and between 125$\degree$E and 70$\degree$W} 
   		\label{fig:concentration_standard_deviation:pacific}
   	\end{subfigure}
    \hfill    
   	\begin{subfigure}[b]{0.495\linewidth}
   		\includegraphics[width=1.0\linewidth]{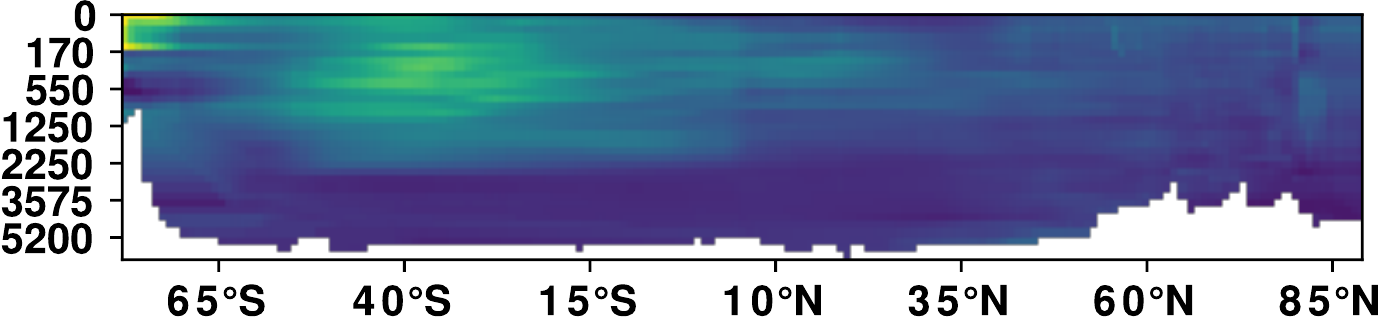}
   		\caption{Atlantic Ocean: averaged over time and between 70$\degree$W and 20$\degree$E}  
   		\label{fig:concentration_standard_deviation:atlantic}
   	\end{subfigure}
	\vspace{-0.5em}
\caption{Climatological standard deviation of phosphate concentration ($\delta$) in ${\rm mmol\,m}^{-3}$.} 
	\label{fig:concentration_standard_deviation}
\end{figure}

As explained in Subsection \ref{subsec:dispersion}, the standard deviations of the measurement results $\eta$ are composed of the standard deviations of the true concentration $\delta$ and the noise $\epsilon$ and are plotted in Figure \ref{fig:noise_standard_deviation} and Figure \ref{fig:concentration_standard_deviation}. The standard deviation of the true concentration describes the climatological variability. In contrast, the standard deviation of the noise covers the short scale variabilities which include the variability within the grid boxes in specific years as well as measurement inaccuracies.

The average estimated standard deviation were 0.10 ${\rm mmol\,m}^{-3}$ for the true concentration and 0.05 ${\rm mmol\,m}^{-3}$ for the noise. Hence, the difference between the measurement results and the climatological mean arose to about two thirds from climatological variabilities and to about one third from short scale variabilities.

The standard deviations of the noise usually were below 0.10 ${\rm mmol\,m}^{-3}$ except around the eastern part of Russia, near the west and south of South America and the west coastal region of Southern Africa, where they were between 0.25 to 0.45 ${\rm mmol\,m}^{-3}$, 0.15 to 0.35 ${\rm mmol\,m}^{-3}$ and 0.10 to 0.30 ${\rm mmol\,m}^{-3}$, respectively, in the upper few hundred meters.

As the standard deviation of the noise includes the variability within a grid box in a specific year, a high standard deviation of the noise indicates that a finer resolution could provide more accurate climatological information and in contrast a low standard deviation indicates that the resolution could be reduced without loosing climatological information. A higher resolution thus seems to make sense for some coastal regions, whereas regions deep in the ocean and far away from coasts could also be resolved with a lower resolution. This indicates that a more non-uniform resolution may be advantageous and should be considered in any further analysis.

The standard deviations averaged over all but the depth are for the noises about half of that for the measurement results. The average monthly absolute change is usually 0.01 ${\rm mmol\,m}^{-3}$ lower for the noises than for the measurement results. In the Pacific Ocean and the Atlantic Ocean averaged over time and longitude only the area around the eastern part of Russia stands out for the noises.

Usually the standard deviations of the true concentrations are approximately 0.05 ${\rm mmol\,m}^{-3}$ lower than the standard deviations of the measurement results except for areas where the standard deviations of the noise is high and thus the difference as well. This relationship arises because the variances of the measurement results are the sum of the variances of the true concentration and the noise.

We also quantified the variabilities of the true concentration $\delta$ and the noise $\epsilon$ by interquartile ranges. However, the plotted results look quite similar to the ones of the standard deviation and are therefore not included.

\subsubsection*{Relative Standard Deviations}

The relative standard deviation is the standard deviation divided by the expected value and therefore allows to assess the variability relative to the typical size. Since the expected values of the noise $\epsilon$ are assumed to be zero, it only makes sense to look at the relative standard deviations of the true concentration $\delta$ and the measurement results $\eta$. The expected value of these two is the climatological mean.

\begin{figure}[t!]
	\centering
   	\begin{subfigure}[b]{0.495\linewidth}
   		\includegraphics[width=1.0\linewidth]{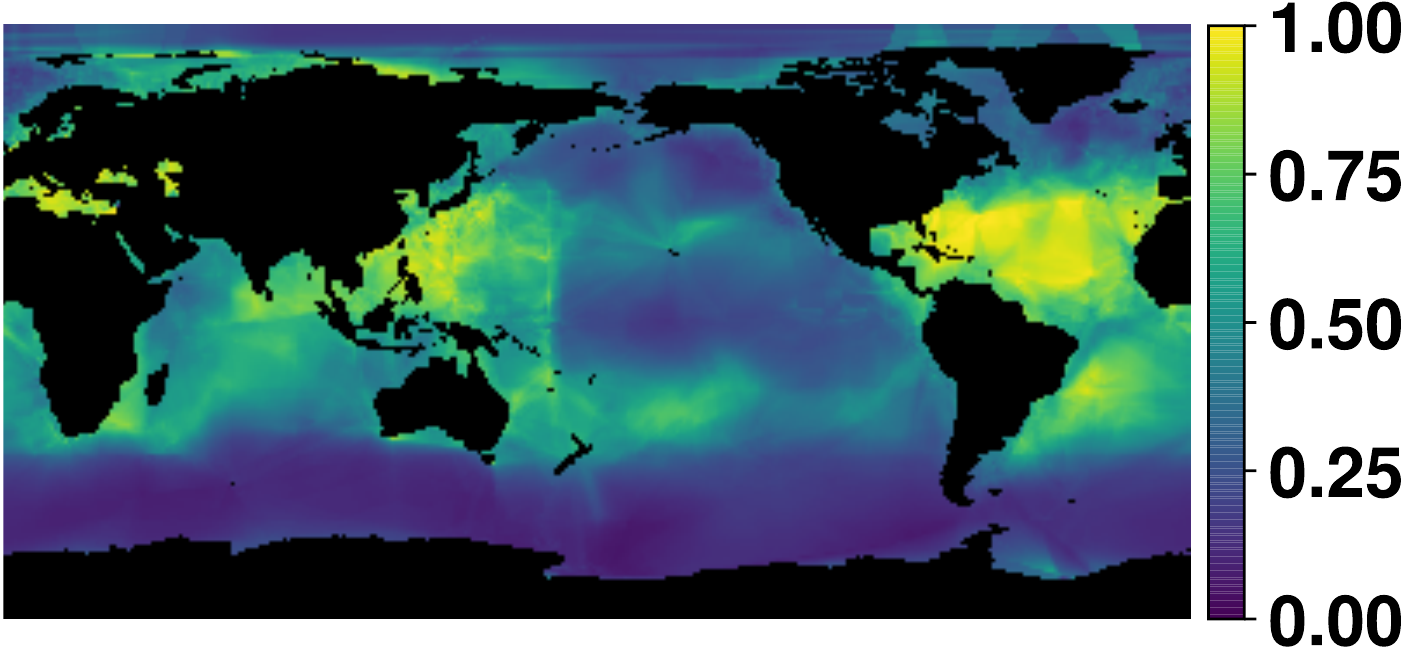}
   		\caption{water surface: averaged over time and 0 to 25 ${\rm m}$ depth}  
   		\label{fig:relative_standard_deviation:surface}
   	\end{subfigure}
   	\hfill
   	\begin{subfigure}[b]{0.245\linewidth}
   		\includegraphics[width=1.0\linewidth, height=0.9\linewidth]{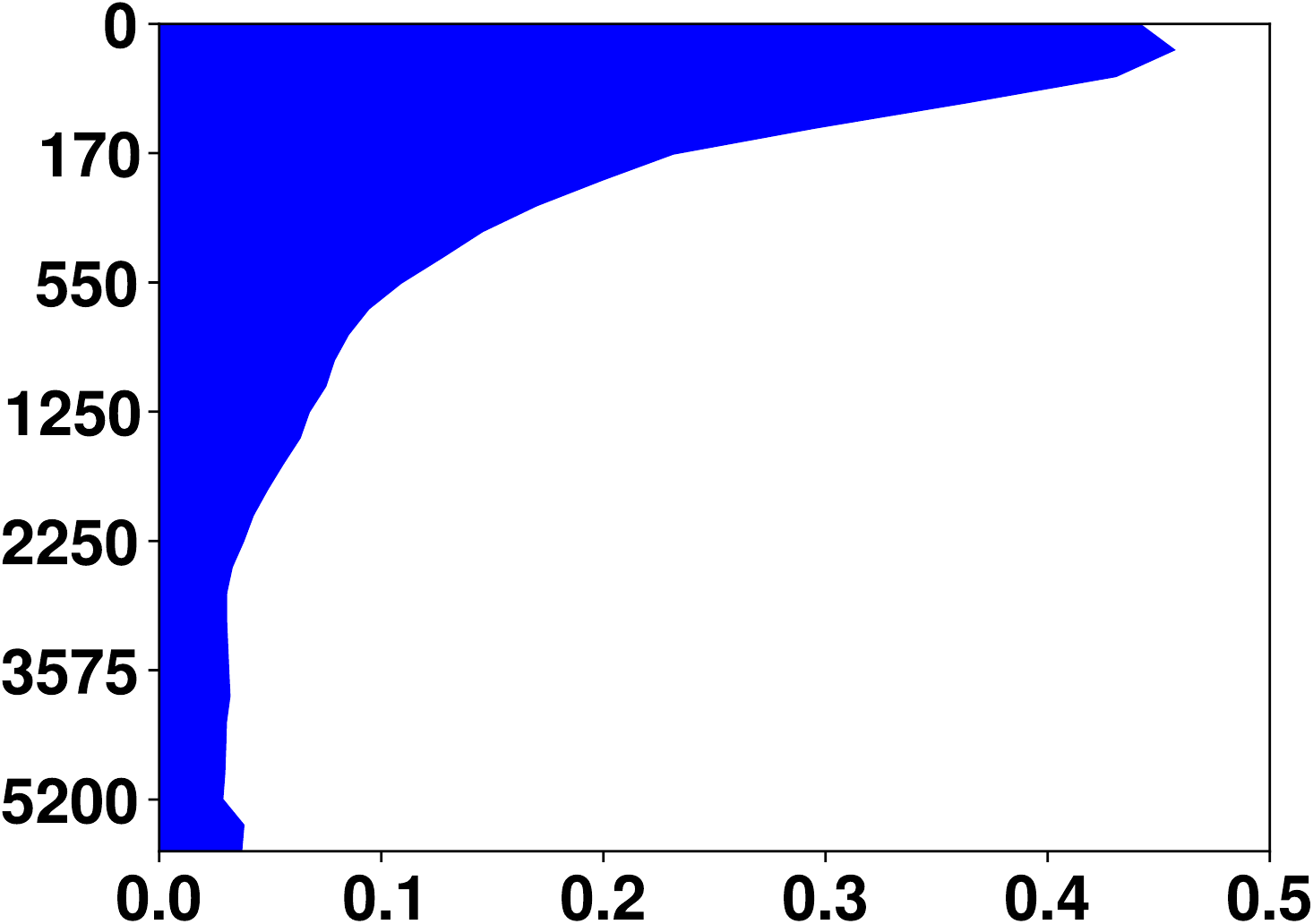}
   		\caption{averaged over all but depth}
   		\label{fig:relative_standard_deviation:depth}
   	\end{subfigure}
   	\hfill
   	\begin{subfigure}[b]{0.245\linewidth}
   		\includegraphics[width=1.0\linewidth, height=0.9\linewidth]{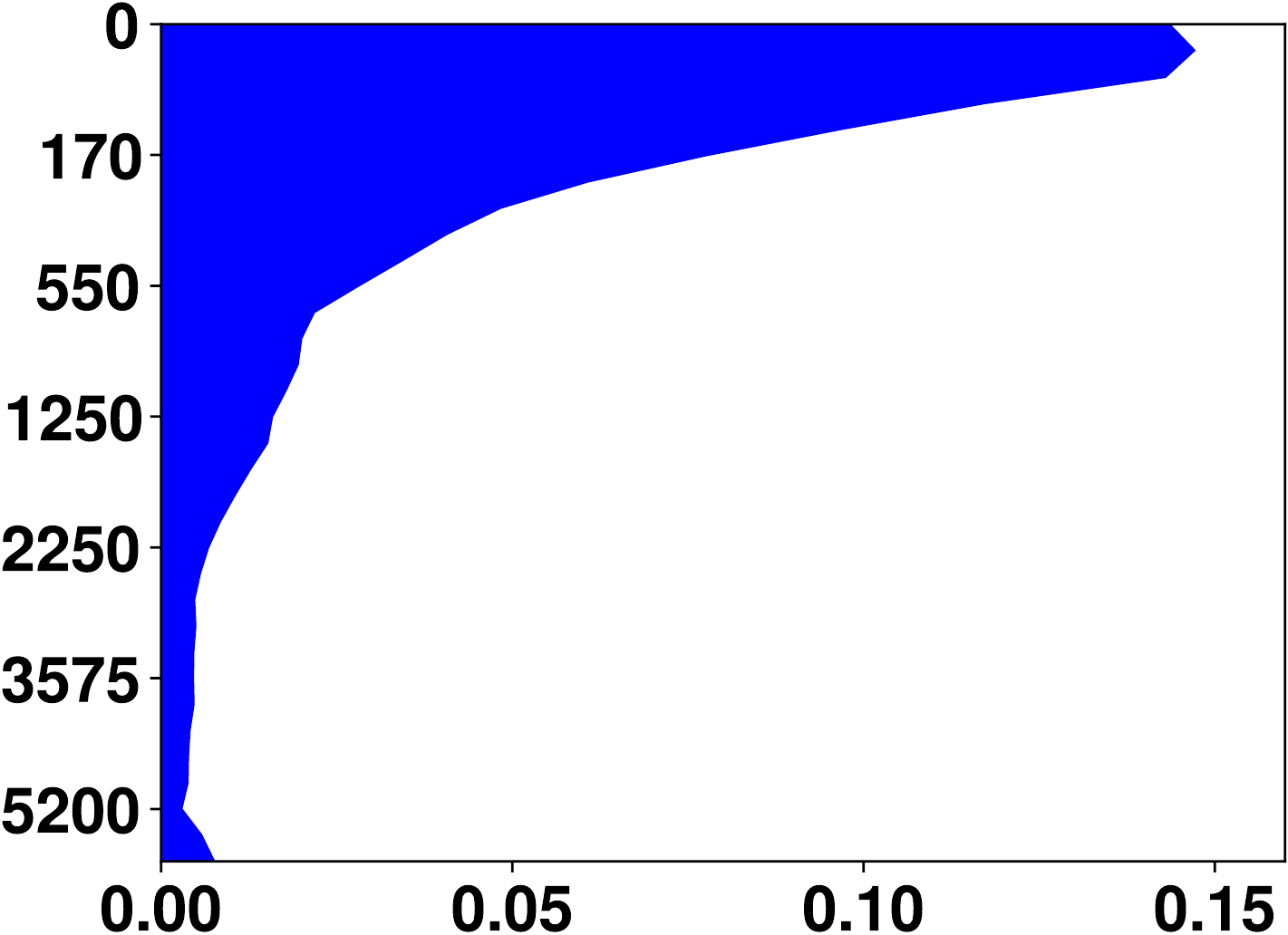}
   		\caption{average monthly change} 
   		\label{fig:relative_standard_deviation:time_diff}
   	\end{subfigure}

   	\begin{subfigure}[b]{0.495\linewidth}
   		\includegraphics[width=1.0\linewidth]{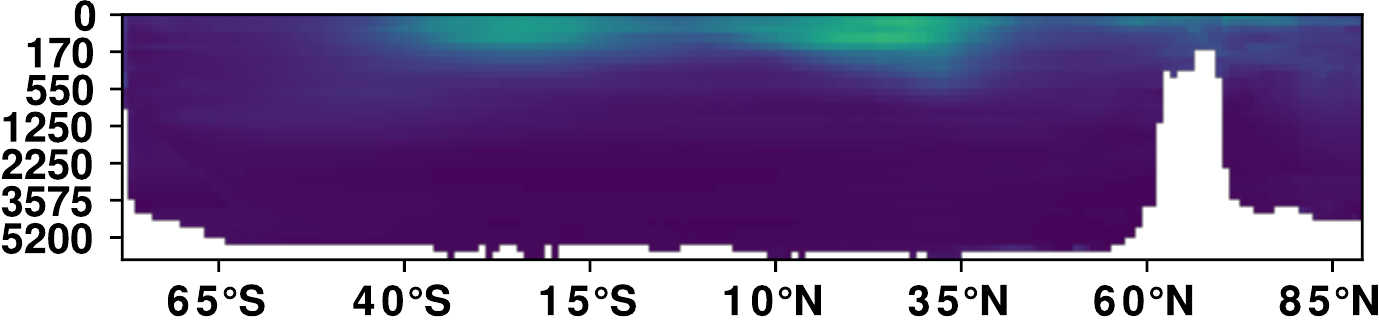}
   		\caption{Pacific Ocean: averaged over time and between 125$\degree$E and 70$\degree$W} 
   		\label{fig:relative_standard_deviation:pacific}
   	\end{subfigure}
    \hfill    
   	\begin{subfigure}[b]{0.495\linewidth}
   		\includegraphics[width=1.0\linewidth]{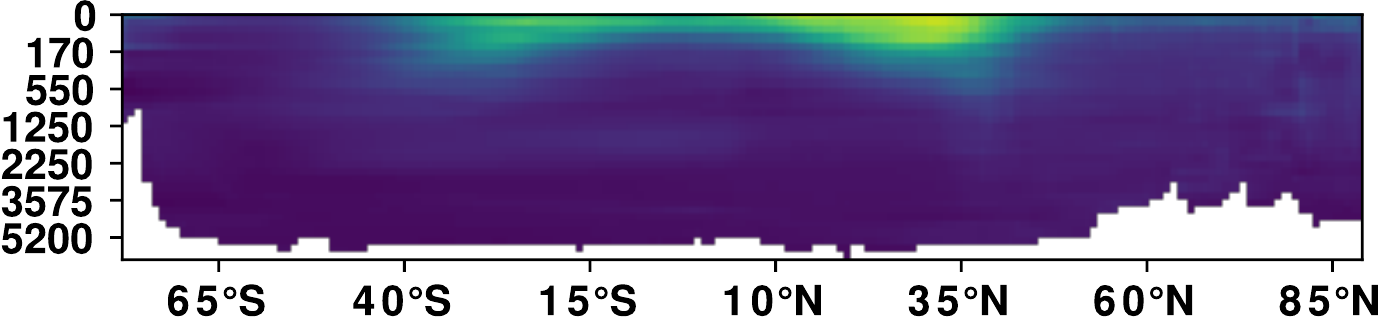}
   		\caption{Atlantic Ocean: averaged over time and between 70$\degree$W and 20$\degree$E}  
   		\label{fig:relative_standard_deviation:atlantic}
   	\end{subfigure}
	\vspace{-0.5em}
\caption{Relative standard deviation of phosphate measurements ($\eta$).} 
	\label{fig:relative_standard_deviation}
\end{figure}

\begin{figure}[t!]
	\centering
   	\begin{subfigure}[b]{0.495\linewidth}
   		\includegraphics[width=1.0\linewidth]{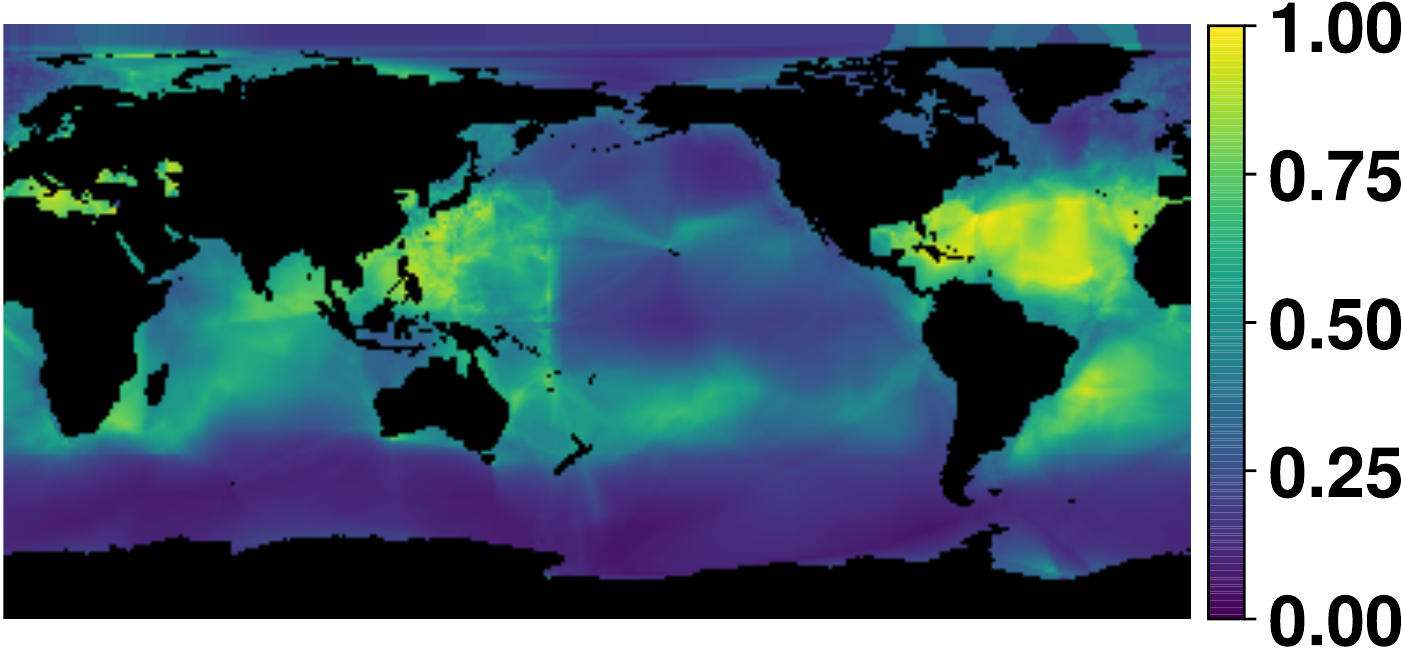}
   		\caption{water surface: averaged over time and 0 to 25 ${\rm m}$ depth}  
   		\label{fig:concentration_relative_standard_deviation:surface}
   	\end{subfigure}
   	\hfill
   	\begin{subfigure}[b]{0.245\linewidth}
   		\includegraphics[width=1.0\linewidth, height=0.9\linewidth]{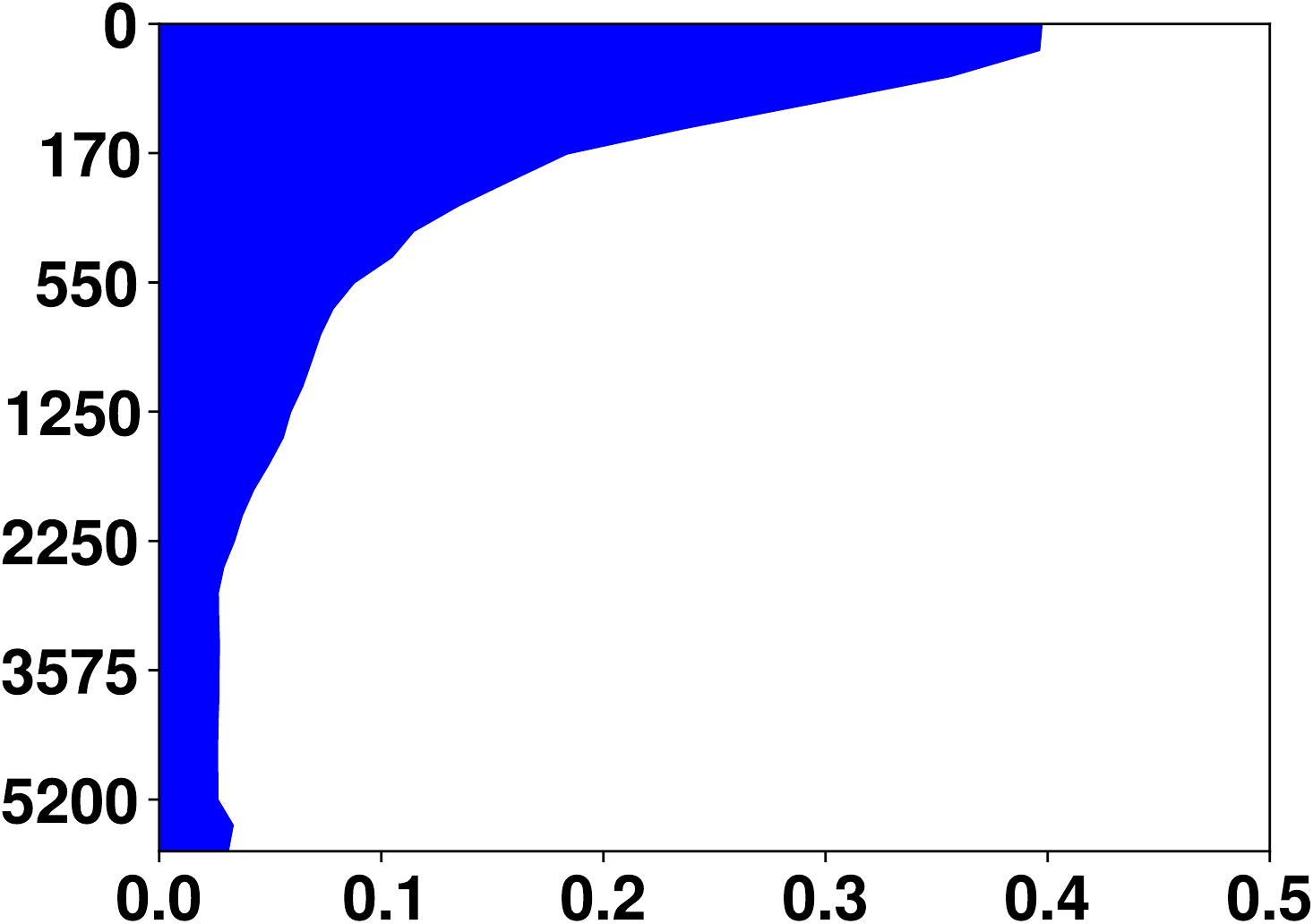}
   		\caption{averaged over all but depth}
   		\label{fig:concentration_relative_standard_deviation:depth}
   	\end{subfigure}
   	\hfill
   	\begin{subfigure}[b]{0.245\linewidth}
   		\includegraphics[width=1.0\linewidth, height=0.9\linewidth]{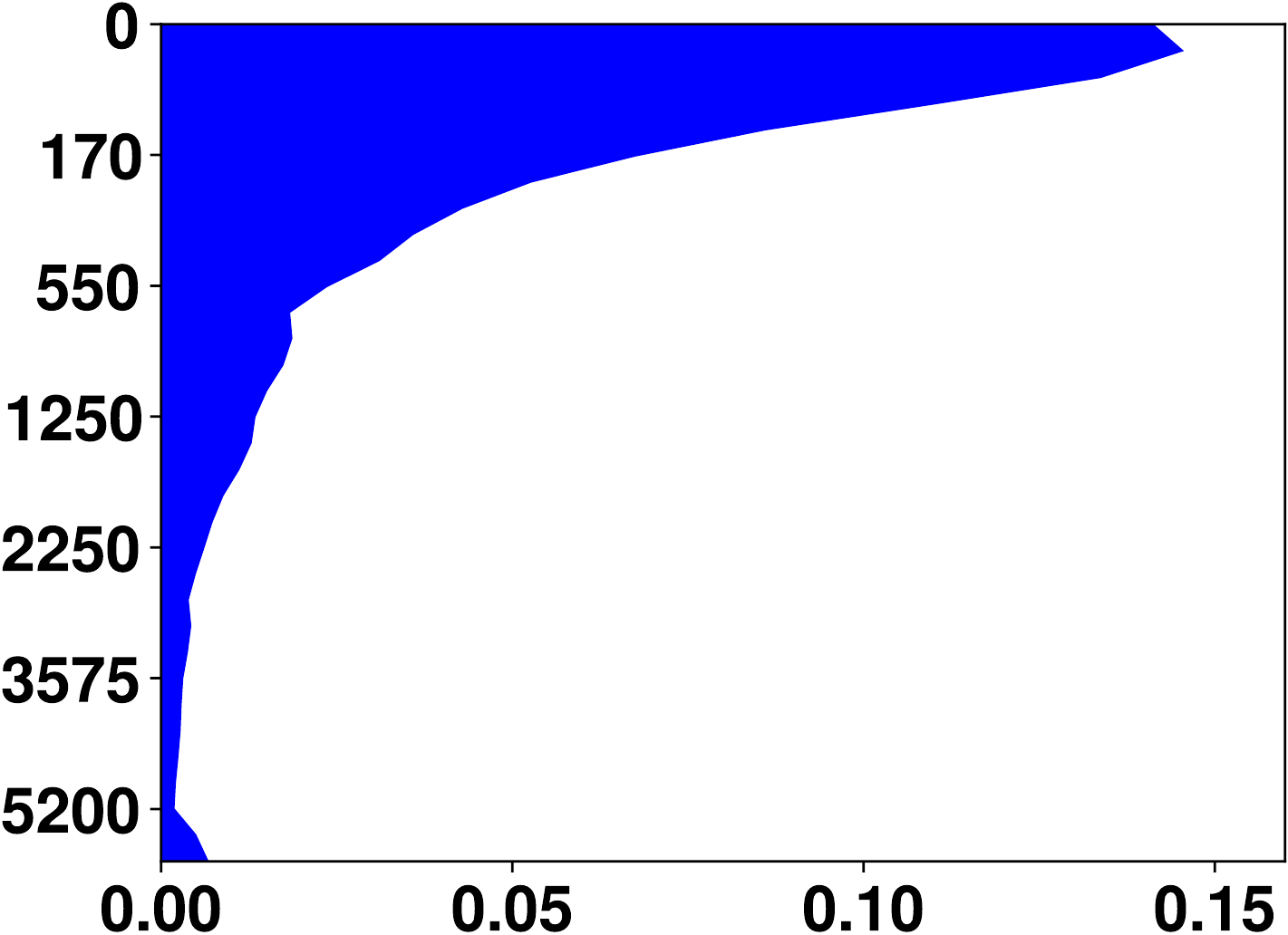}
   		\caption{average monthly change} 
   		\label{fig:concentration_relative_standard_deviation:time_diff}
   	\end{subfigure}

   	\begin{subfigure}[b]{0.495\linewidth}
   		\includegraphics[width=1.0\linewidth]{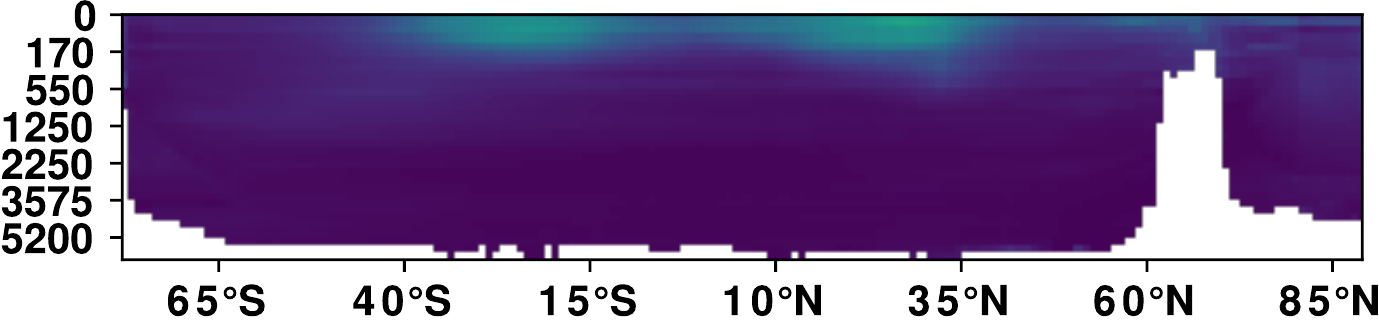}
   		\caption{Pacific Ocean: averaged over time and between 125$\degree$E and 70$\degree$W} 
   		\label{fig:concentration_relative_standard_deviation:pacific}
   	\end{subfigure}
    \hfill    
   	\begin{subfigure}[b]{0.495\linewidth}
   		\includegraphics[width=1.0\linewidth]{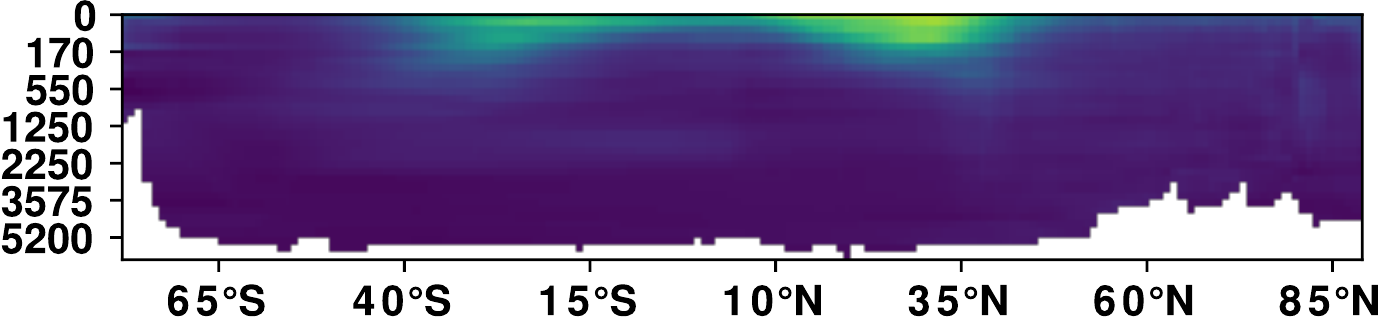}
   		\caption{Atlantic Ocean: averaged over time and between 70$\degree$W and 20$\degree$E}  
   		\label{fig:concentration_relative_standard_deviation:atlantic}
   	\end{subfigure}
	\vspace{-0.5em}
\caption{Relative standard deviation of climatological phosphate concentration ($\delta$).} 
	\label{fig:concentration_relative_standard_deviation}
\end{figure}

The standard deviations for the true concentrations and the measurement results are quite similar as explained earlier. For this reason, the relative standard deviations of the true concentration and the measurement results are quite similar, too, as shown in Figure \ref{fig:relative_standard_deviation} and \ref{fig:concentration_relative_standard_deviation}. The only differ notable near the eastern part of Russia where the standard deviations of the noise are very high.

The average relative standard deviation of the measurement results is 0.07 and that of the true concentration is 0.06. The highest values of the relative standard deviations of the measurement results are mostly between 0.4 and 1. This is mainly near the surface between 40$\degree$S and 45$\degree$N everywhere expect in the east of the Pacific Ocean. Elsewhere near the surface, the relative standard deviations are usually below 0.4 and in the Southern Ocean even below 0.2. Usually, high relative standard deviations result from low climatological means as well as low relative standard deviations from high climatological means.

The average relative standard deviation of the measurement results depending on the depth is 0.45 near the surface and decreases fast with growing depth. At 500 ${\rm m}$ depth it is 0.13, and below 2000 ${\rm m}$ depth it is below 0.05. The average absolute difference after one month is 0.15 near the surface and decreases in a similar way. At 500 ${\rm m}$ depth it is 0.04, and below 2000 ${\rm m}$ depth it is below 0.01. Hence, the climatological variability is negligible after a few hundred meters of depth. 

In the Pacific Ocean and the Atlantic Ocean only the previously mentioned areas between 40$\degree$S and 45$\degree$N and with a depth up to 200 ${\rm m}$ stand out with higher values.

The quartile coefficients of dispersion of the true concentration $\delta$ and the measurement results $\eta$ were also used to quantify its variabilities as an alternative to its relative standard deviations. However, the results look quite similar and are thus not illustrated here.

\subsection{Statistical Dependencies} \label{subsec: po4: statistical dependencies}

We quantified the statistical dependencies regarding $\delta$, $\epsilon$ and $\eta$ with covariances and correlations. It suffices to estimate the covariance or correlation of the true concentration $\delta$. The other covariances and correlations result from these estimates (compare Subsection \ref{subsection: statistical dependencies}).

\subsubsection*{Pointwise Correlations}

The calculated estimates depend on the required number of years where measurement results are available. We set up three estimates: The first covers measurements within at least 35 years, second within at least 40 years and for the final at least 45 years resulting in $1.9 \times 10^{10}$, $3.3 \times 10^9$ and $2.2 \times 10^8$ pointwise estimates, respectively. Estimating the correlation, we assumed the standard deviation of the noise $\epsilon$ to be at least 0.1 ${\rm mmol\,m}^{-3}$, which corresponds to a coarse measurements accuracy. Otherwise some unrealistic low sample standard deviations would compromise our correlation estimation.

\begin{figure}[t]
	\centering
	\begin{subfigure}[b]{0.32\linewidth}
		\includegraphics[width=1.0\linewidth,height=0.8\linewidth]{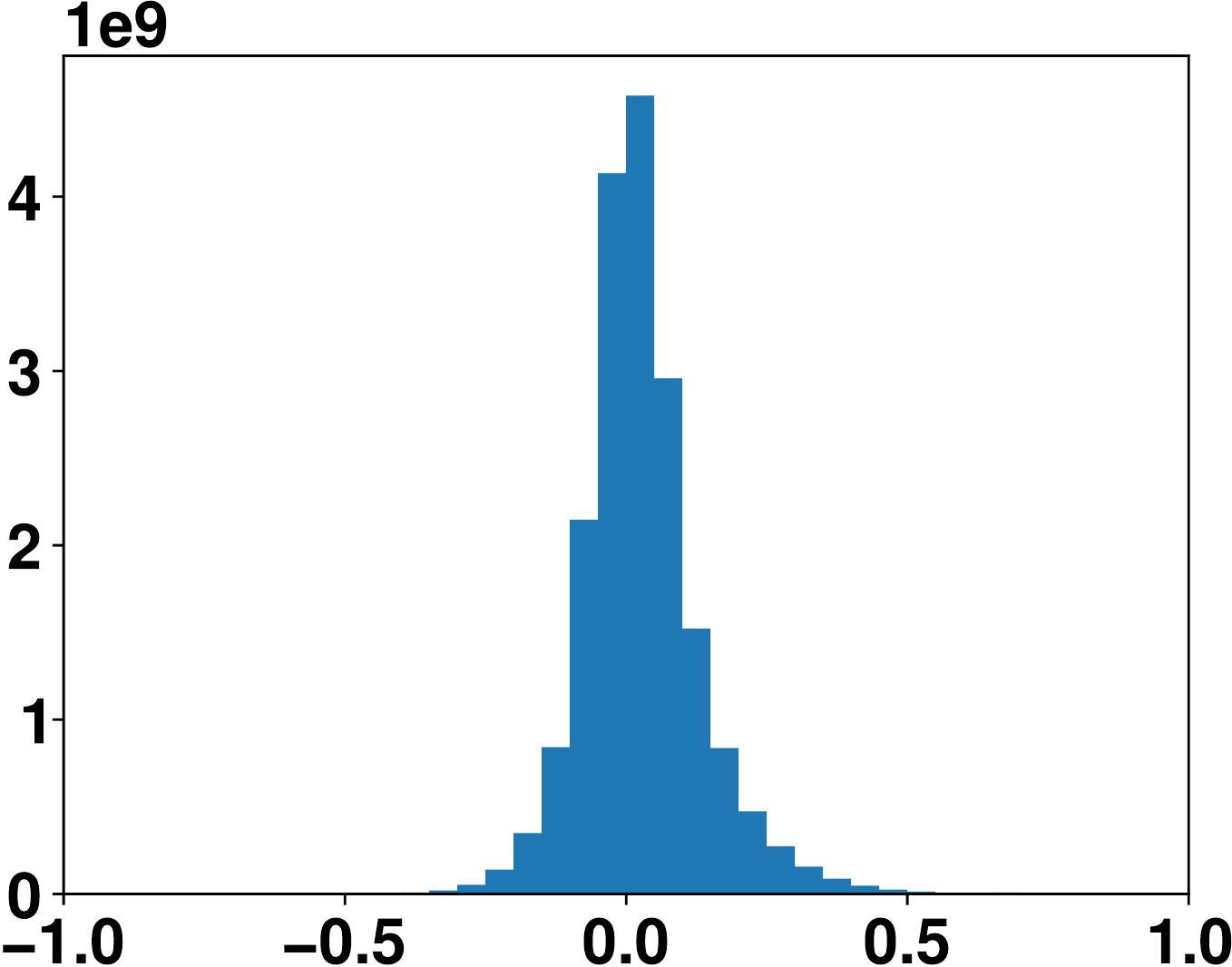}
		\caption{35 years required} 
		\label{fig:correlation:histogram:35}
	\end{subfigure}
	\hfill
	\begin{subfigure}[b]{0.32\linewidth}
		\includegraphics[width=1.0\linewidth,height=0.8\linewidth]{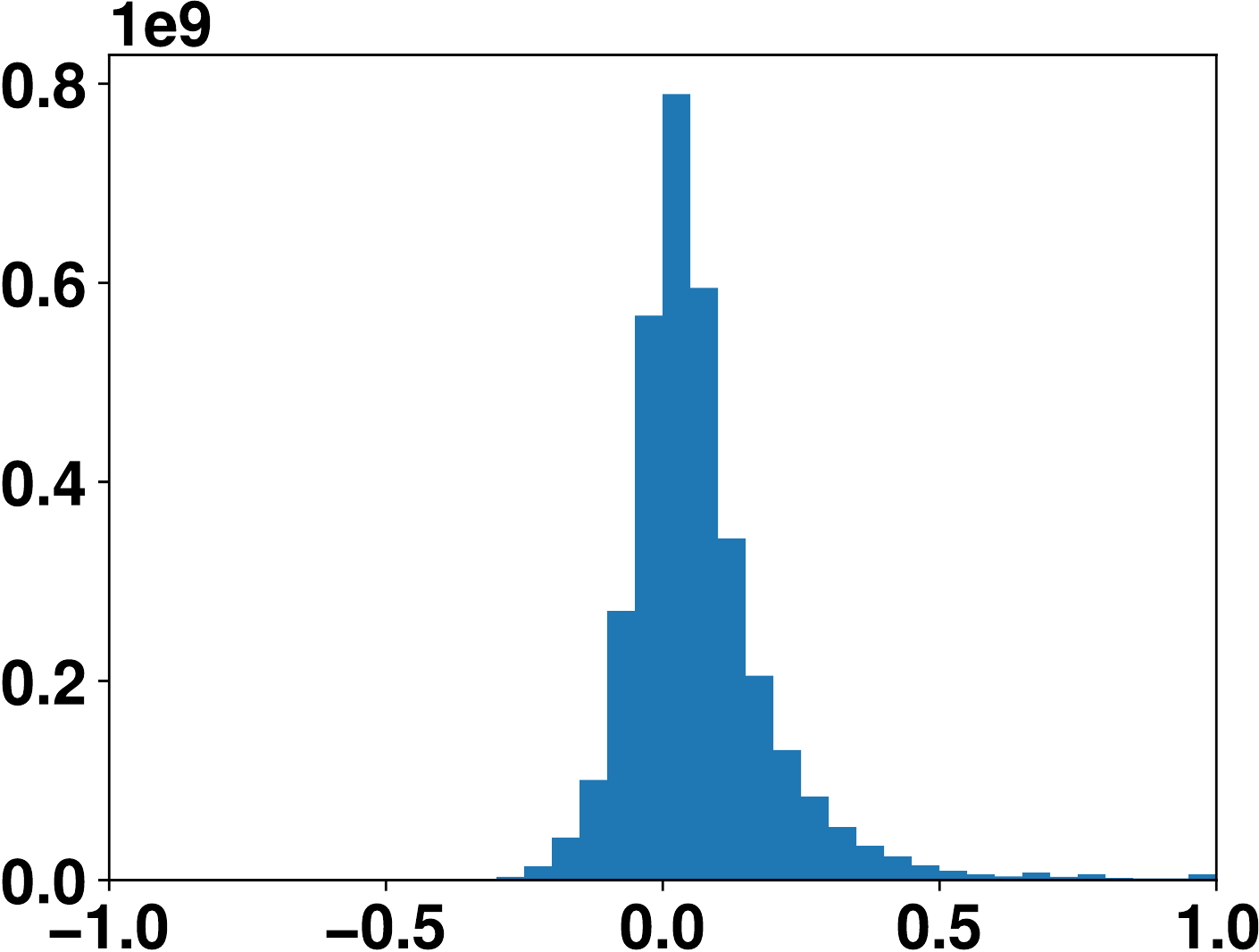}
		\caption{40 years required} 
		\label{fig:correlation:histogram:40}
	\end{subfigure}
	\hfill
	\begin{subfigure}[b]{0.32\linewidth}		
		\includegraphics[width=1.0\linewidth,height=0.8\linewidth]{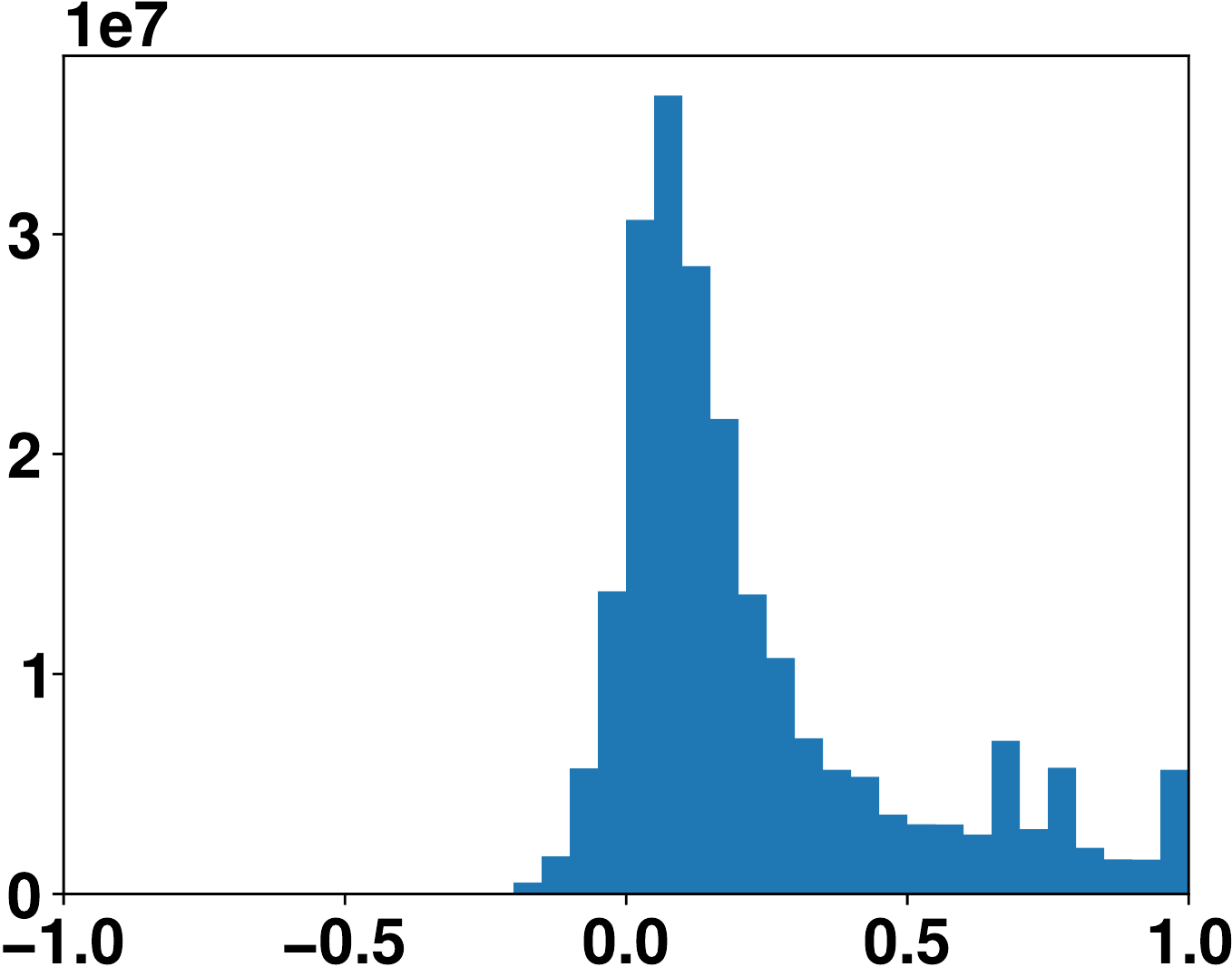}
		\caption{45 years required} 
		\label{fig:correlation:histogram:45}
	\end{subfigure}
	\vspace{-0.5em}
	\caption{Number of estimated correlations. (0.05 used as bin size in the histograms)} 
	\label{fig:correlation:histogram}
\end{figure}

The number of estimated correlation values is shown in Figure \ref{fig:correlation:histogram} where estimates less than 0.01 in absolute value were not considered. The figure shows clearly that most of the estimated correlations are small in amount supporting our approach to assume that not estimated correlations are zero. High absolute values in the estimates are rare. Positive estimates are slightly more frequent than negative ones. Of course, these results depend on the spatial and temporal distribution of the measurement points and would possibly look different with a more uniform distribution of the measurement points.

\subsubsection*{Correlation Matrix}

To generate a well-conditioned positive definite correlation matrix, we considered 4.1 million measurement points.

\begin{figure}[H]
	\centering
   	\hfill
	\begin{subfigure}[b]{0.495\linewidth}
        \centering
		\includegraphics[width=0.55\linewidth,height=0.55\linewidth]{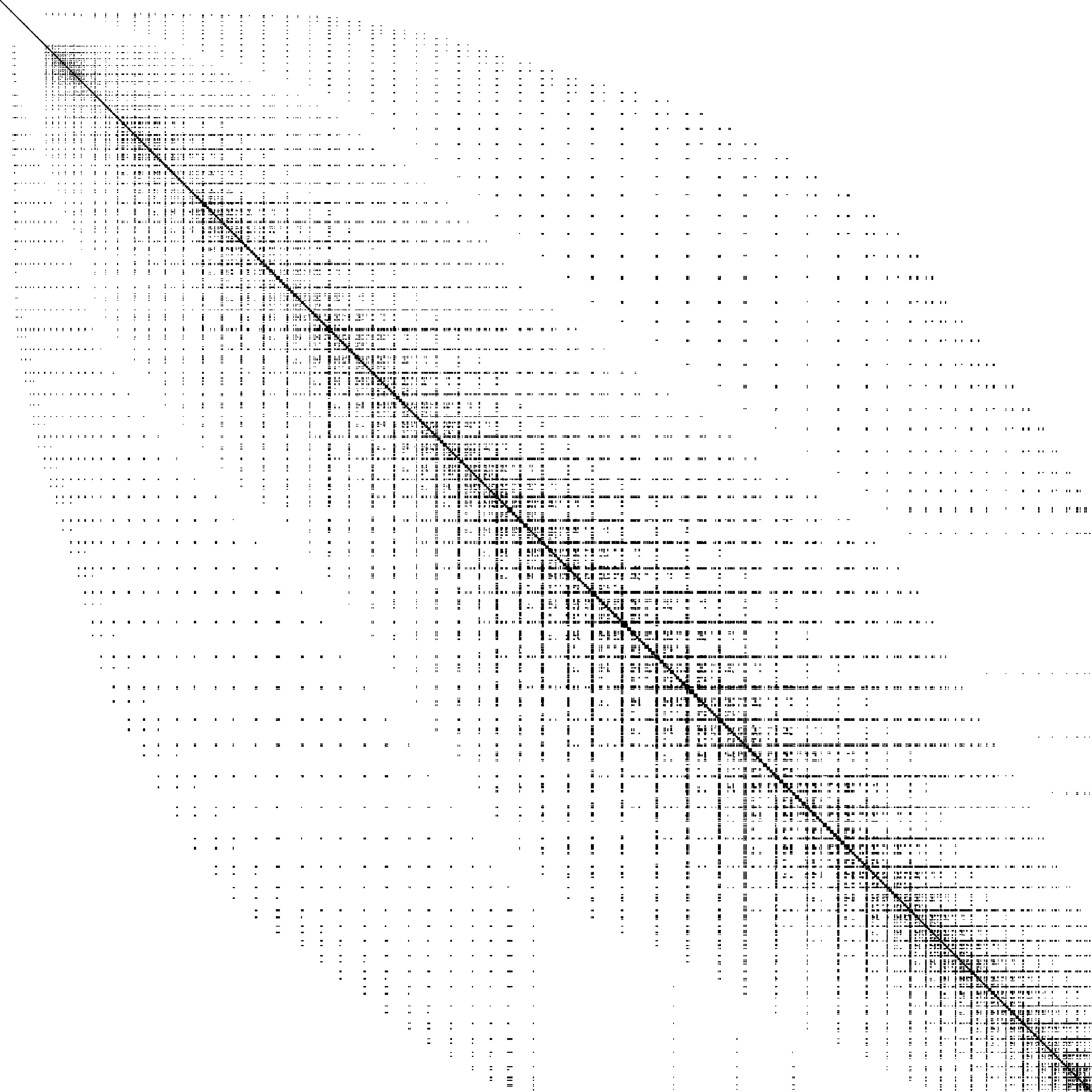}
		\caption{unpermuted} 
		\label{fig:correlation:sparsity:unpermuted}
   	\end{subfigure}
	\hfill
	\begin{subfigure}[b]{0.495\linewidth}	
        \centering	
		\includegraphics[width=0.55\linewidth,height=0.55\linewidth]{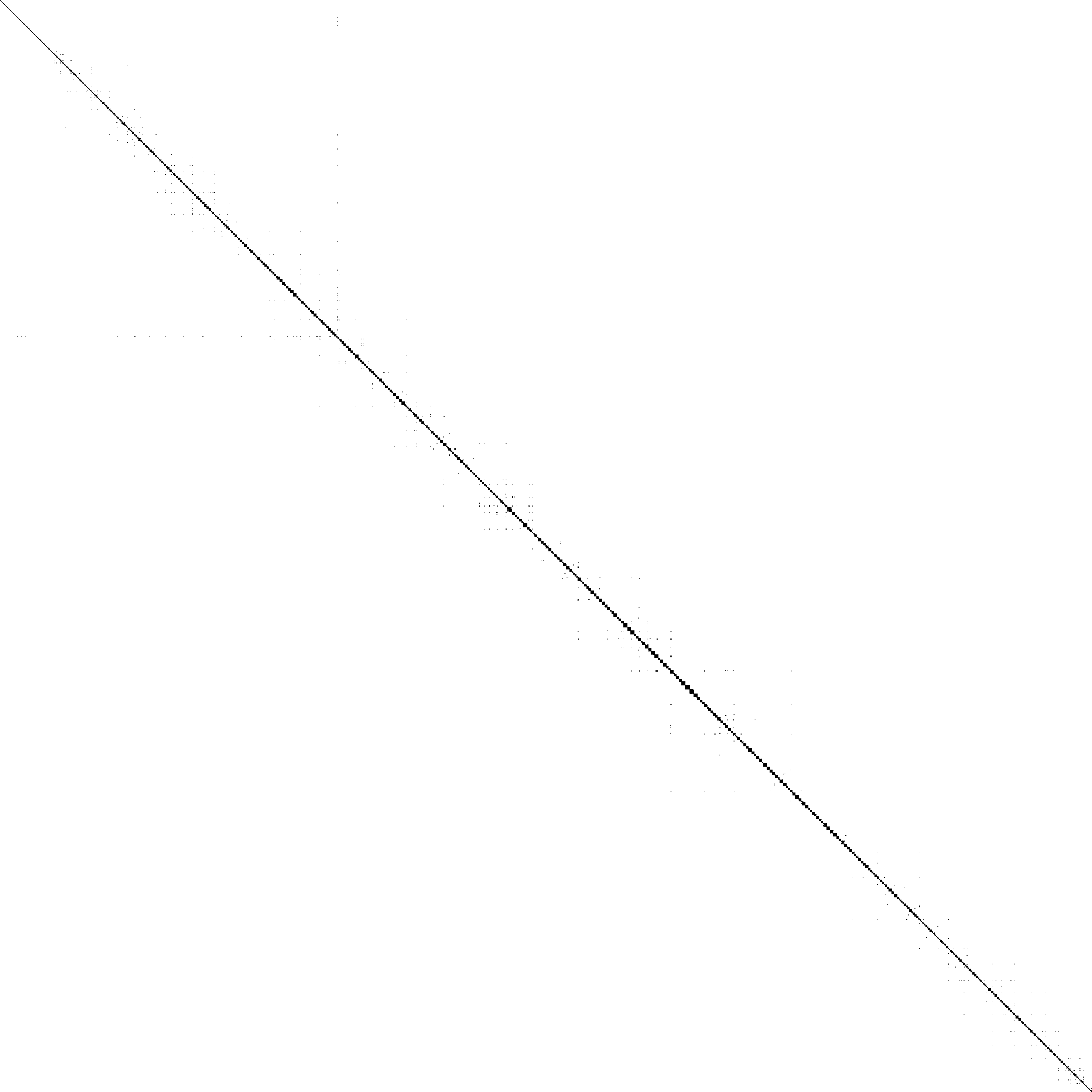}
		\caption{permuted} 
		\label{fig:correlation:sparsity:permuted}
	\end{subfigure}
   	\hfill
	\vspace{-0.5em}
	\caption{Sparsity pattern of correlation matrix.} 
	\label{fig:correlation:sparsity}
\end{figure}

In average, every estimate was reduced in absolute value by 0.078 to generate this matrix. In total 67\% of the estimates were modified. By saving the $LDL^\top$ decomposition instead of the generated correlation matrix itself, the amount of entries could be reduced by 36\%. The sparsity pattern of the unmodified correlation matrix and the permuted well-conditioned correlation matrix are plotted in Figure \ref{fig:correlation:sparsity}.

\subsubsection*{Correlations Dependencies on Distances}

We analyzed if the estimated correlation depends solely on the distance between the related measurement points (compare Subsection \ref{subsection: statistical dependencies}).

We sorted all estimated pointwise correlations in groups in which the associated measuring points differ by the same value. For each of these group with at least ten correlations, we calculated the interquartile range, resulting in more than $3 \times 10^5$ values in total. If the estimated correlations would depend only on the distance between the associated measuring points, most of these interquartile ranges must be close to zero. However, $50\%$ are greater or equal than $0.05$, $25\%$ are greater than $0.10$, and $5\%$ are even greater than $0.20$. Hence, it is unlikely that the estimated correlations can be described by a function depending only on the distance between the corresponding measurement points when using 0.05 as a threshold or even 0.10 as a less restrictive threshold.

\begin{figure}[H]
	\centering
   	\begin{subfigure}[b]{0.24\linewidth}
      	\includegraphics[width=1.0\linewidth,height=0.8\linewidth]{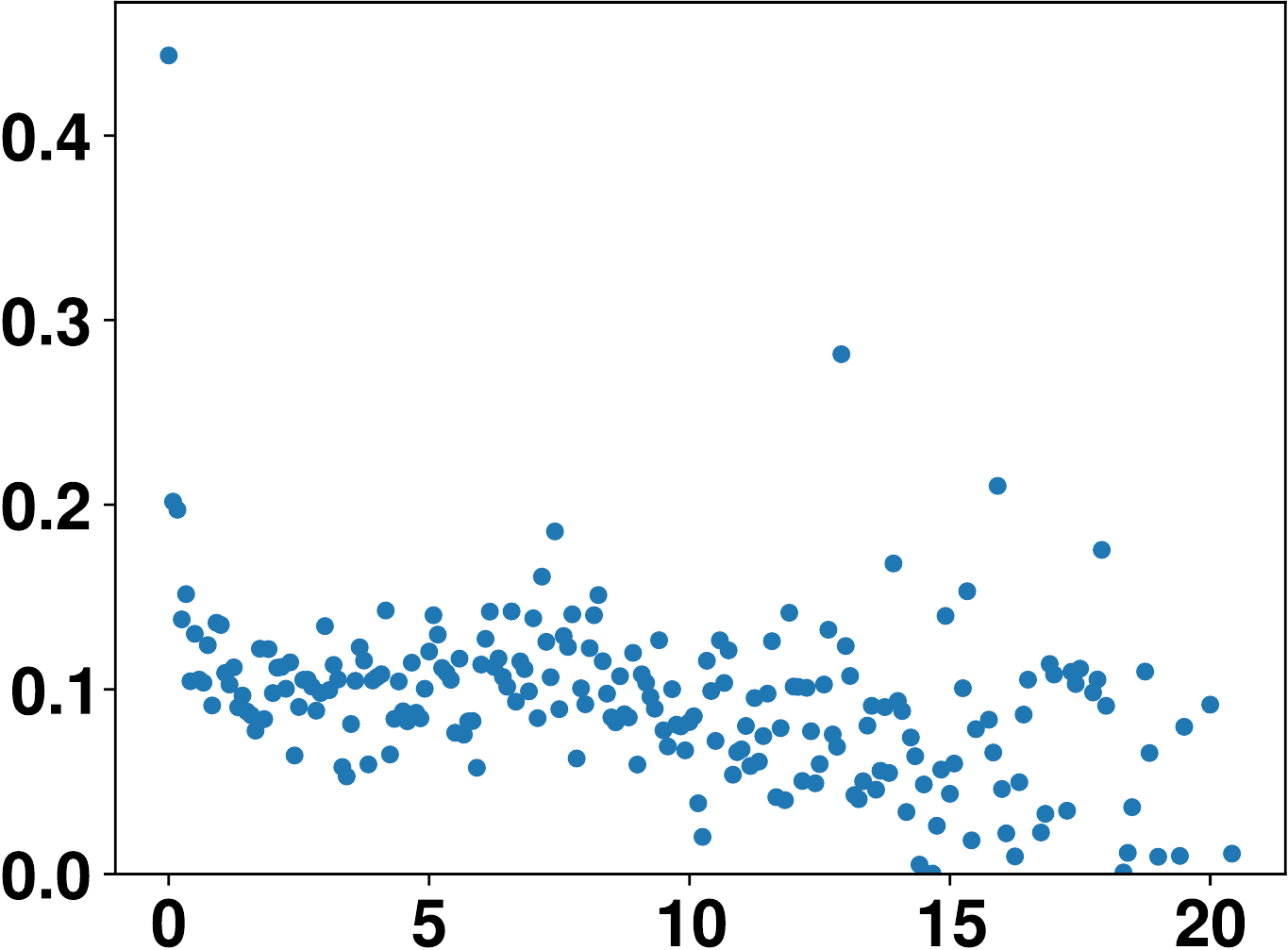}
   		\caption{time} 
   		\label{fig:correlation:iqr:1}
   	\end{subfigure}
   	\hfill
   	\begin{subfigure}[b]{0.24\linewidth}
      	\includegraphics[width=1.0\linewidth,height=0.8\linewidth]{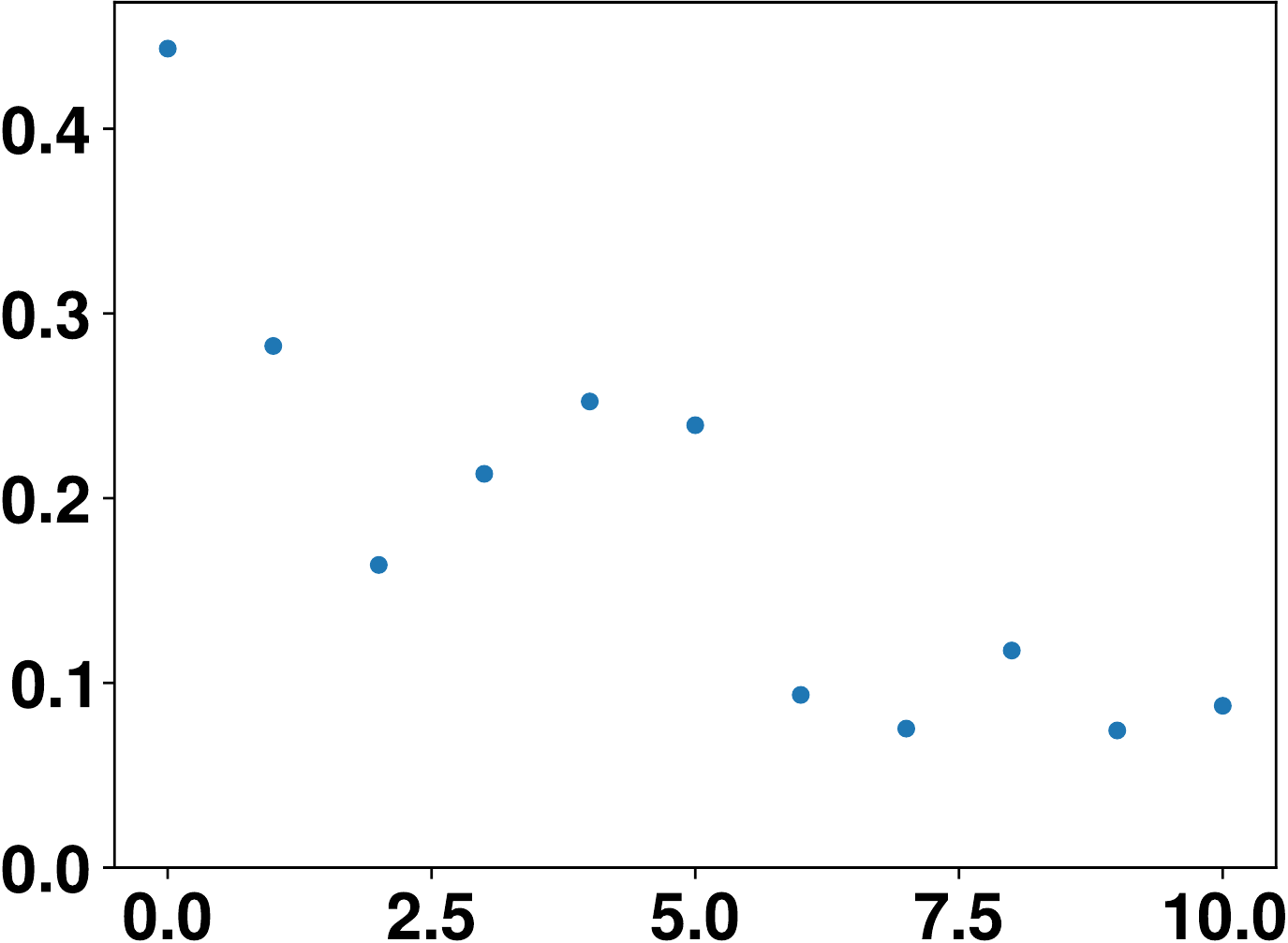}
   		\caption{longitude} 
   		\label{fig:correlation:iqr:2}
   	\end{subfigure}
   	\hfill
   	\begin{subfigure}[b]{0.24\linewidth}
      	\includegraphics[width=1.0\linewidth,height=0.8\linewidth]{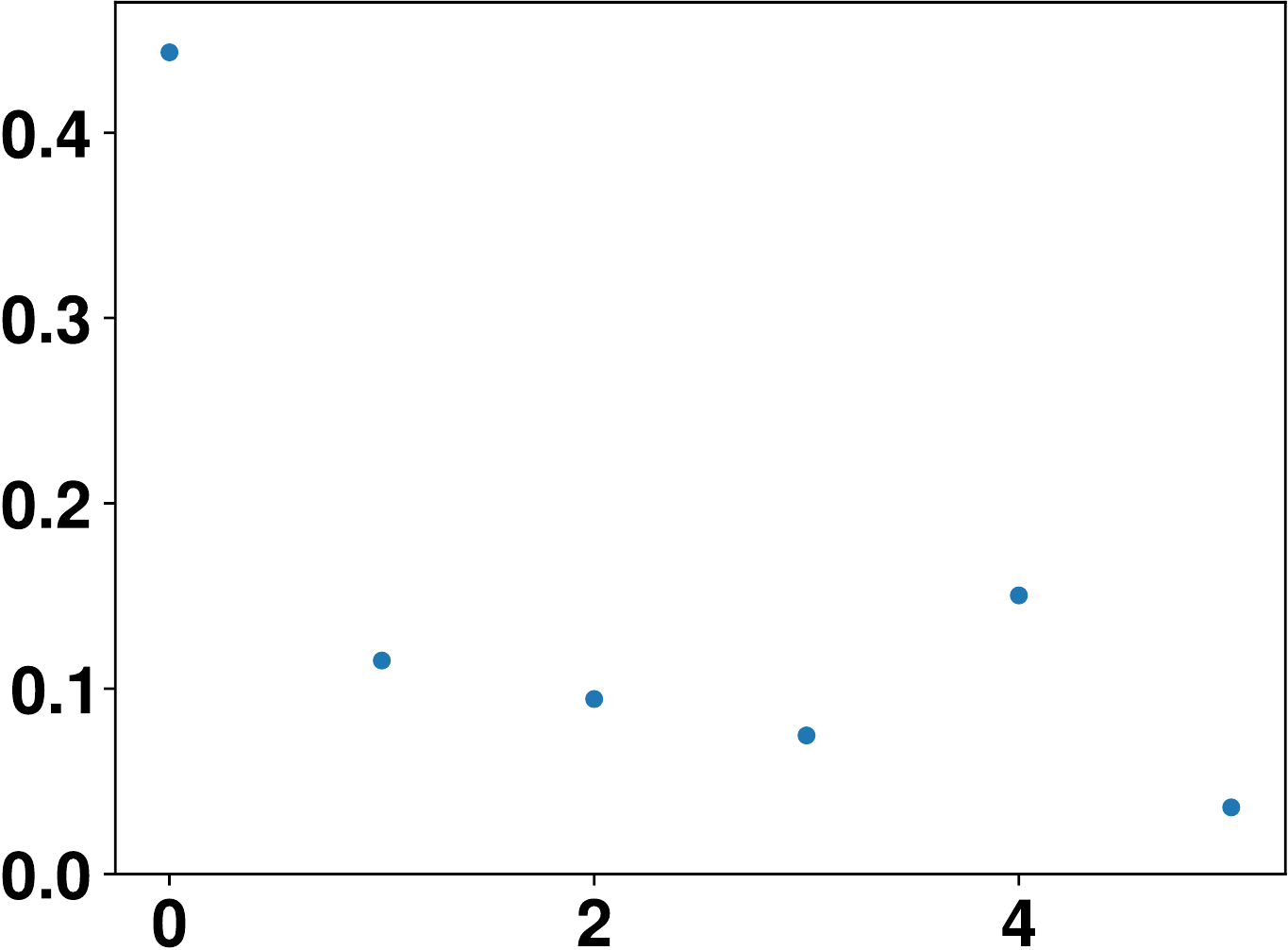}
   		\caption{latitude} 
   		\label{fig:correlation:iqr:3}
   	\end{subfigure}
   	\hfill
   	\begin{subfigure}[b]{0.24\linewidth}
      	\includegraphics[width=1.0\linewidth,height=0.8\linewidth]{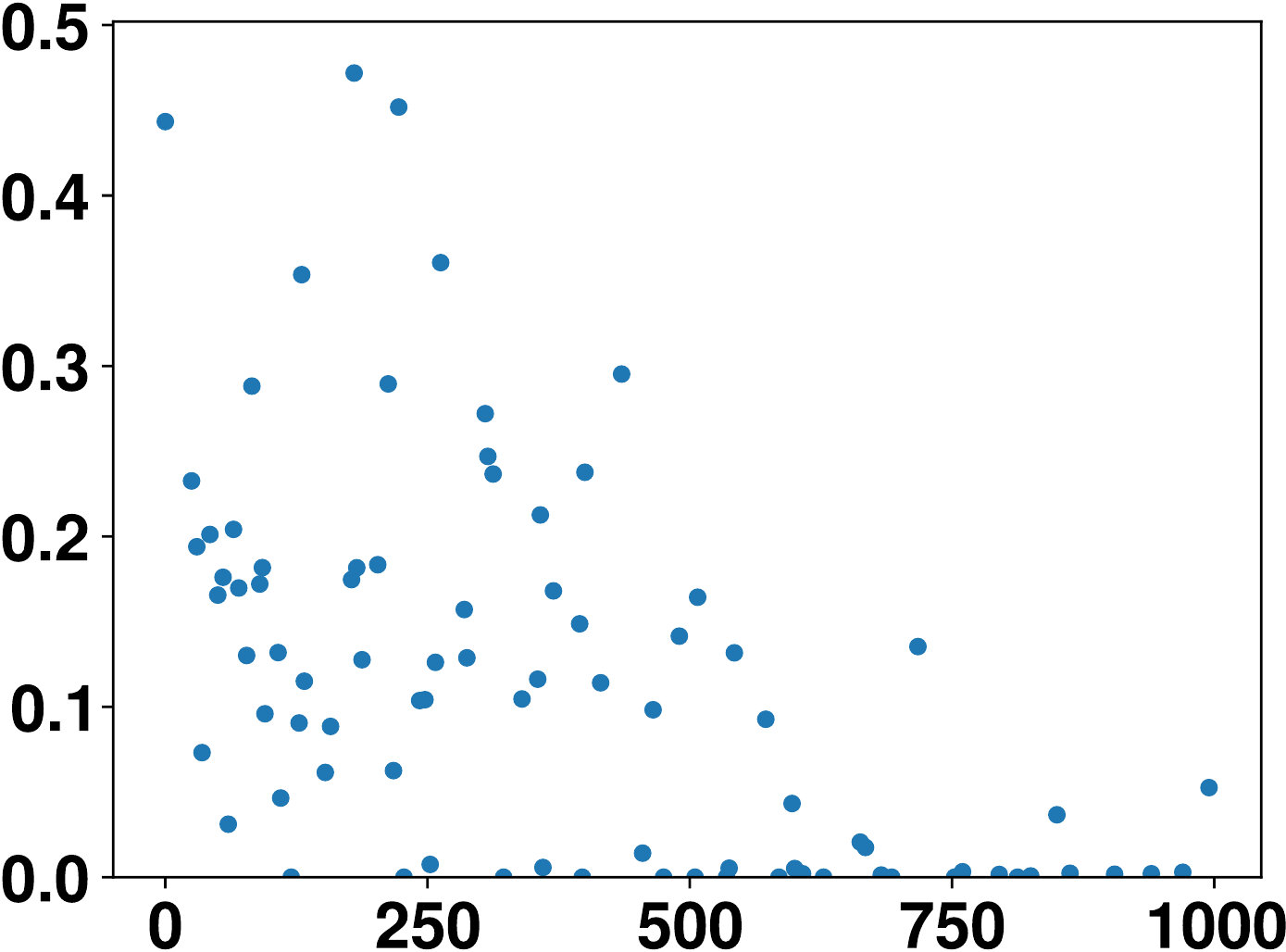}
   		\caption{depth} 
   		\label{fig:correlation:iqr:4}
   	\end{subfigure}
   	\vspace{-0.5em}
   	\caption{Interquartile ranges of estimated correlations associated with measuring points which differ in a single direction.} 
   	\label{fig:correlation:iqr}
\end{figure}

The calculated interquartile ranges, associated with measuring points which differ in a single direction, are plotted in Figure \ref{fig:correlation:iqr}. They show, especially with regard to depth, a solely distance-related dependence is unlikely.

\begin{figure}[H]
   	\begin{subfigure}[b]{0.24\linewidth}
   		\includegraphics[width=1.0\linewidth,height=0.8\linewidth]{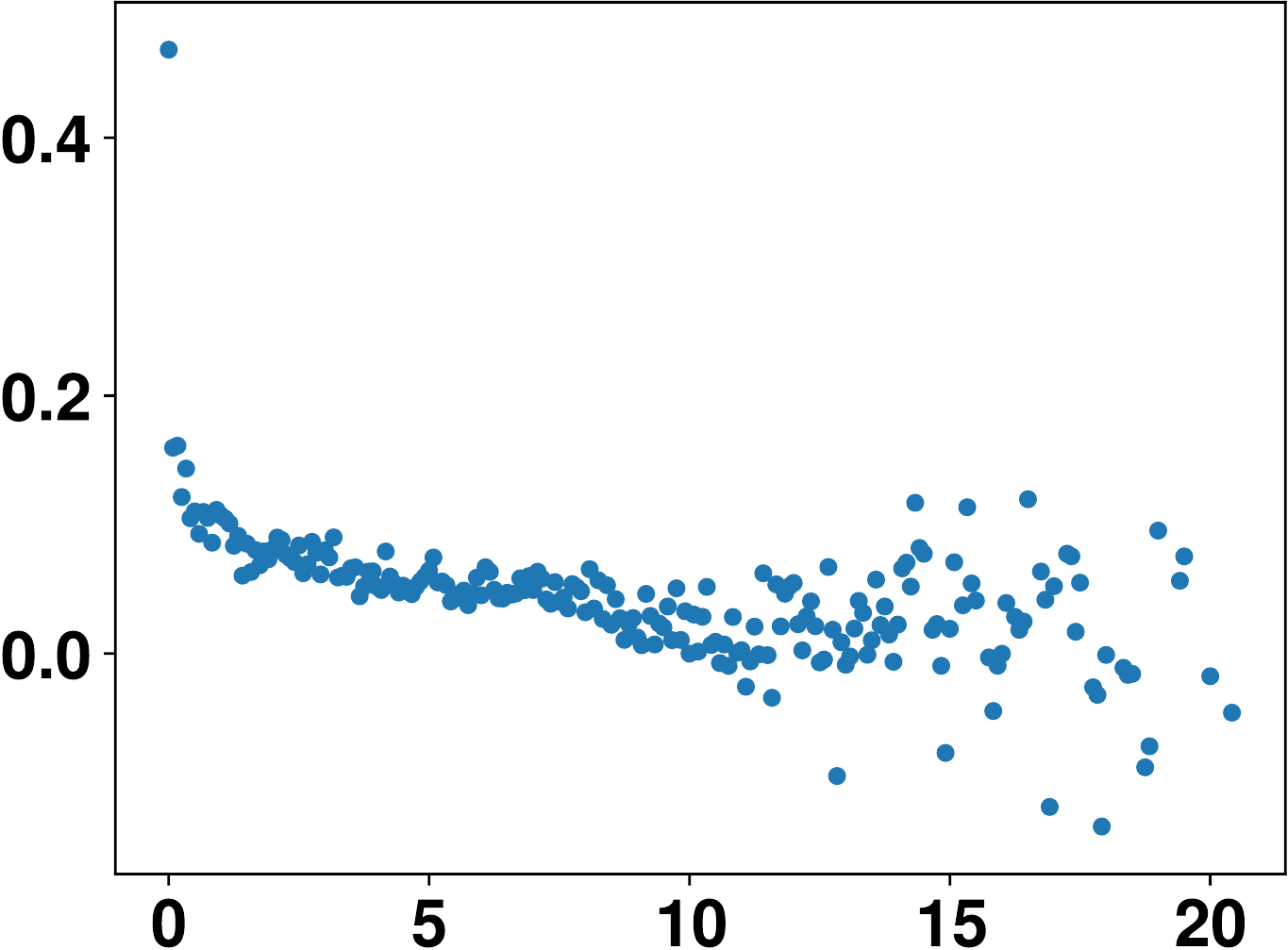}
   		\caption{time} 
   		\label{fig:correlation:mean:0}
   	\end{subfigure}
   	\hfill
   	\begin{subfigure}[b]{0.24\linewidth}
       		\includegraphics[width=1.0\linewidth,height=0.8\linewidth]{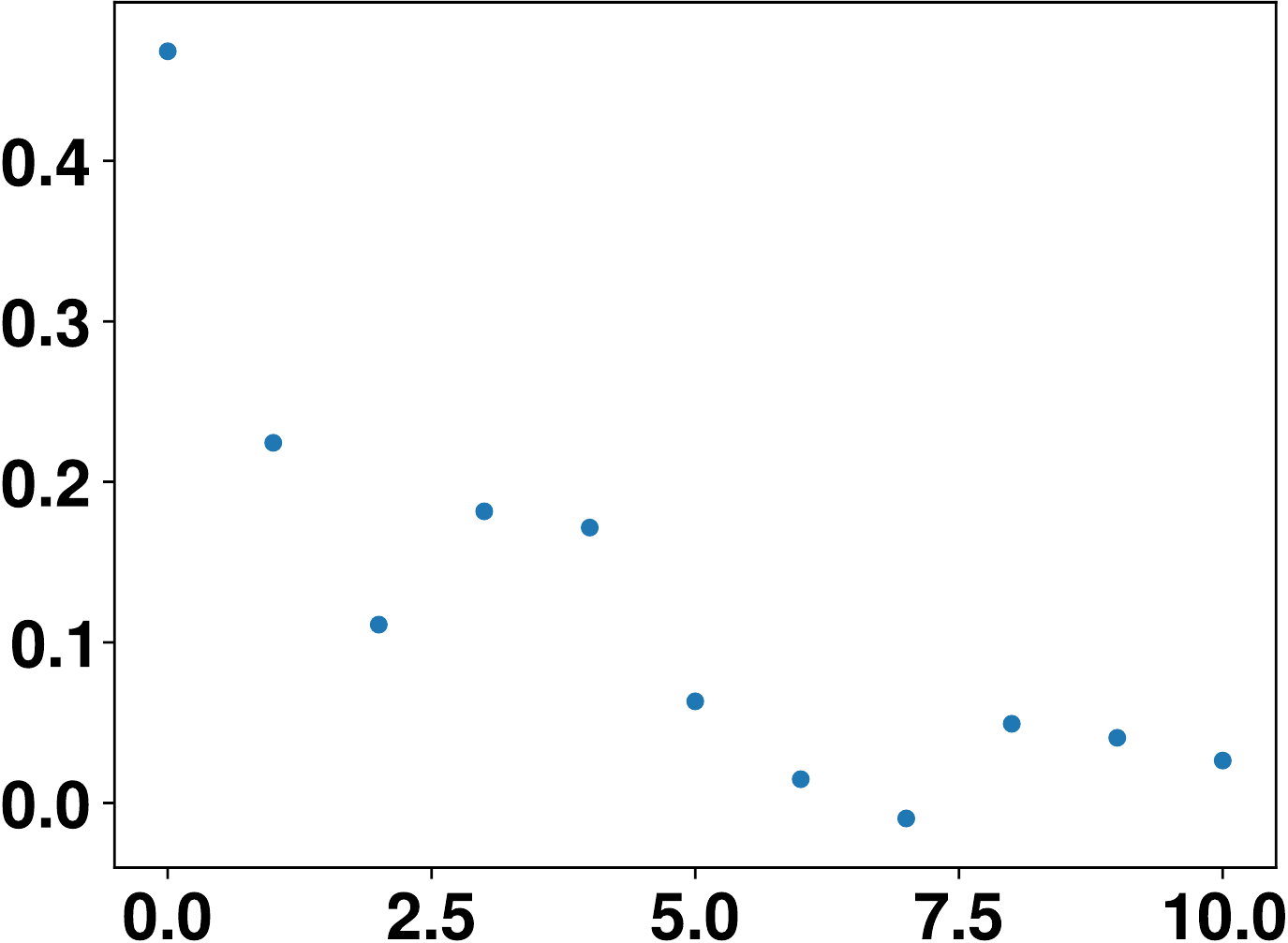}
   		\caption{longitude} 
   		\label{fig:correlation:mean:1}
   	\end{subfigure}
   	\hfill
   	\begin{subfigure}[b]{0.24\linewidth}
       		\includegraphics[width=1.0\linewidth,height=0.8\linewidth]{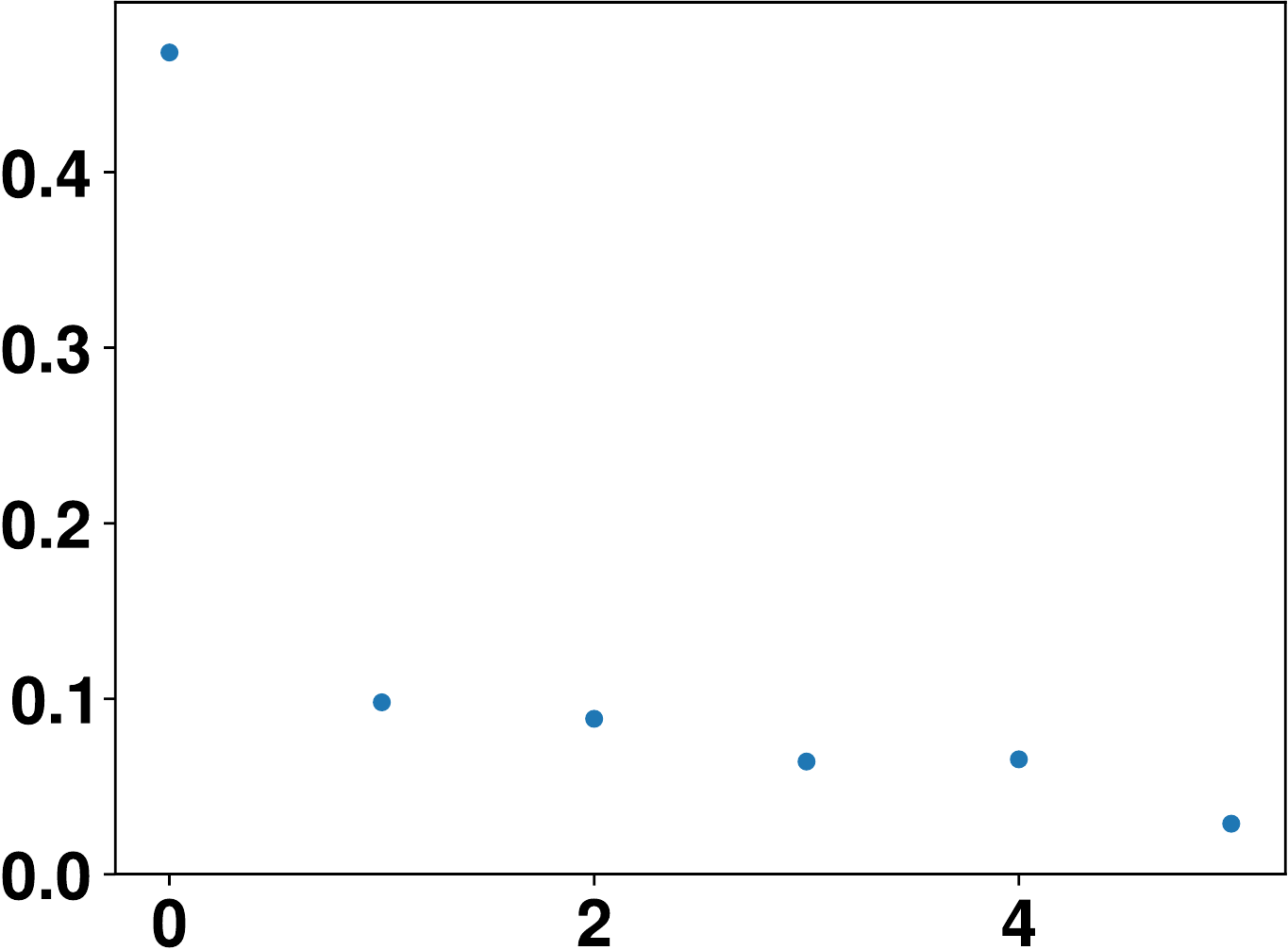}
   		\caption{latitude} 
   		\label{fig:correlation:mean:2}
   	\end{subfigure}
   	\hfill
   	\begin{subfigure}[b]{0.24\linewidth}
       		\includegraphics[width=1.0\linewidth,height=0.8\linewidth]{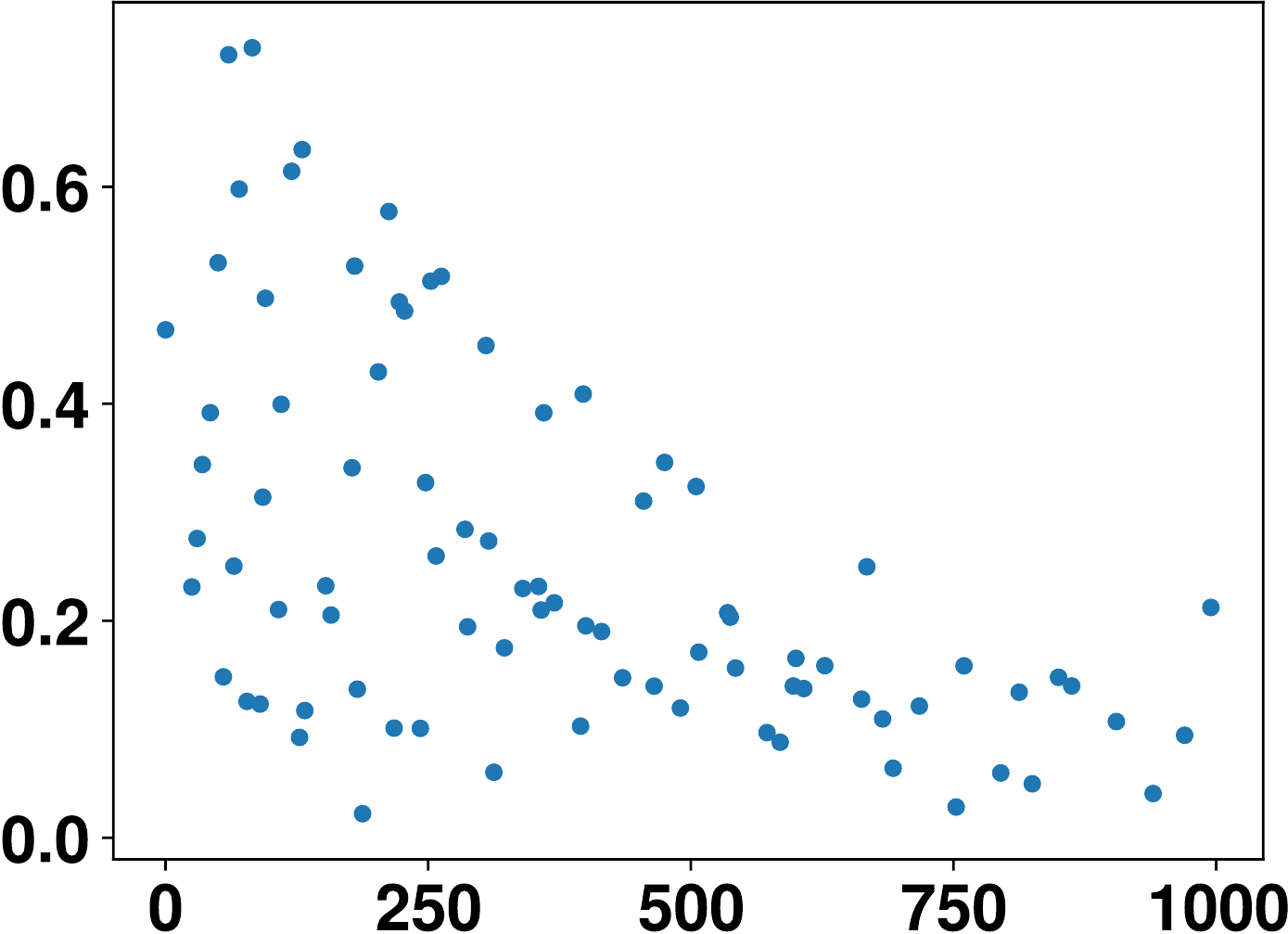}
   		\caption{depth} 
   		\label{fig:correlation:mean:3}
   	\end{subfigure}
	\vspace{-0.5em}
   	\caption{Means of estimated correlations associated with measuring points which differ in a single direction.} 
	\label{fig:correlation:mean}
\end{figure}

The means of the groups were plotted in Figure \ref{fig:correlation:mean} for measuring points which differ in a single direction. They must be close to each other at points where the associated distances are close to each other, if the estimated correlations can be described by a continuous function that depends only on the distances. The graphs do not support this assumption, especially for points which differ by the depth or by long times.

It should be noted that considerably less estimates for points that differ only in longitude or only in latitude are available than for points that differ only in time or only in depth. Hence, the results regarding the longitude and latitude lack significance.

The plots also show  that the correlation tend to decrease in terms of absolute value with increasing distance between the measurement points.

\subsection{Probability Distributions} \label{subsec: po4: probability distribution}

\begin{figure}[b!]
    \captionsetup[sub]{}

   	\begin{subfigure}[b]{0.24\linewidth}
   		\includegraphics[width=1.0\linewidth,height=0.8\linewidth]{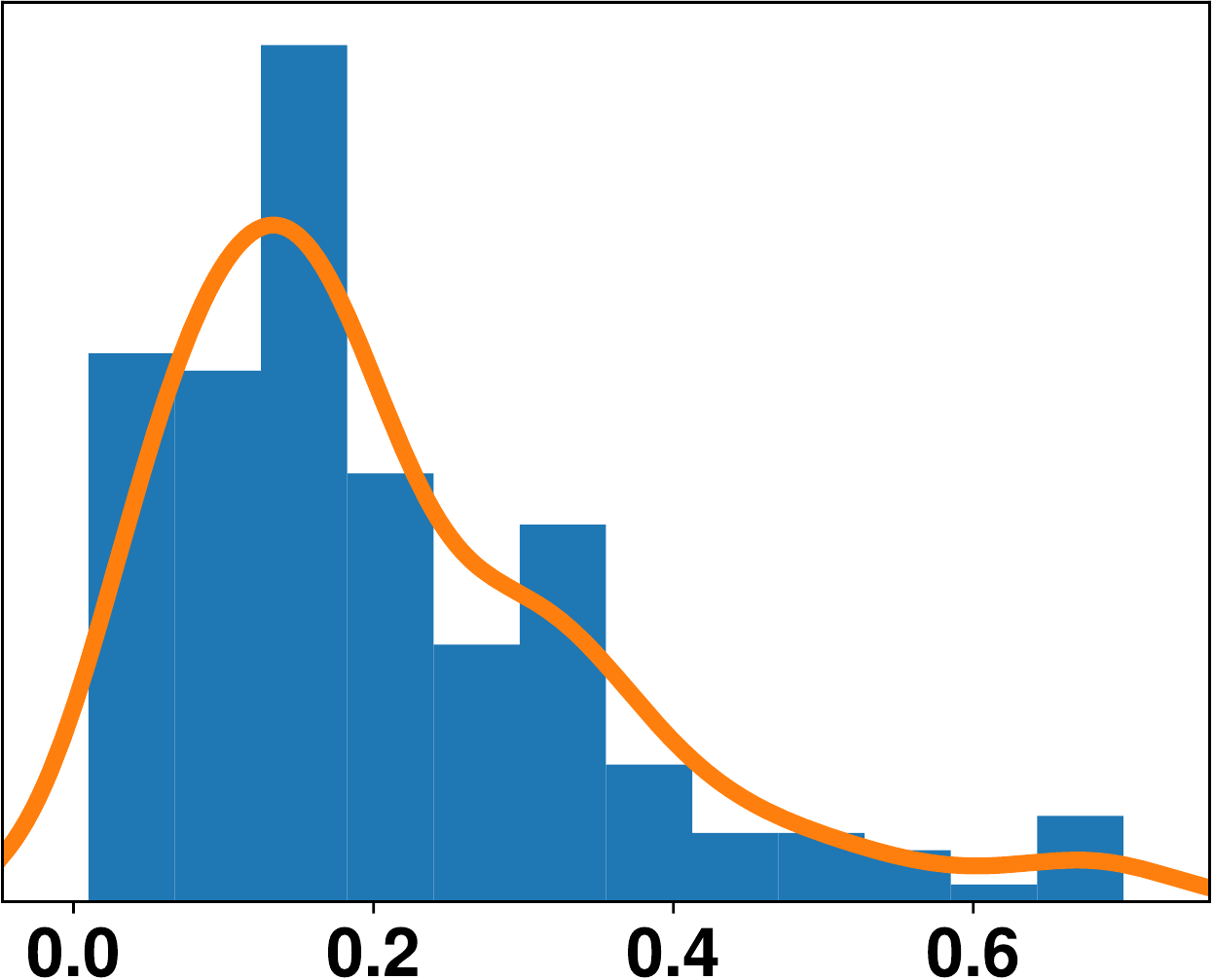}
   		\caption{Mar 1951, N33\degree-34\degree, E136\degree-137\degree, 0-25m} 
   	\end{subfigure}
   	\hfill
   	\begin{subfigure}[b]{0.24\linewidth}
   		\includegraphics[width=1.0\linewidth,height=0.8\linewidth]{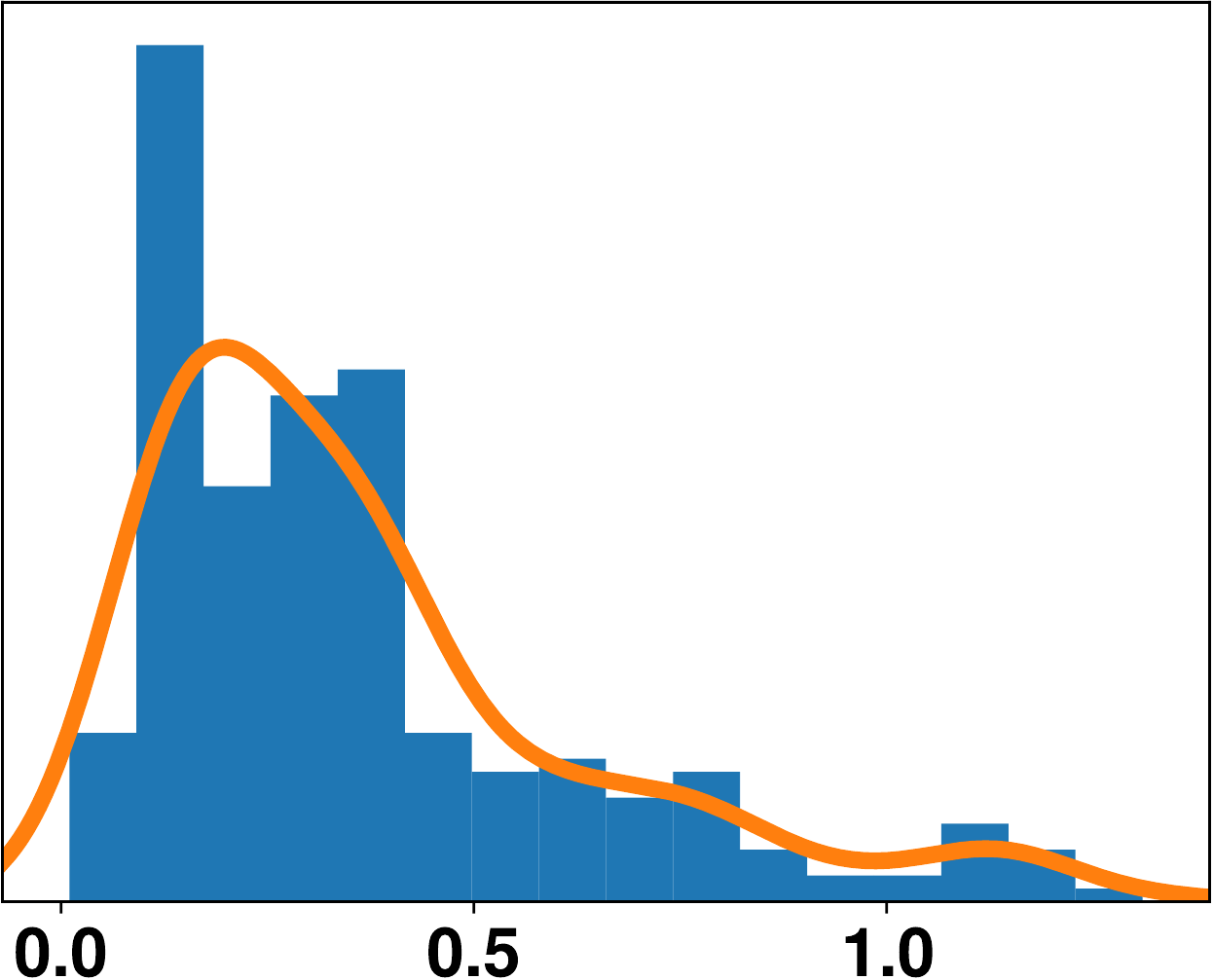}
   		\caption{Aug 1964, N55\degree-56\degree, E16-17\degree, 50-85m} 
   	\end{subfigure}
   	\hfill
   	\begin{subfigure}[b]{0.24\linewidth}
   		\includegraphics[width=1.0\linewidth,height=0.8\linewidth]{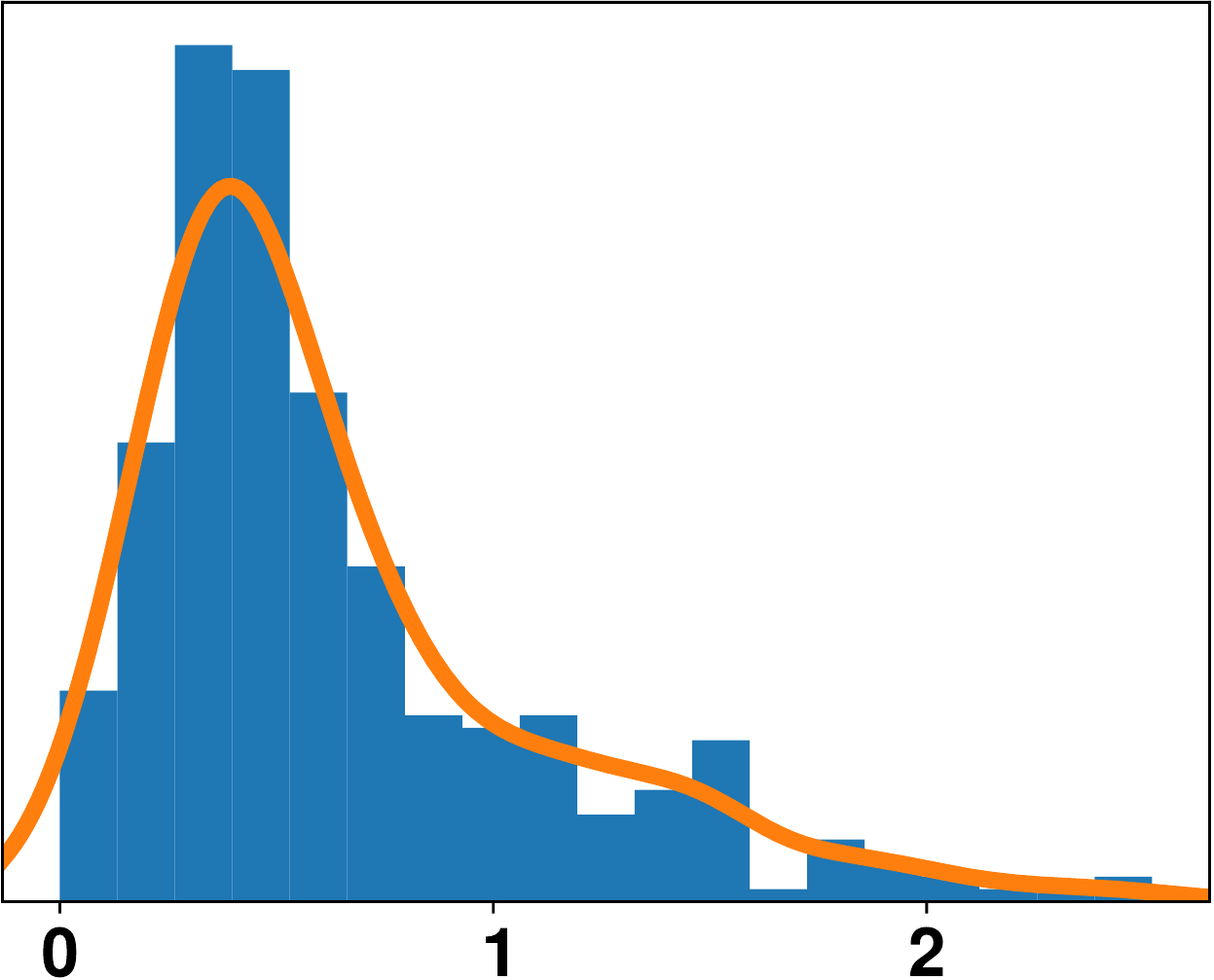}
  		\caption{Aug 1997, N55\degree-56\degree, E14-15\degree, 0-25m} 
   	\end{subfigure}
   	\hfill
   	\begin{subfigure}[b]{0.24\linewidth}
   		\includegraphics[width=1.0\linewidth,height=0.8\linewidth]{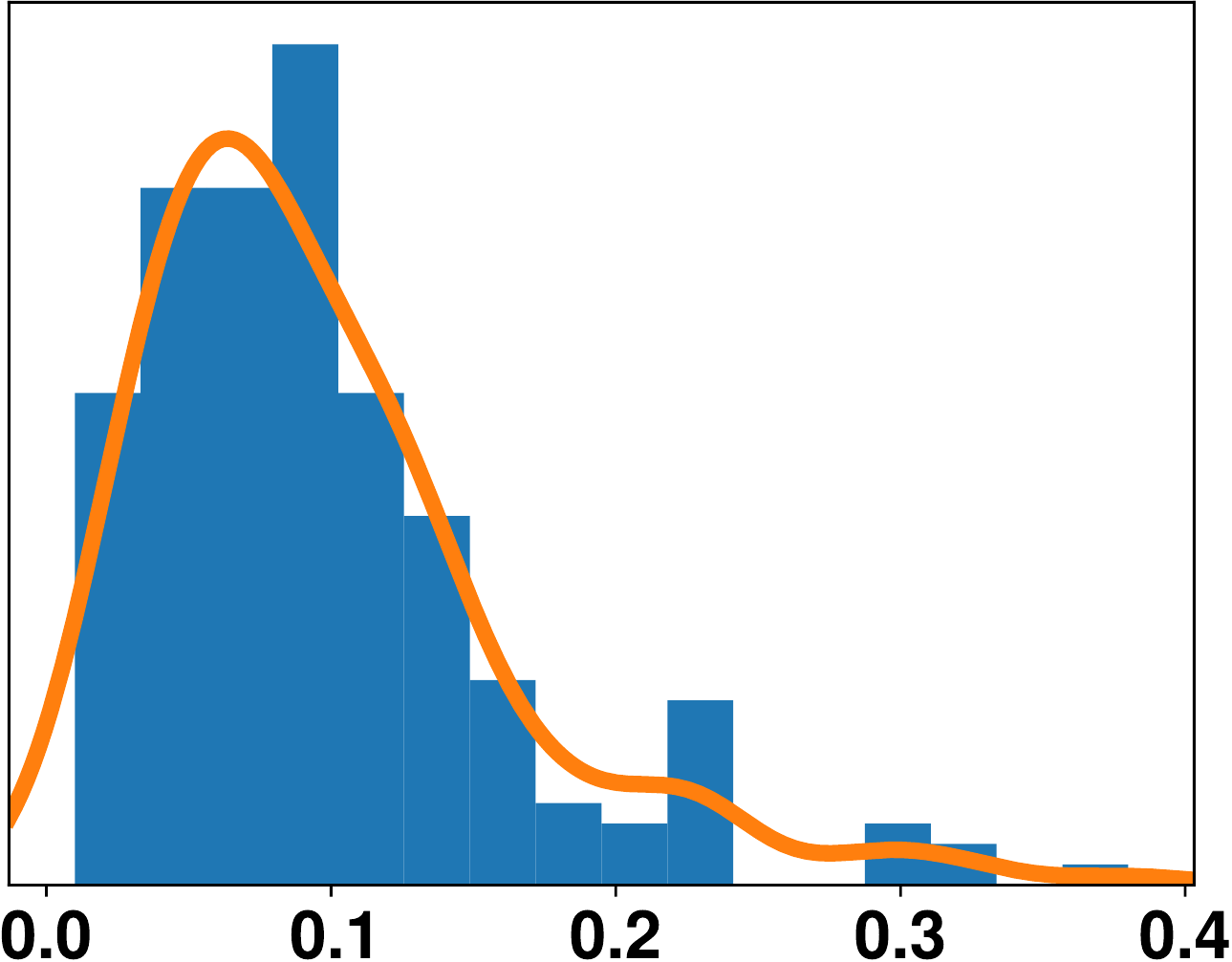}
   		\caption{Sep 1971, N30\degree-31\degree, W88-87\degree, 0-25m} 
   	\end{subfigure}
       
   	\begin{subfigure}[b]{0.24\linewidth}
   		\includegraphics[width=1.0\linewidth,height=0.8\linewidth]{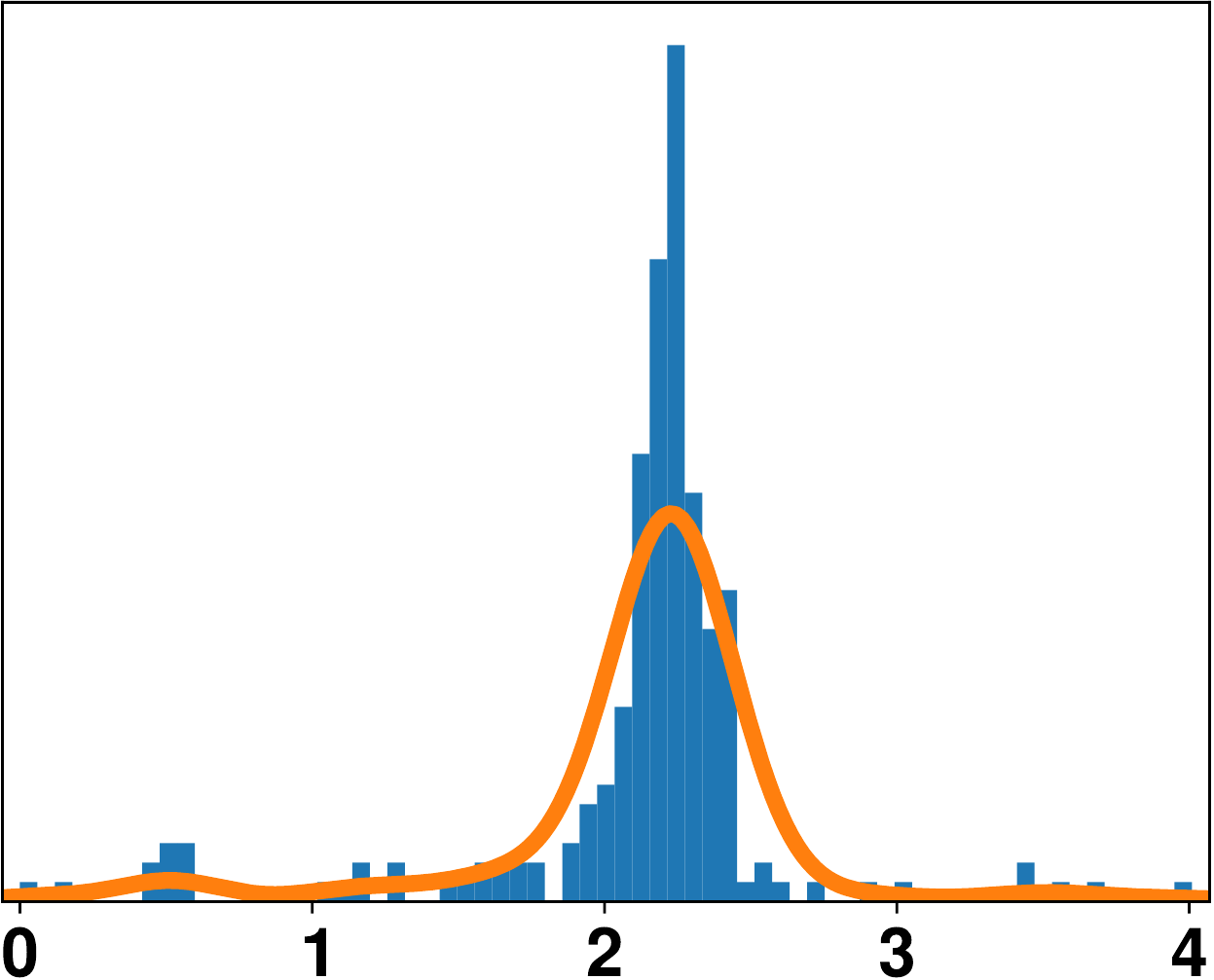}
   		\caption{Mar 1959, N47\degree-48\degree, W123\degree-122\degree, 0-25m} 
   	\end{subfigure}
   	\hfill
   	\begin{subfigure}[b]{0.24\linewidth}
   		\includegraphics[width=1.0\linewidth,height=0.8\linewidth]{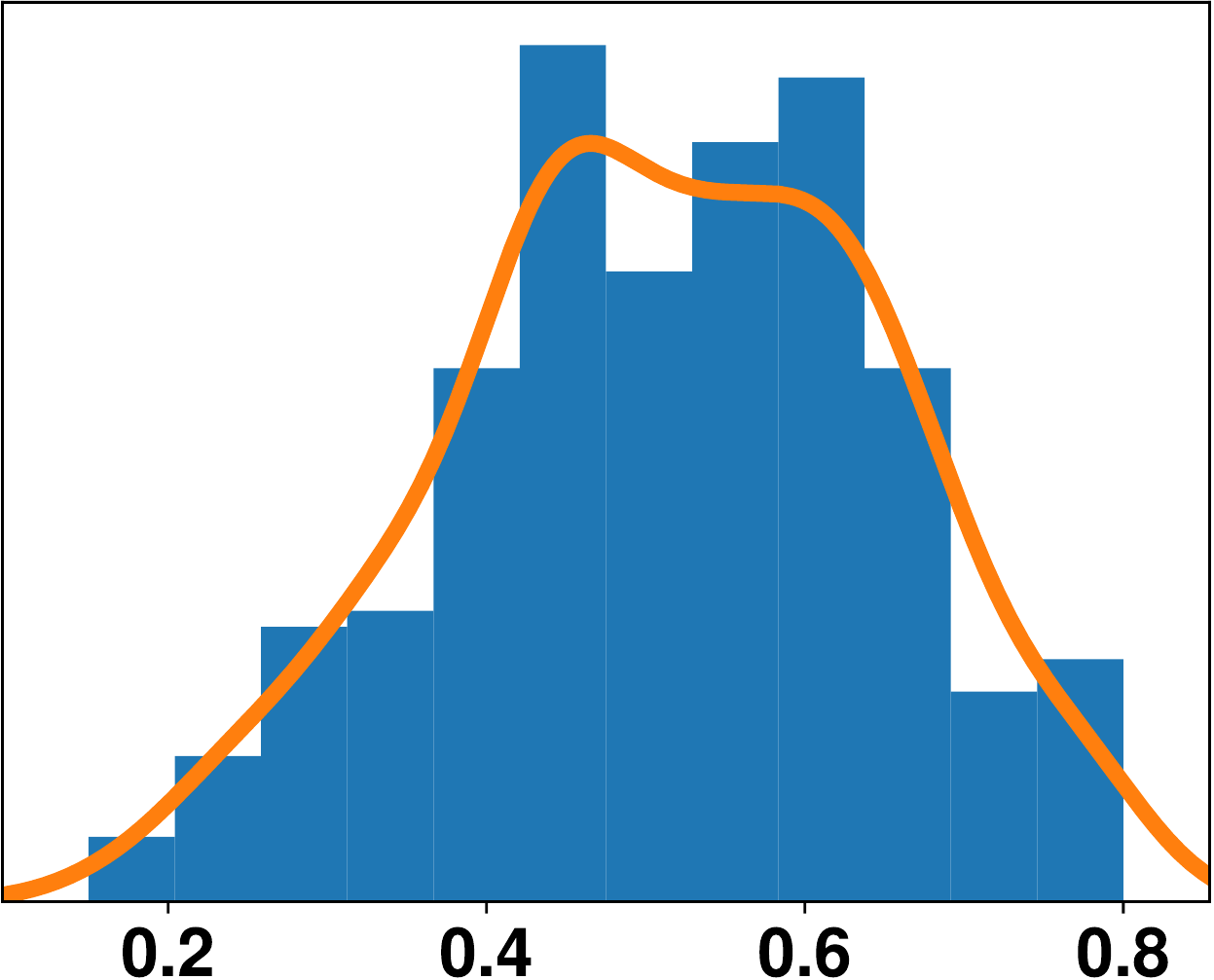}
   		\caption{July 1972, N20\degree-21\degree, W18\degree-17\degree, 0-25m} 
   	\end{subfigure}
   	\hfill
   	\begin{subfigure}[b]{0.24\linewidth}
  		\includegraphics[width=1.0\linewidth,height=0.8\linewidth]{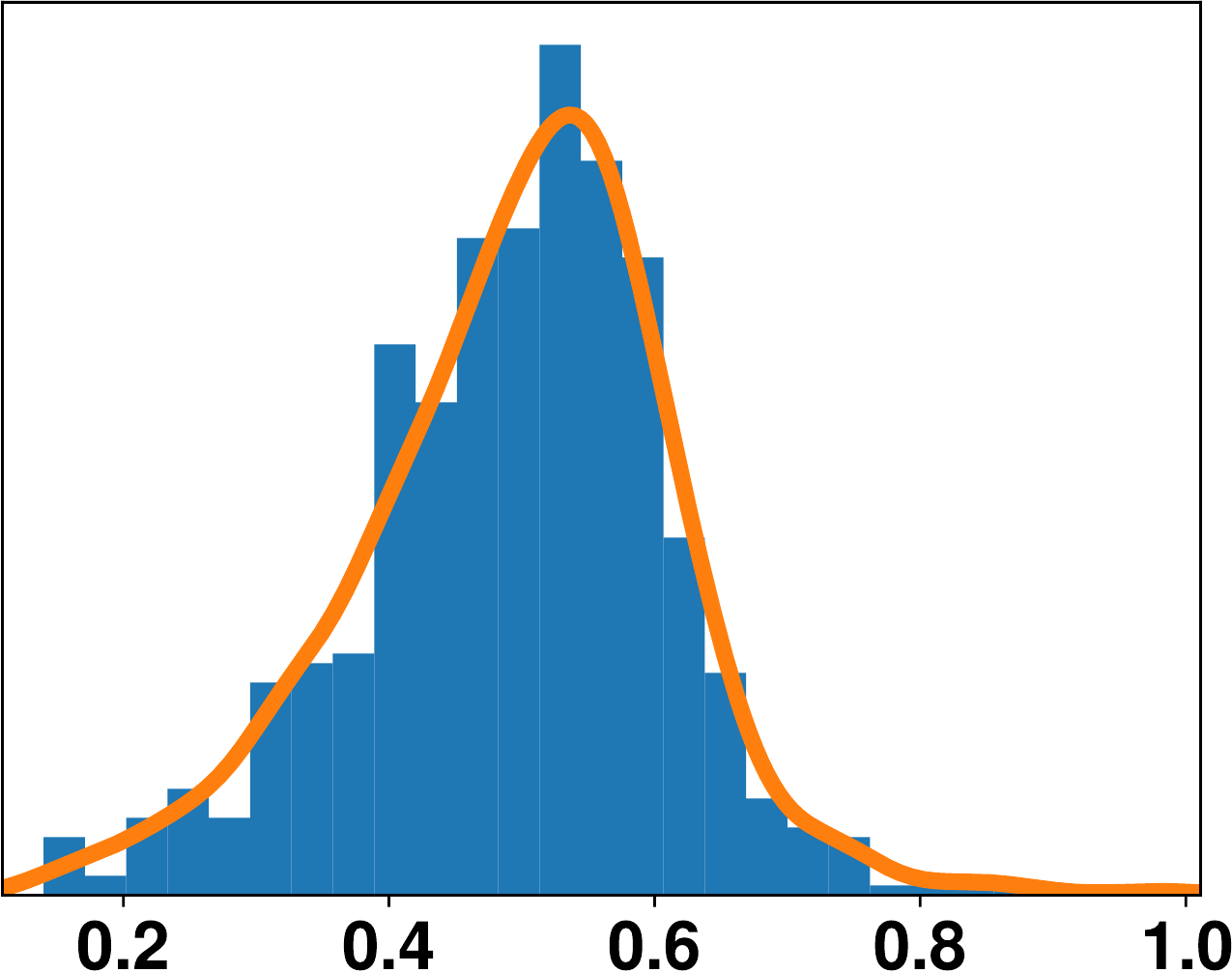}
   		\caption{Apr 1986, N56\degree-57\degree, E18\degree-19\degree, 25-50m} 
   	\end{subfigure}
   	\hfill
   	\begin{subfigure}[b]{0.24\linewidth}
   		\includegraphics[width=1.0\linewidth,height=0.8\linewidth]{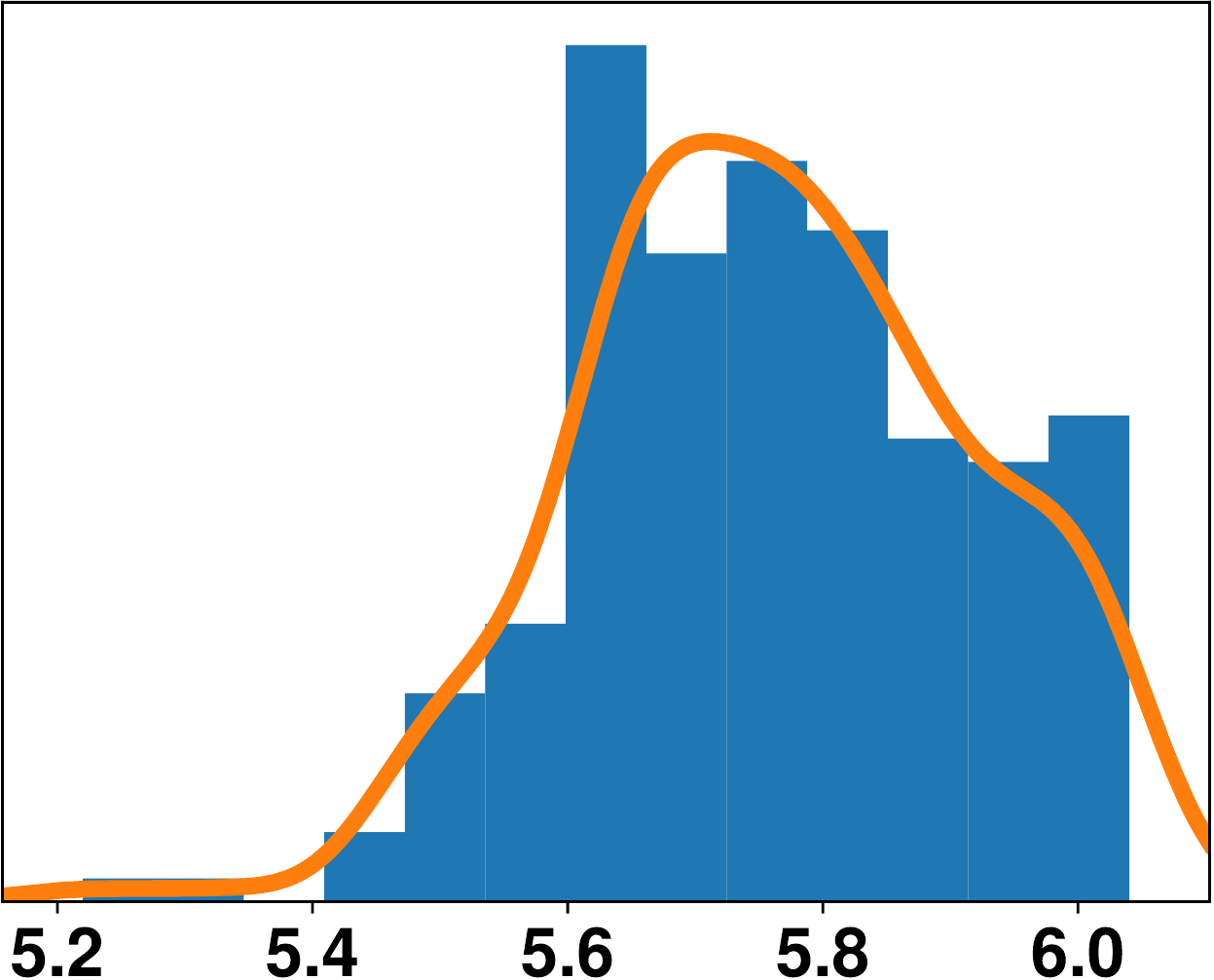}
   		\caption{July 1988, N43\degree-44\degree, E34\degree-35\degree, 290-360m} 
   	\end{subfigure}
       
 	\begin{subfigure}[b]{0.24\linewidth}
 		\includegraphics[width=1.0\linewidth,height=0.8\linewidth]{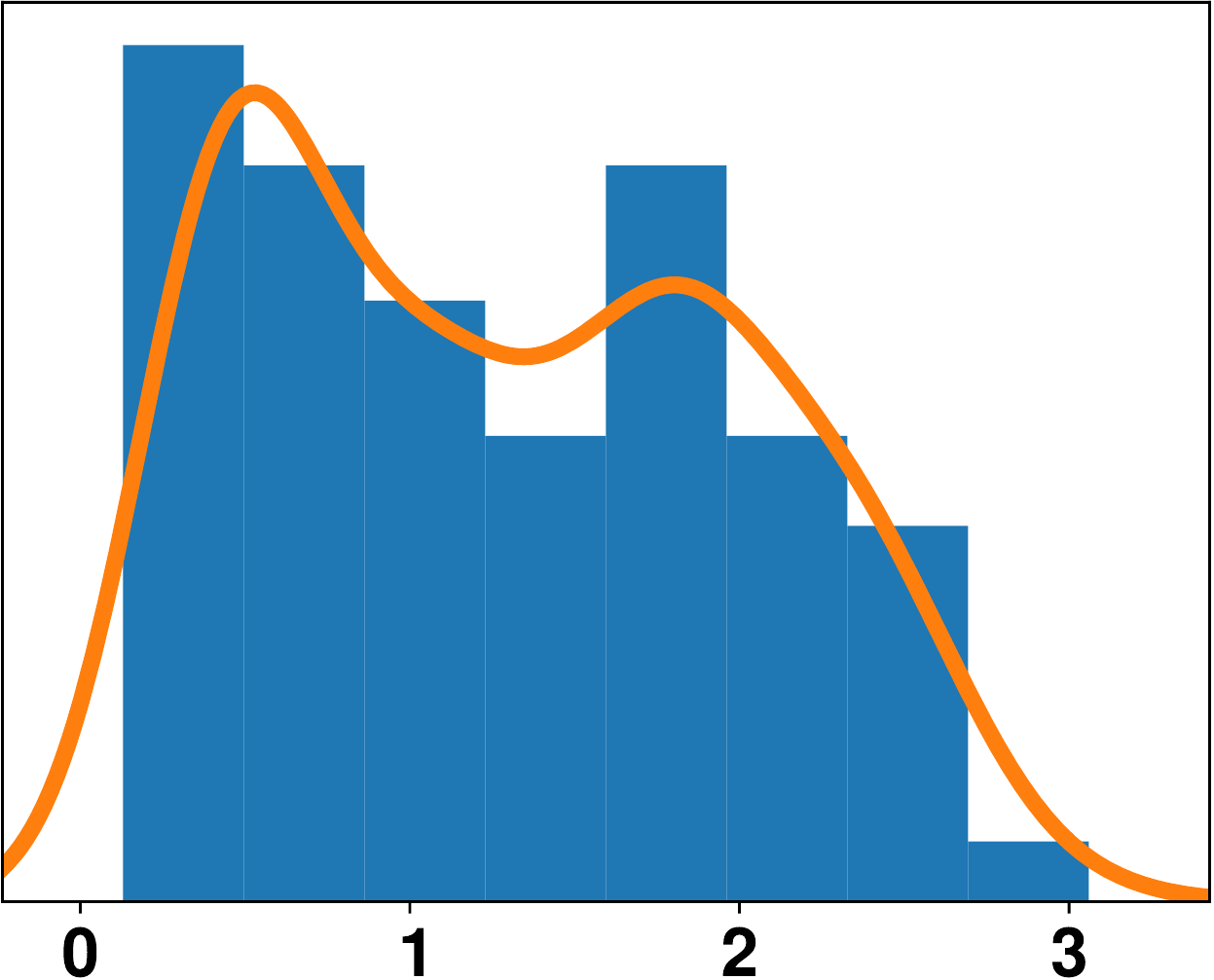}
   		\caption{June 1970, N44\degree-45\degree, W125-124\degree, 0-25m} 
 	\end{subfigure}
 	\hfill
 	\begin{subfigure}[b]{0.24\linewidth}
  		\includegraphics[width=1.0\linewidth,height=0.8\linewidth]{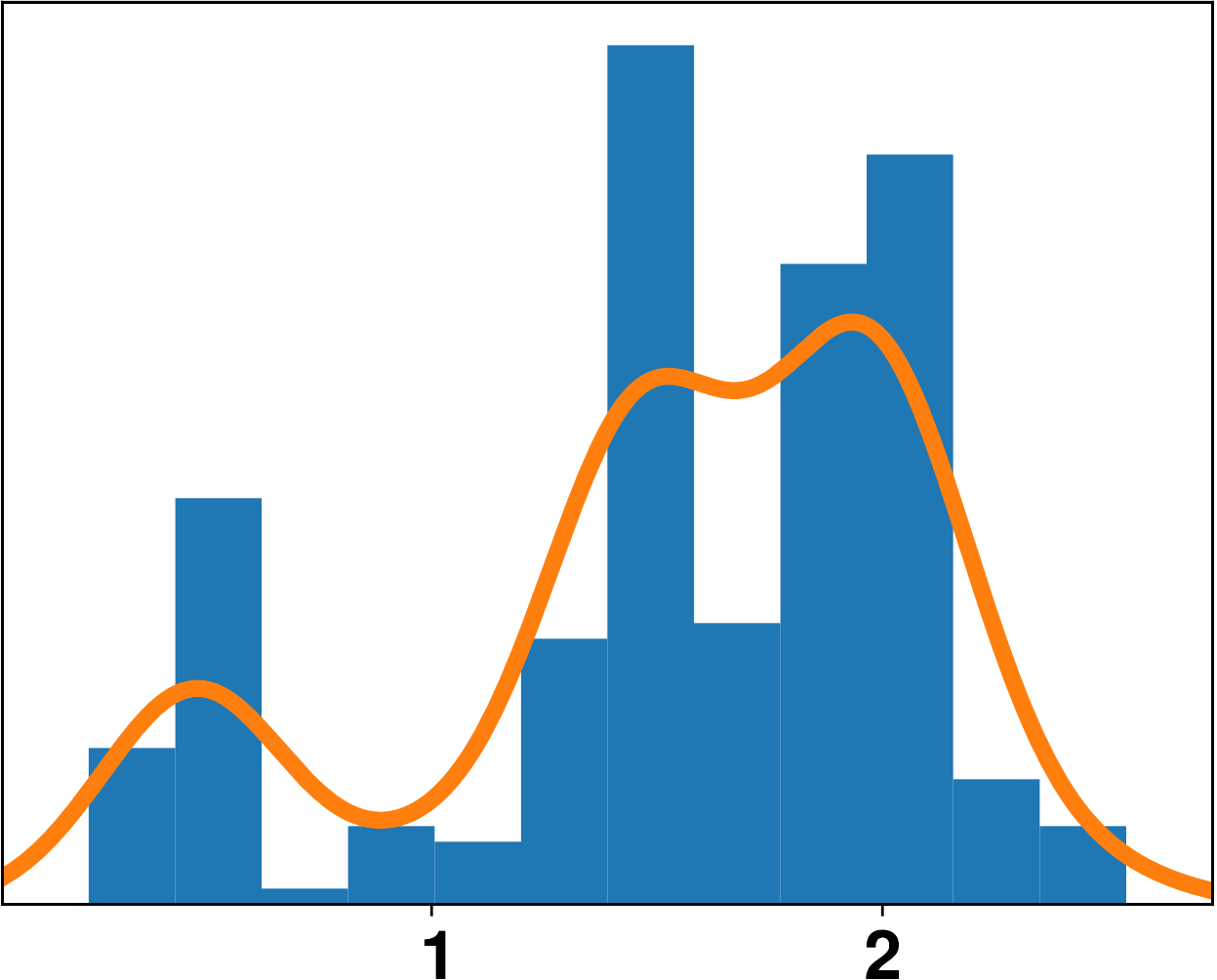}
   		\caption{Apr 1975, N55\degree-56\degree, E15-E16\degree, 50-85m} 
 	\end{subfigure}
 	\hfill
 	\begin{subfigure}[b]{0.24\linewidth}
   		\includegraphics[width=1.0\linewidth,height=0.8\linewidth]{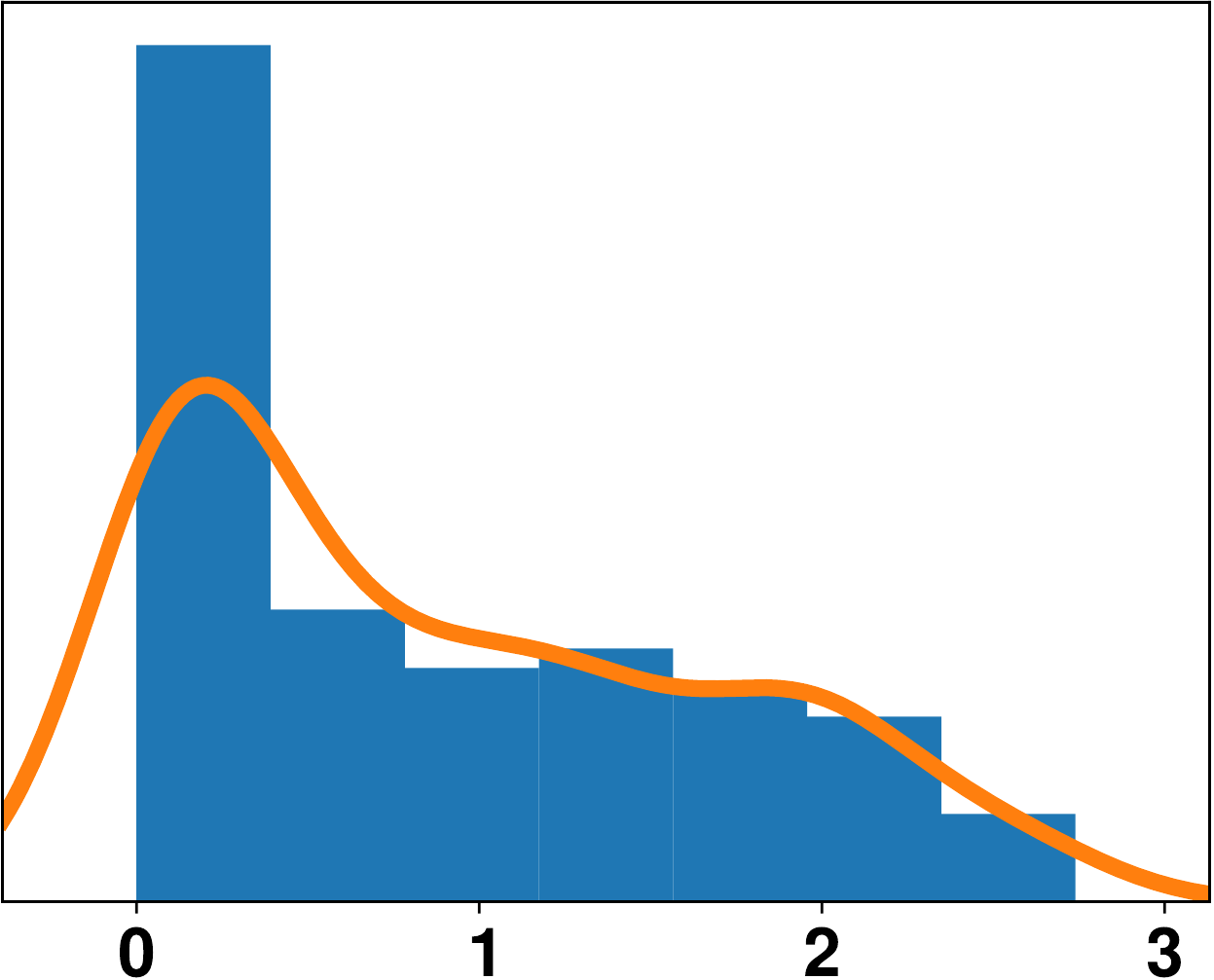}
 		\caption{June 1982, N47\degree-48\degree, W125-124\degree, 0-25m} 
 	\end{subfigure}
 	\hfill
 	\begin{subfigure}[b]{0.24\linewidth}
   		\includegraphics[width=1.0\linewidth,height=0.8\linewidth]{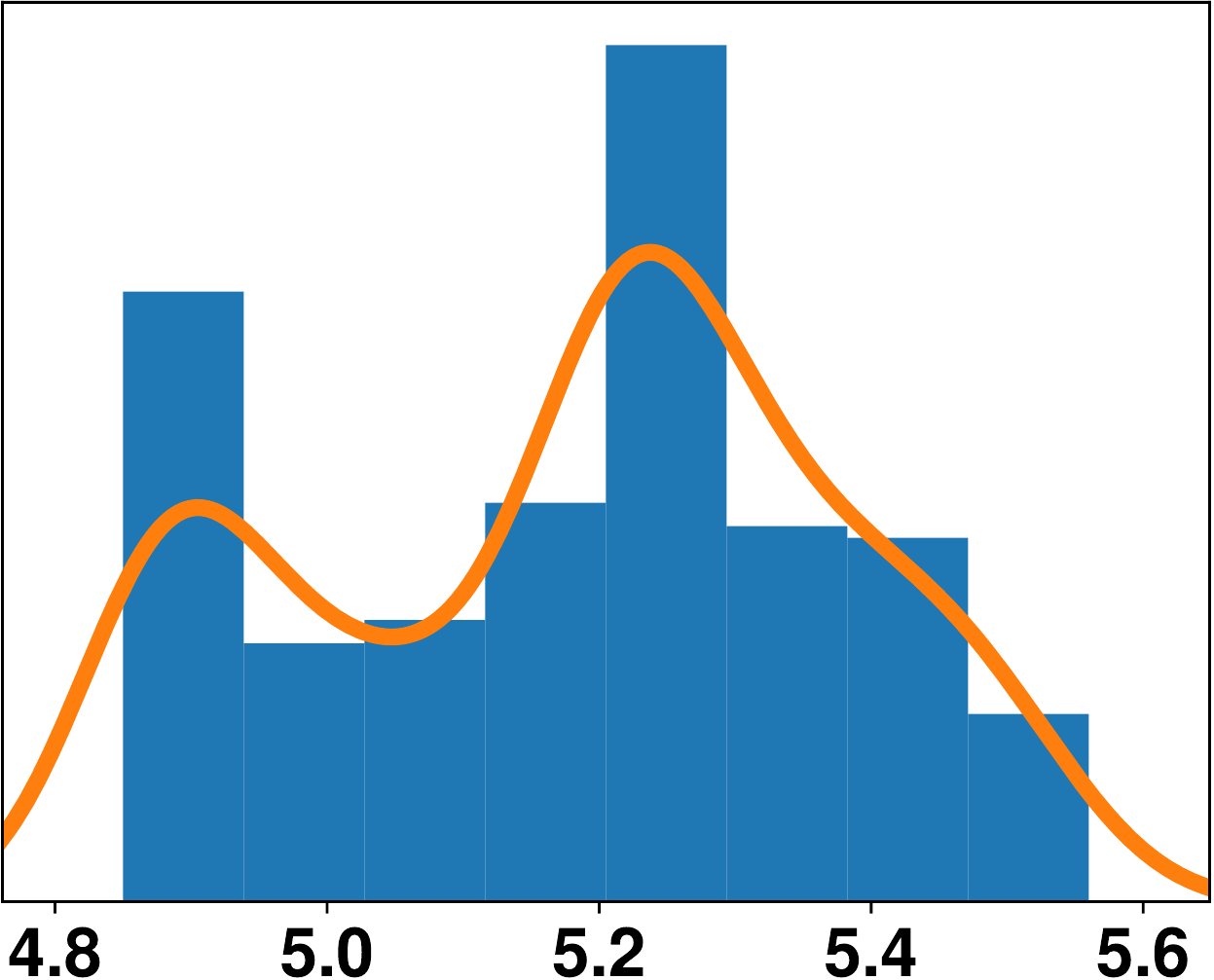}
   		\caption{June 1988, N42\degree-43\degree, E32-33\degree, 220-290m} 
 	\end{subfigure}
	\vspace{-0.5em}
\caption{Selected histograms and kernel density estimations for the measurement results ($\eta$). Most of them look like the ones in the first and second row, i.e., log-normal or normal distributed. A few look like the ones in the last row which do not look like either of the these two distributions.} 
	\label{fig:histograms:measurement_results}
\end{figure}
\begin{figure}[b!]
    \captionsetup[sub]{}

   	\begin{subfigure}[b]{0.24\linewidth}
   		\includegraphics[width=1.0\linewidth,height=0.8\linewidth]{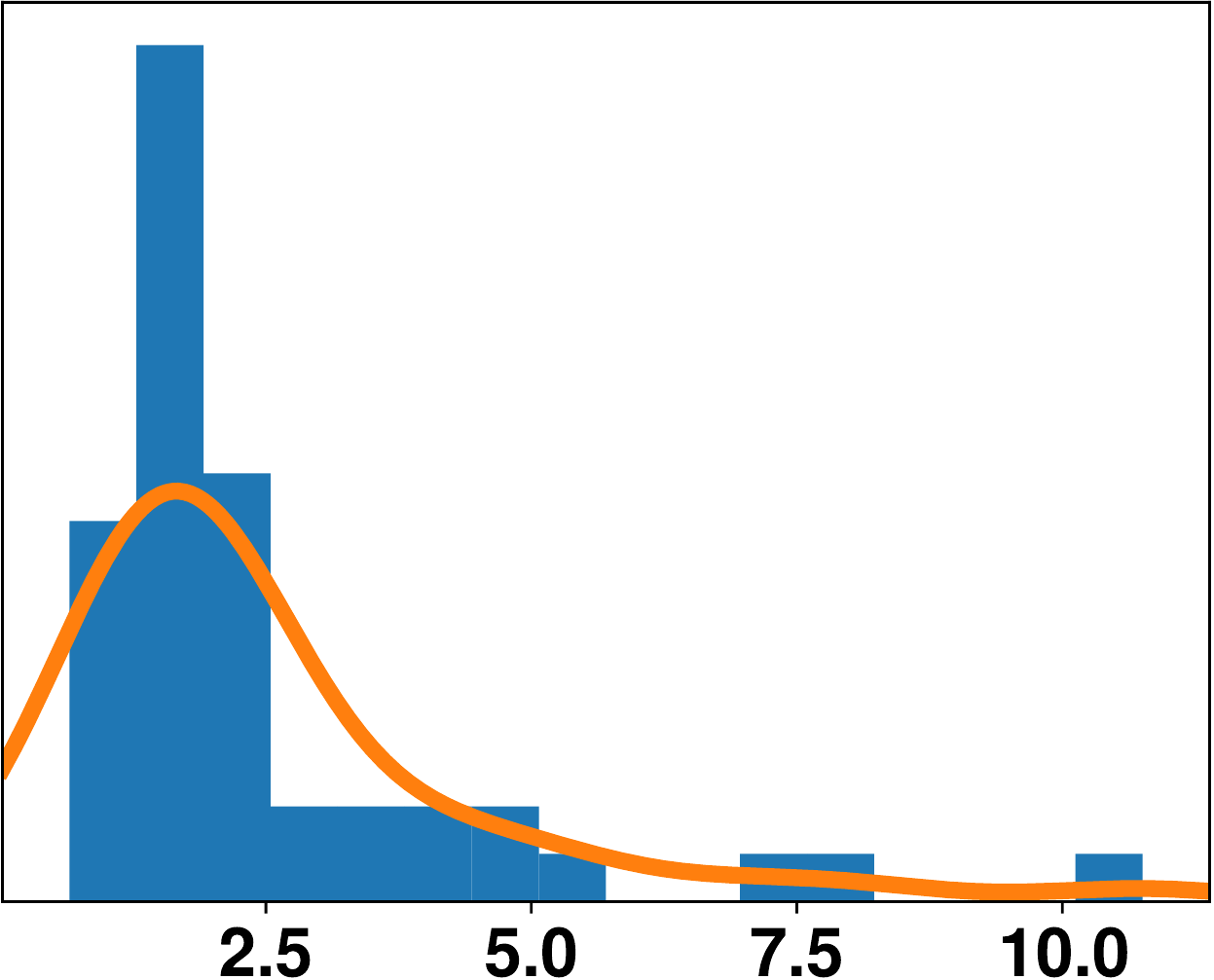}
   		\caption{Mar, N55\degree-56\degree, E15\degree-16\degree, 85-120m} 
   	\end{subfigure}
  	\hfill
  	\begin{subfigure}[b]{0.24\linewidth}
  		\includegraphics[width=1.0\linewidth,height=0.8\linewidth]{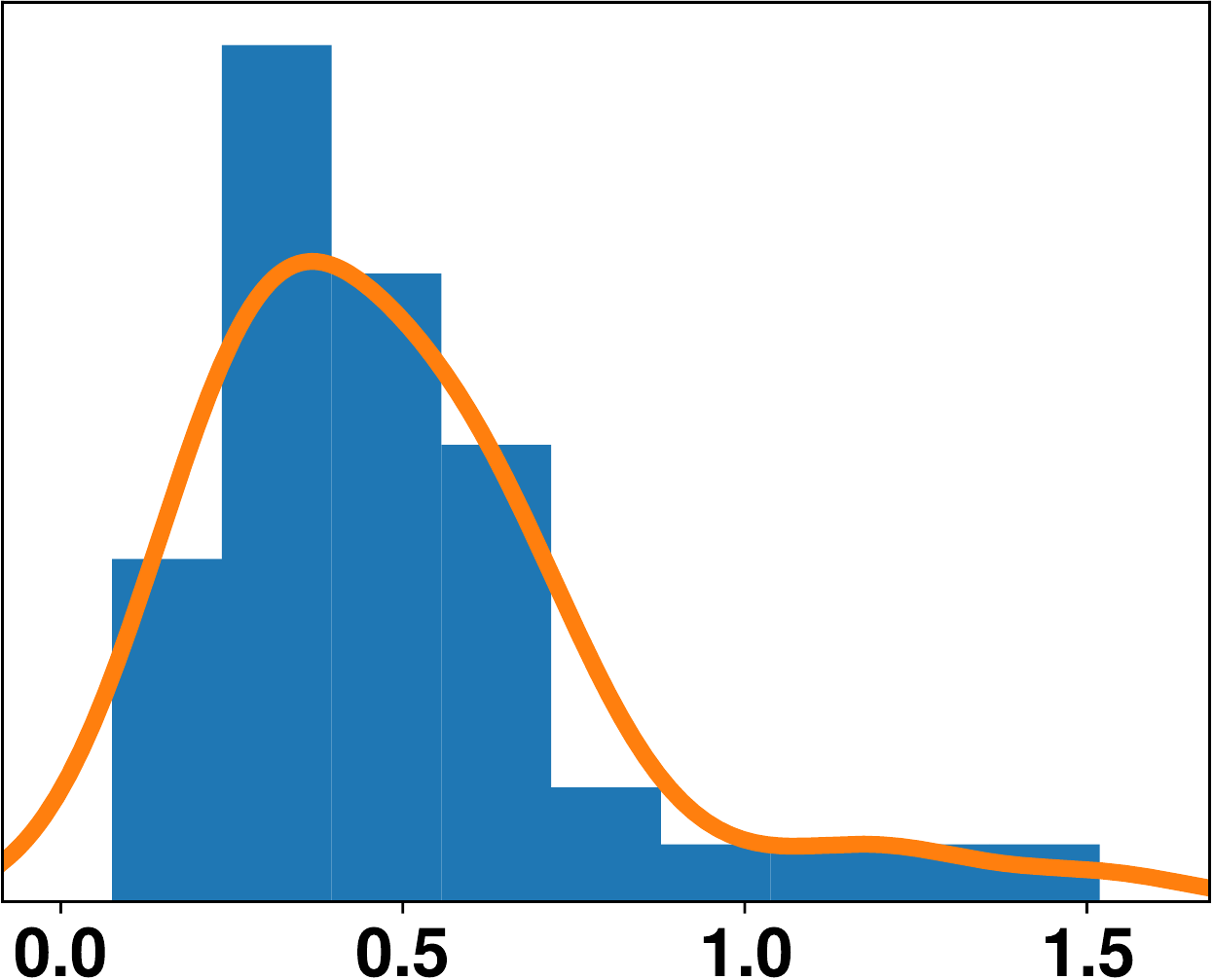}
 		\caption{Apr, N55\degree-56\degree, E14-15\degree, 25-50m} 
  	\end{subfigure}
   	\hfill
   	\begin{subfigure}[b]{0.24\linewidth}
   		\includegraphics[width=1.0\linewidth,height=0.8\linewidth]{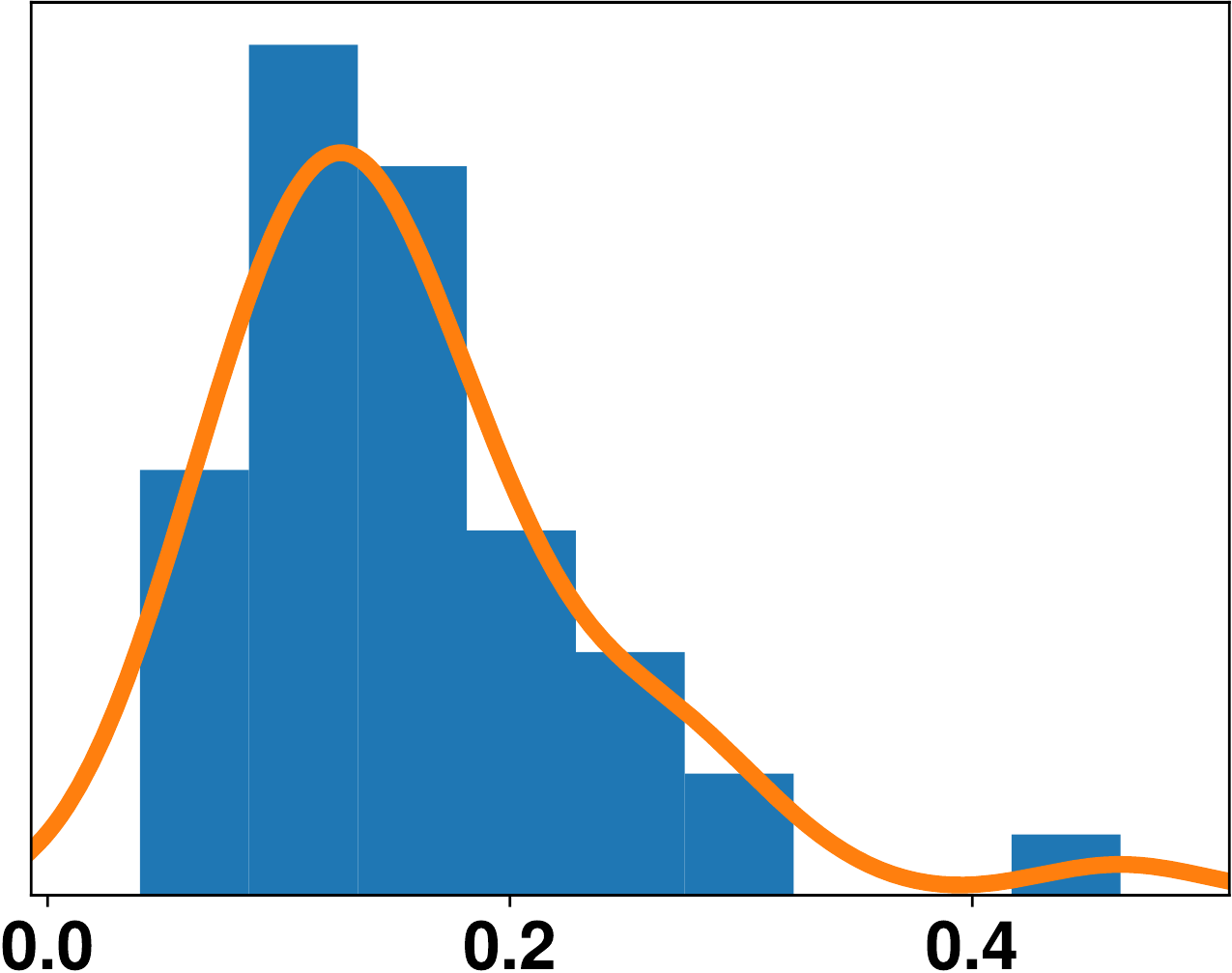}
   		\caption{May, N59\degree-60\degree, E21-22\degree, 0-25m} 
   	\end{subfigure}
   	\hfill
   	\begin{subfigure}[b]{0.24\linewidth}
   		\includegraphics[width=1.0\linewidth,height=0.8\linewidth]{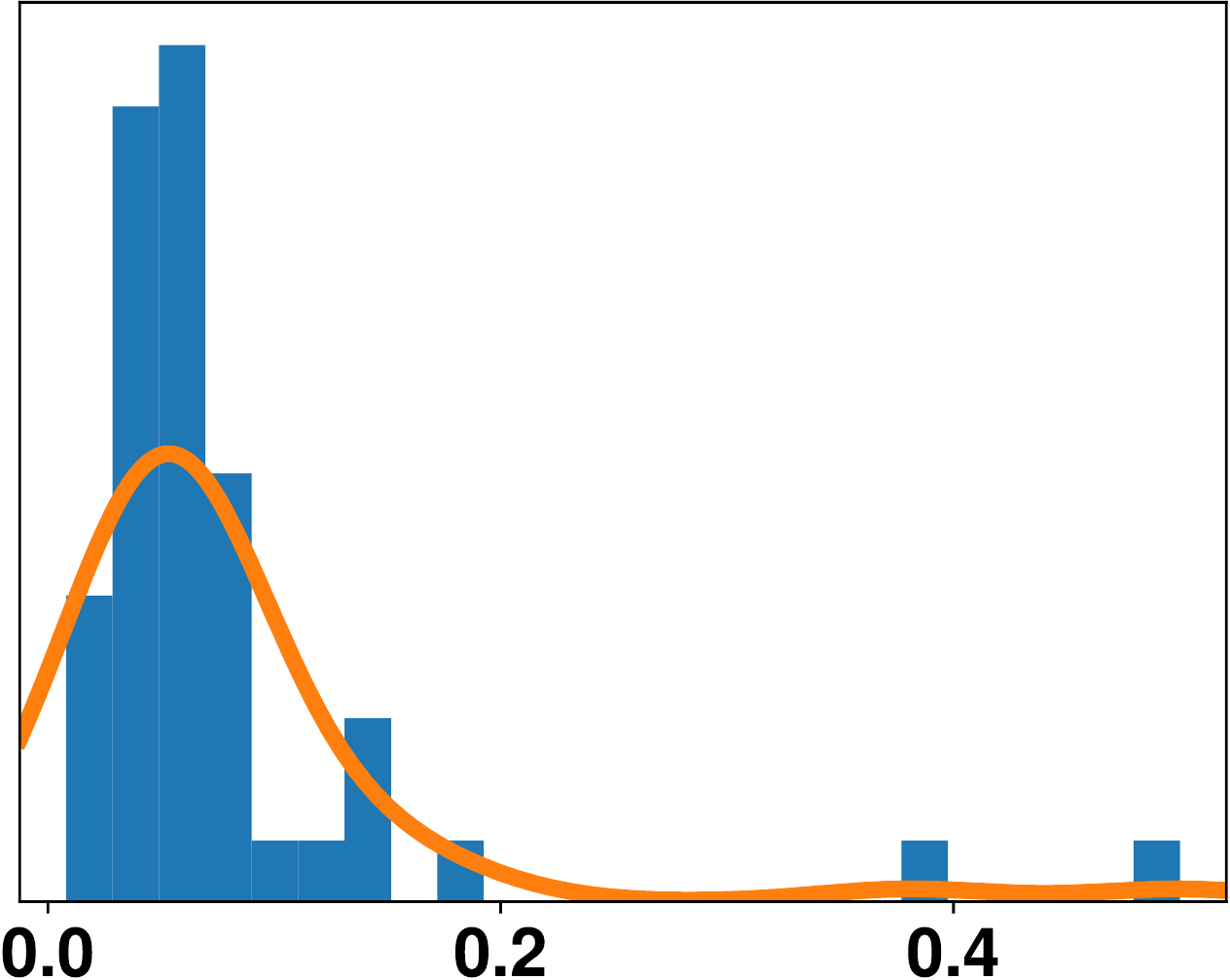}
   		\caption{Aug, N56\degree-57\degree, E19-20\degree, 0-25m} 
   	\end{subfigure}
       
   	\begin{subfigure}[b]{0.24\linewidth}
   		\includegraphics[width=1.0\linewidth,height=0.8\linewidth]{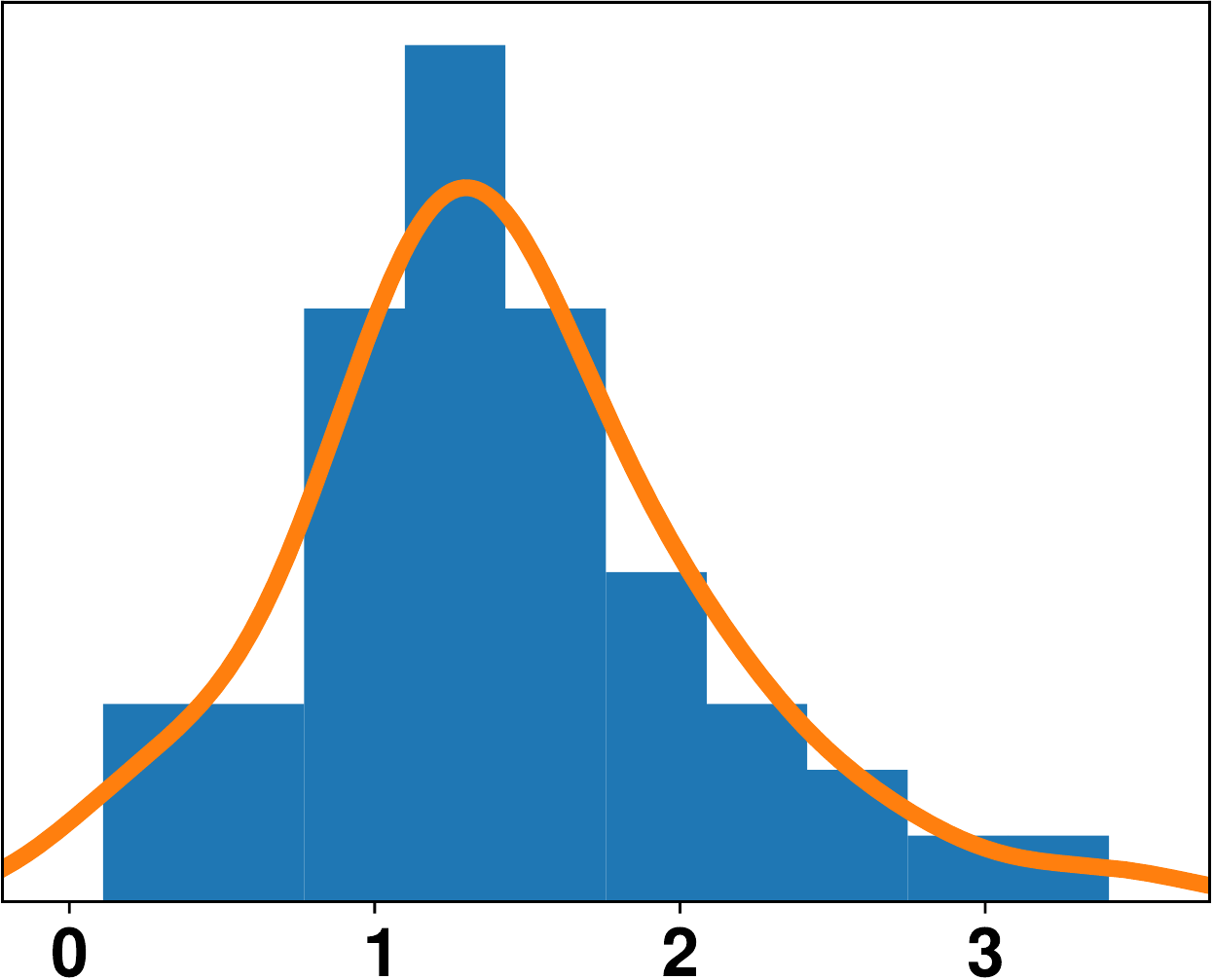}
   		\caption{Mar, N55\degree-56\degree, E15\degree-16\degree, 50-85m} 
   	\end{subfigure}
   	\hfill
   	\begin{subfigure}[b]{0.24\linewidth}
   		\includegraphics[width=1.0\linewidth,height=0.8\linewidth]{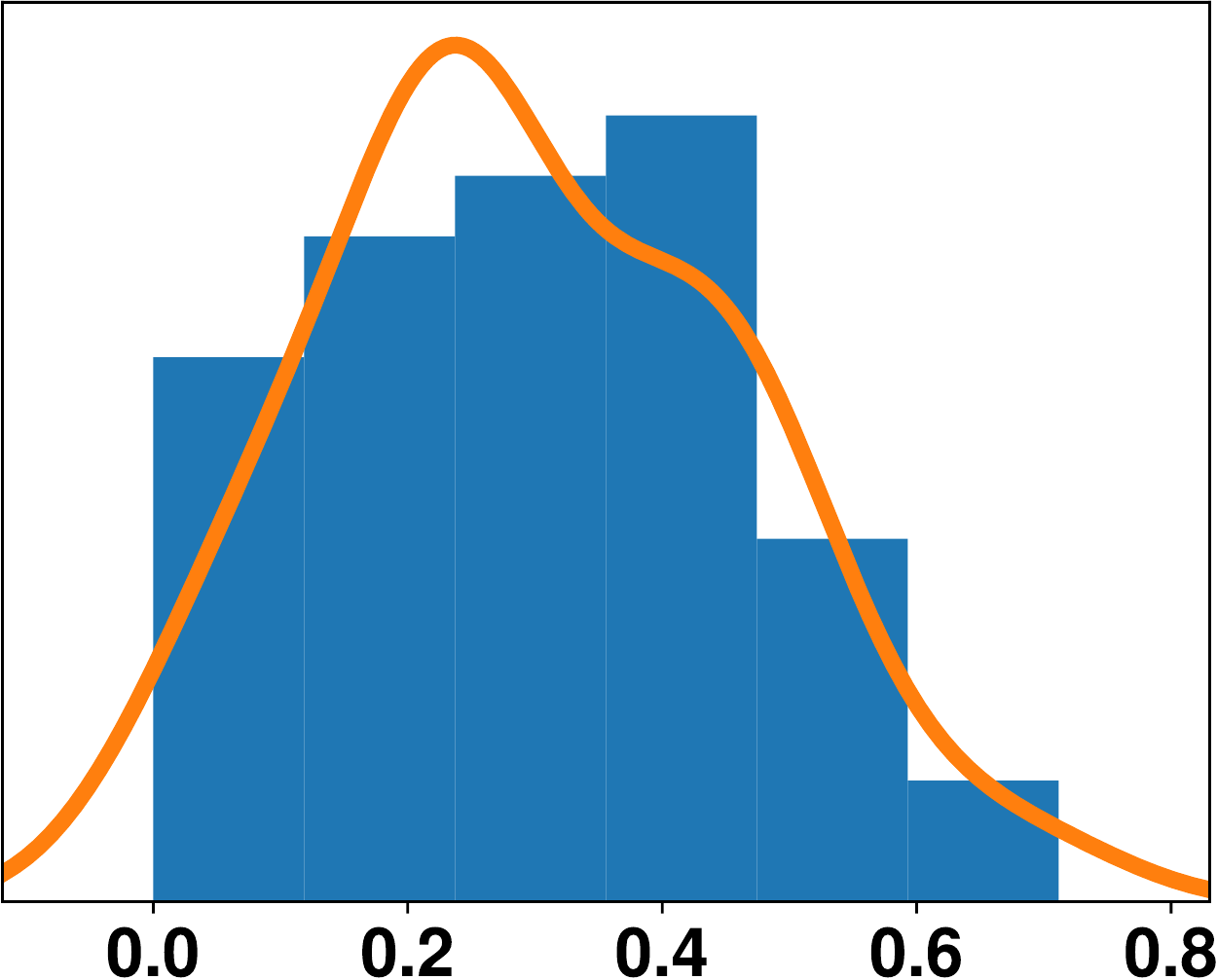}
   		\caption{May, N55\degree-56\degree, E15\degree-16\degree, 25-50m} 
   	\end{subfigure}
   	\hfill
   	\begin{subfigure}[b]{0.24\linewidth}
   		\includegraphics[width=1.0\linewidth,height=0.8\linewidth]{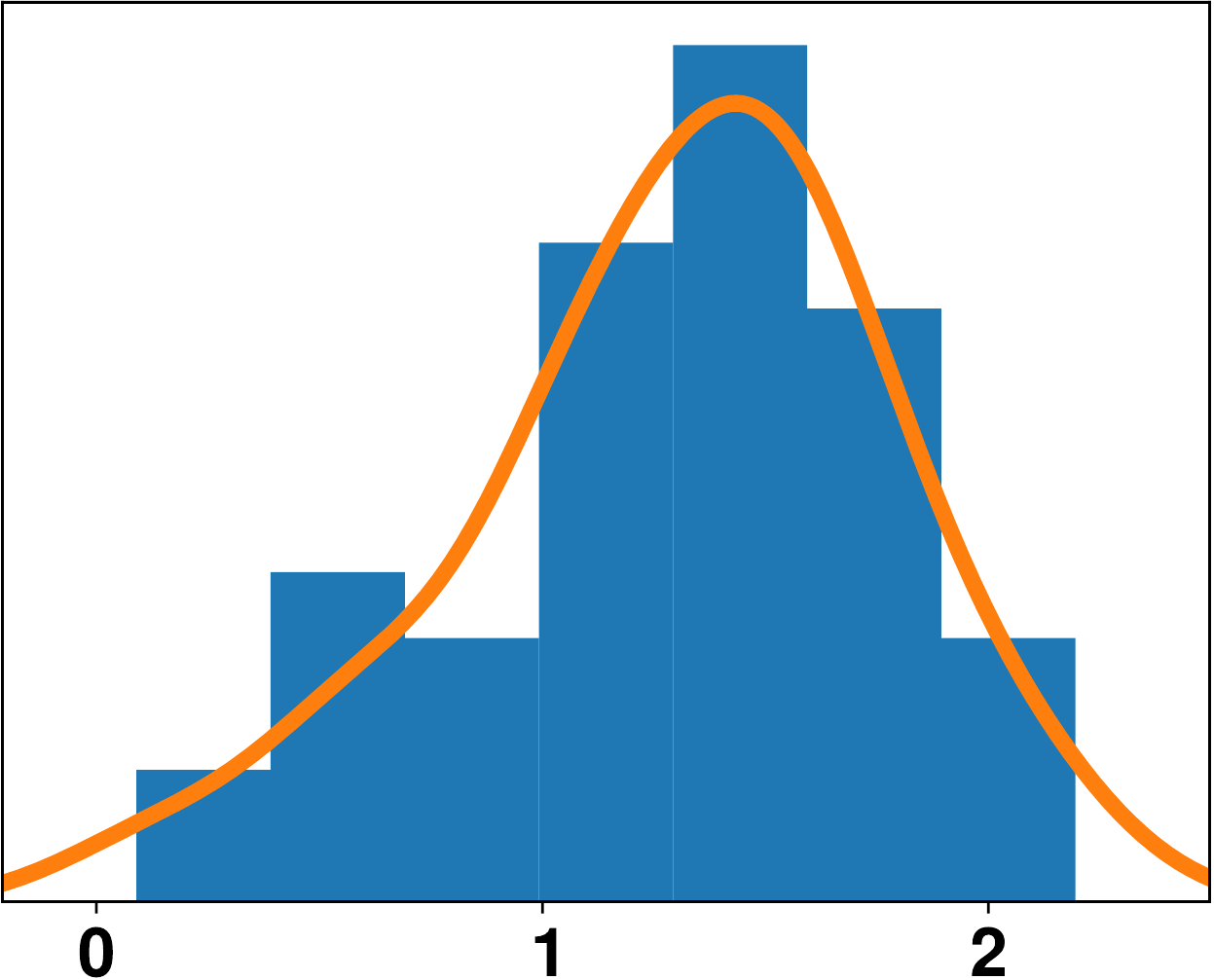}
   		\caption{Aug, N58\degree-59\degree, E19\degree-20\degree, 50-85m} 
   	\end{subfigure}
   	\hfill
   	\begin{subfigure}[b]{0.24\linewidth}
  		\includegraphics[width=1.0\linewidth,height=0.8\linewidth]{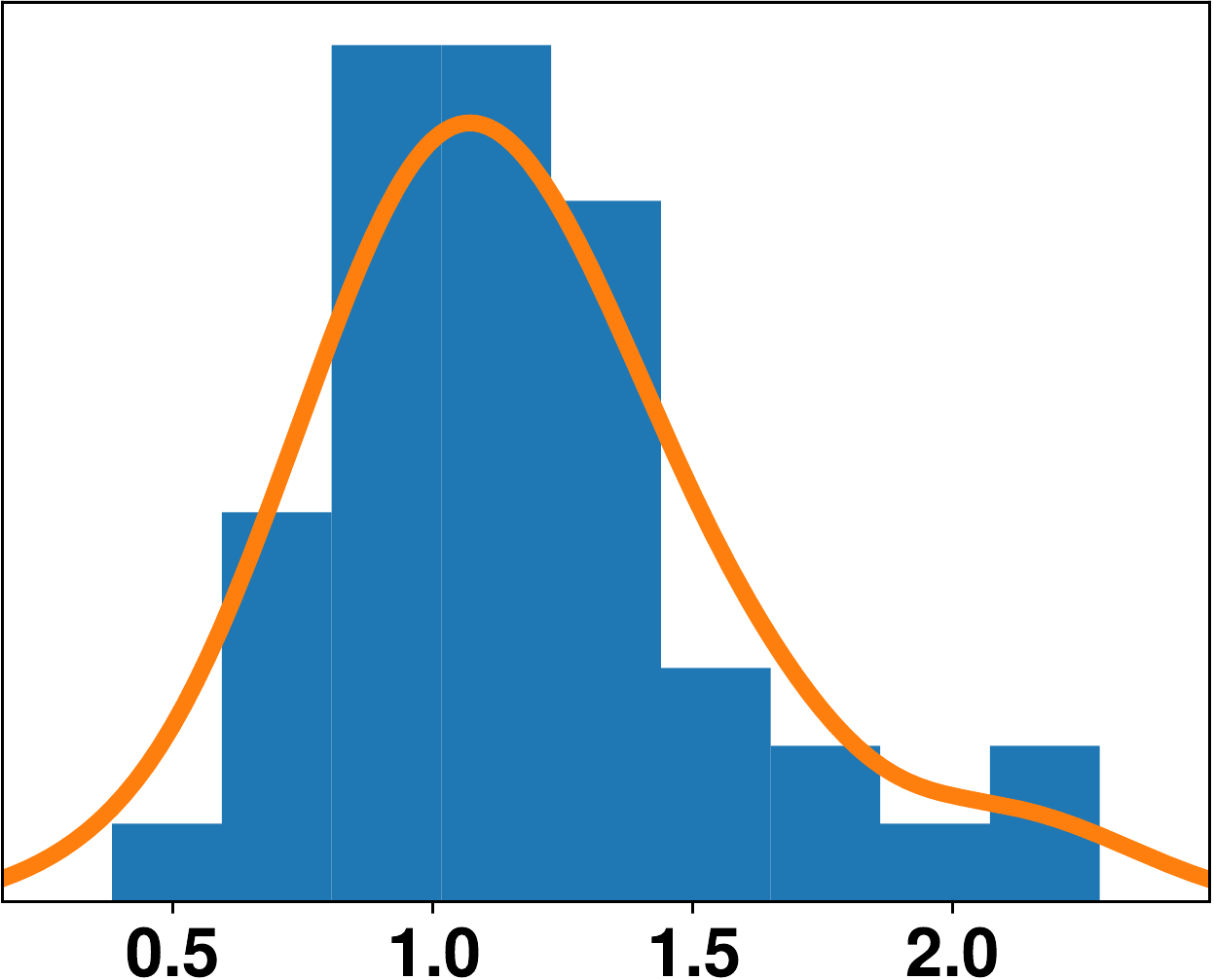}
   		\caption{Nov, N57\degree-58\degree, E20\degree-21\degree, 50-85m} 
   	\end{subfigure}
       
 	\begin{subfigure}[b]{0.24\linewidth}
  		\includegraphics[width=1.0\linewidth,height=0.8\linewidth]{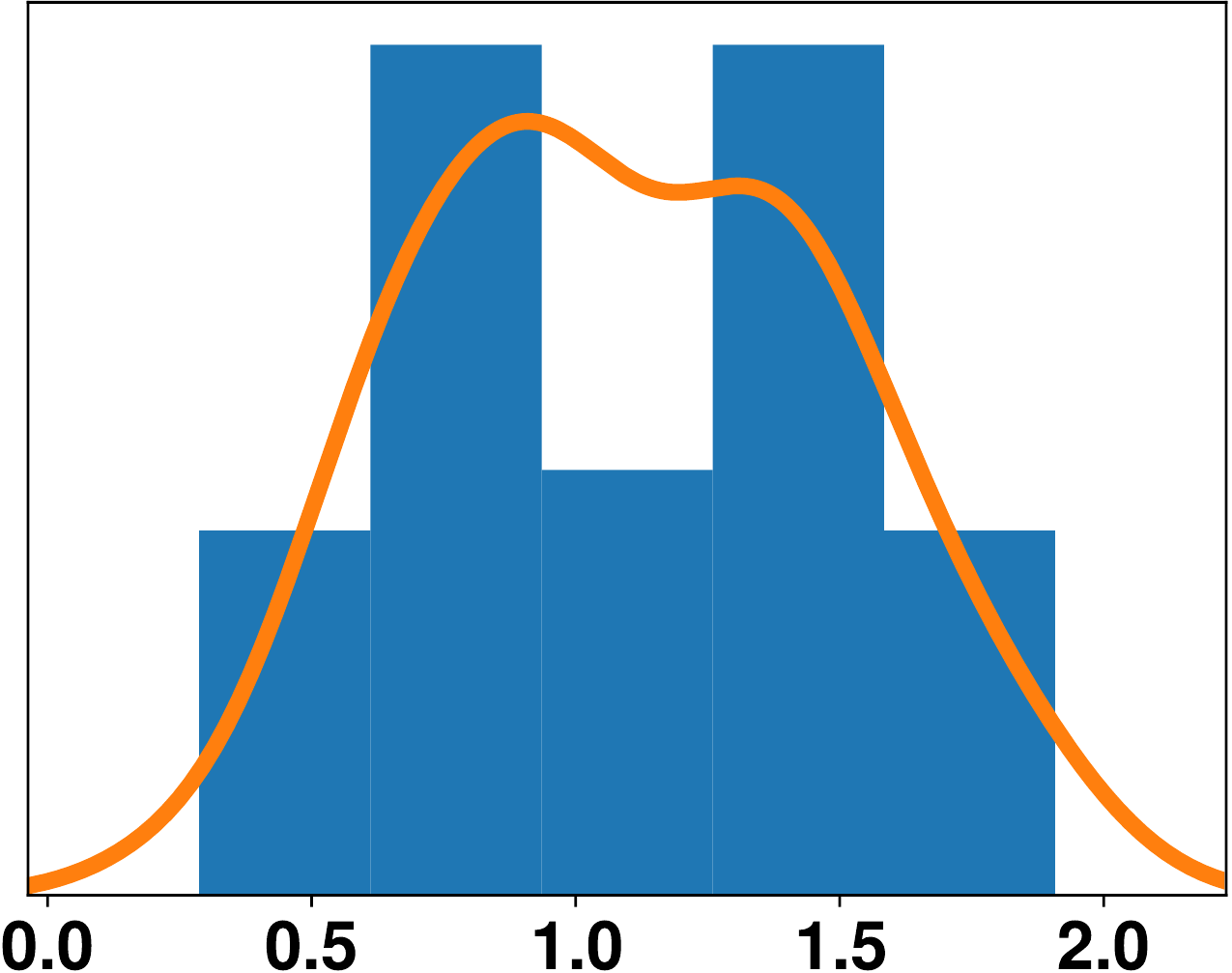}
   		\caption{May, N57\degree-58\degree, E20-E21\degree, 50-85m} 
 	\end{subfigure}
   	\hfill
   	\begin{subfigure}[b]{0.24\linewidth}
   		\includegraphics[width=1.0\linewidth,height=0.8\linewidth]{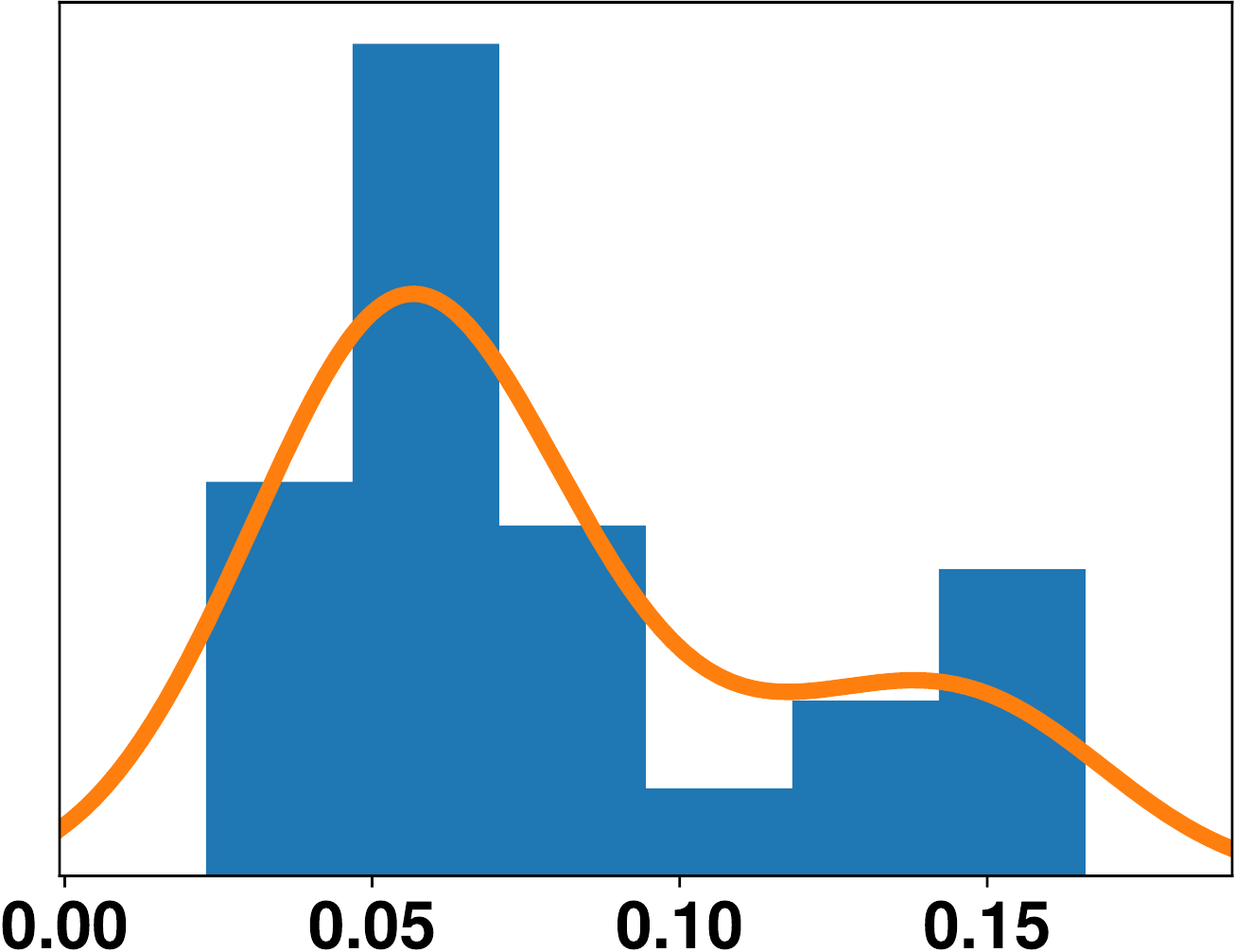}
   		\caption{Aug, N52\degree-53\degree, E20-21\degree, 0-25m} 
   	\end{subfigure}
 	\hfill
 	\begin{subfigure}[b]{0.24\linewidth}
   		\includegraphics[width=1.0\linewidth,height=0.8\linewidth]{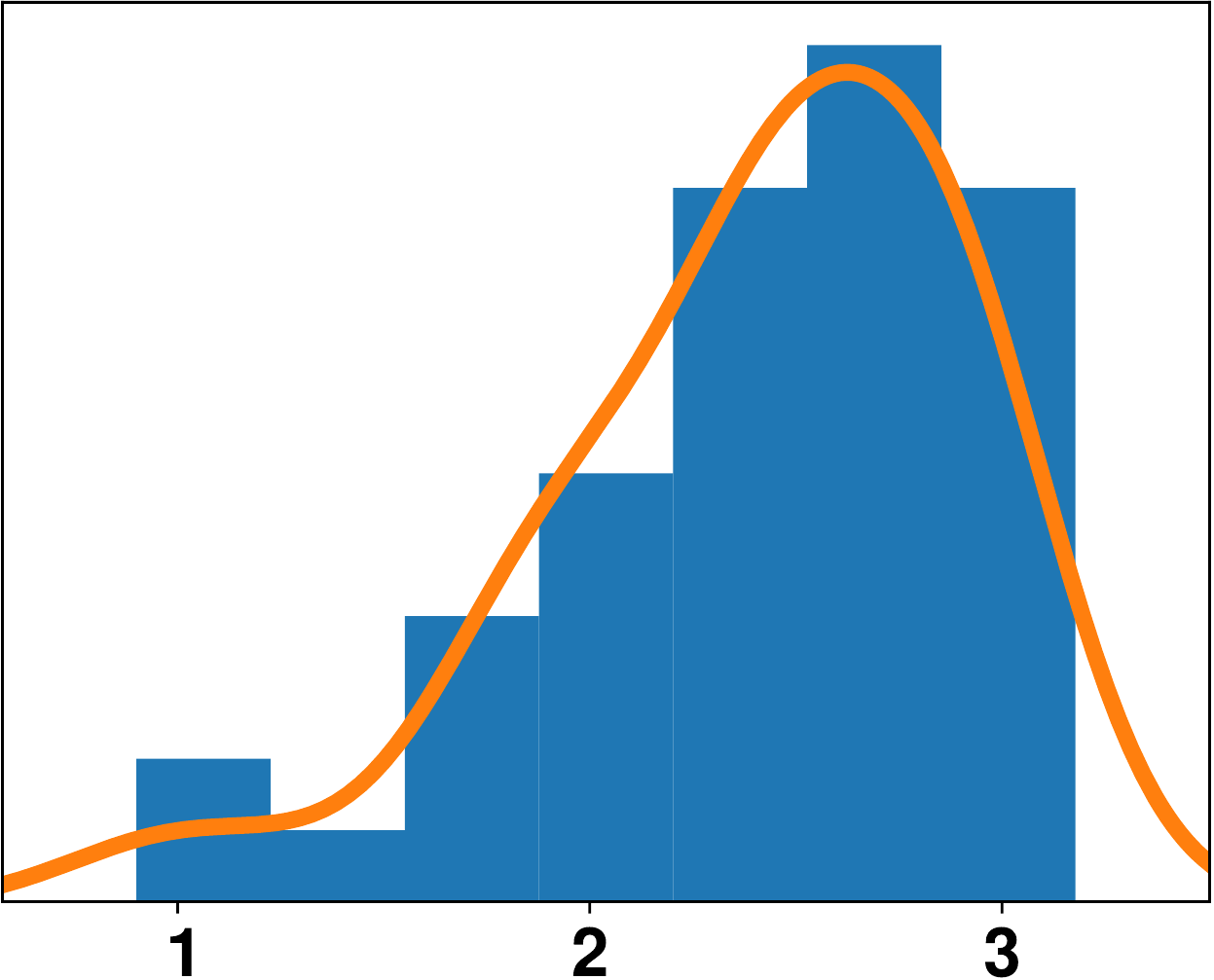}
 		\caption{Nov, N57\degree-58\degree, E20-21\degree, 85-120m} 
 	\end{subfigure}
    \hfill
    \begin{subfigure}[b]{0.24\linewidth}
   		\includegraphics[width=1.0\linewidth,height=0.8\linewidth]{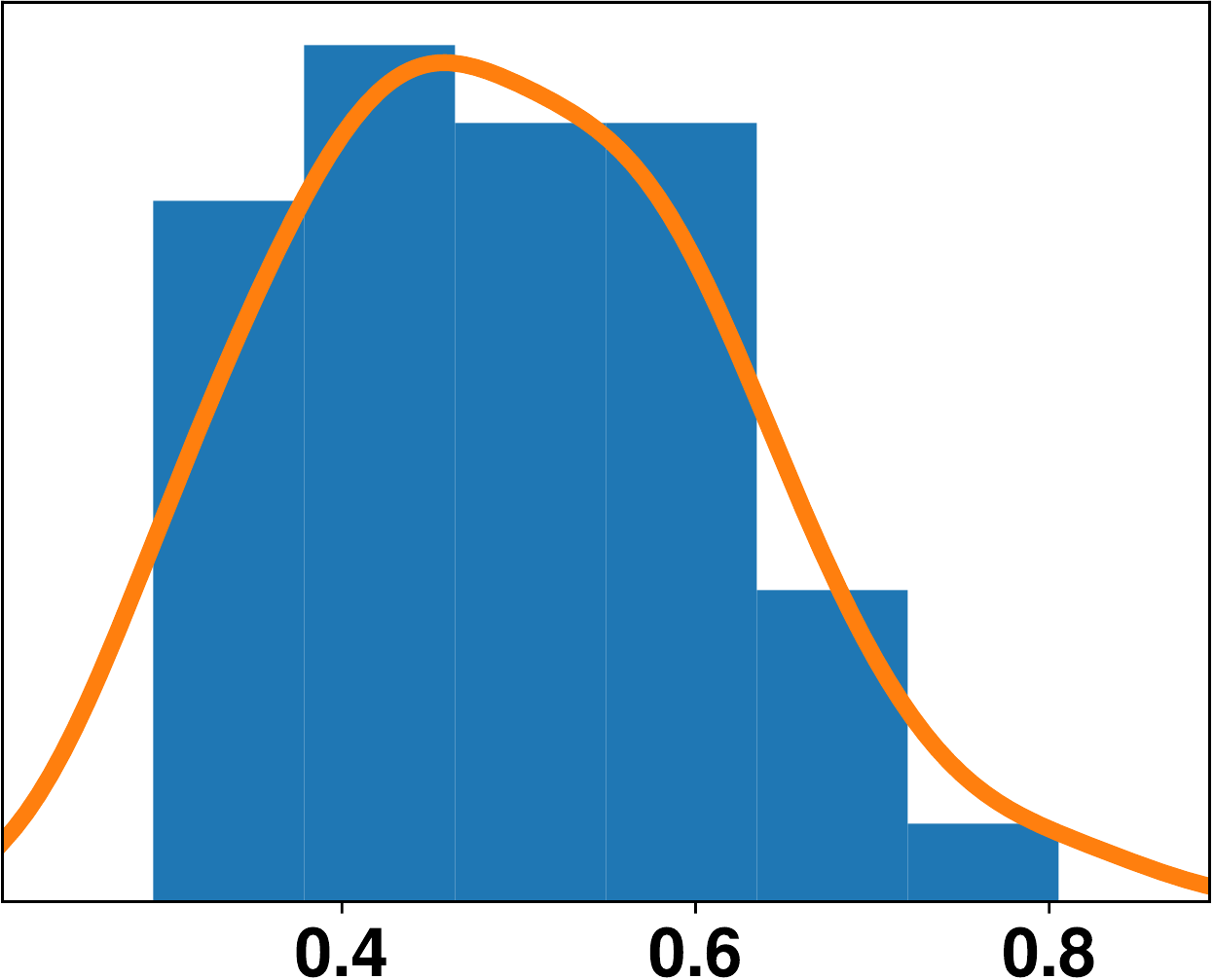}
   		\caption{Dec, N57\degree-58\degree, E11-12\degree, 0-25m} 
   	\end{subfigure}
	\vspace{-0.5em}
\caption{Selected histograms and kernel density estimations for the true concentration ($\delta$). Most of them look like the ones in the first and second row, i.e., log-normal or normal distributed. A few look like the ones in the last row which do not look like either of the these two distributions.} 
	\label{fig:histograms:true_concentrations}
\end{figure}

We investigated the underlying probability distributions of the measurement results $\eta$, the true concentration $\delta$ and the noise $\epsilon$, (compare Subsection \ref{subsec:probability distributions}).

To assess the probability distributions, we produced histograms and kernel density estimations (KDEs). For the histograms, the sizes of the bins were determined by the rule of  \cite{Freedman1981}. For the KDEs, normal kernel were used with bandwidths determined by the rule of \cite{Scott2015}.

Histograms and KDEs with at least hundred values were generated for the measurement results $\eta$ and with at least forty values for the true concentration $\delta$, resulting in around eight hundred and four hundred plots, respectively. When most of the values are close to zero, the plots usually look like log-normal distributions and otherwise usually like normal distributions. However, there is no clear overall trend and some plots look like neither of these distributions.

A selection of these is presented in Figure \ref{fig:histograms:measurement_results} and \ref{fig:histograms:true_concentrations}. In the first rows typical plots which indicate log-normal distributions are presented, in the second rows typical plots which indicate normal distributions and in the third rows some of the rear obscure distributions.

As stated in Subsection \ref{subsec:probability distributions}, the values of the noise $\epsilon$ are just the values of $\eta$ shifted by a constant. Hence, the histograms and KDEs for $\epsilon$ look similar to the ones in Figure \ref{fig:histograms:measurement_results}, just with other values on the horizontal axis.

We also applied the statistical tests mentioned in Subsection \ref{subsec:probability distributions} to test against normal and log-normal distributions. We used a significance level of 1\% and tested only at points where at least 40 values are available. 

Regarding the measurement results $\eta$, the tests rejected in average the normal distribution assumption for 73\% of the cases. The log-normal distribution assumption was rejected in average for 65\% of the cases. For the true concentration $\delta$, the tests rejected in average the normal distribution assumption for 25\% of the cases and the log-normal distribution assumption for 52\% of the cases.

These high number of rejections gave a different impression than the visual inspection, indicating that at some grid boxes the probability distributions might neither be a normal nor a log-normal one. The high proportion of rejections may be explained by the fact that the measurement results have at most three significant digits. Thus they represent at best only heavily rounded realizations of a normal or log-normal distribution. However, these roundings are not taken into account in the tests.

 \conclusions

\label{sec: conclusion}

Phosphate is a key component in understanding the marine ecosystem. Millions of measurement data for phosphate are available at the World Ocean Database but these are not statistically analyzed in such extent as it is done here.

The climatological mean and variability as well as the short scale variability were quantified. The results indicate that it makes sense to increase the resolution of the analysis in (some) coastal areas to obtain more accurate climatological information and to decrease the resolution in areas far away from coasts and deep in the ocean without loosing climatological information.

The correlation of climatological concentrations were estimated as well. They generally decrease with increasing spatial and temporal distance, but do not solely depend on the distance. 

The climatological concentrations and the measurement results seem to be mostly normally or log-normally distributed. However, there is no clear trend and in some cases they do not seem to belong to either distribution. Hence, they do not seem to originate from a single type of distribution. 

This extensive analysis is useful in understanding the marine phosphate concentration. It may also be valuable in the calibration of marine biogeochemical models where our estimated standard deviations and correlations could be incorporated to achieve a more accurate model calibration (cf. \cite[4]{Walter1997} or \cite[2.1.4]{Seber2003}).

The analysis is also helpful in planning new phosphate measurements. A lower short-scale standard deviation would mean that average concentration in this year could be be determined more accurate compared to a higher one. If the long-scale standard deviation is small as well the climatological concentration can also be determined more accurate.

If the measurements should be used to estimate parameters of a model or to determine the most realistic one among several models, optimal experimental design methods (cf. \cite[6]{Walter1997} or \cite[5]{Pronzato2013}), which include statistical properties such as standard deviations and correlations, can be used to determine the places and times of measurements which provide the highest information gain.

The approaches in the statistical analysis are not limited to phosphate but can also be applied to other data which satisfies the assumptions. 

\appendix
\section{Spatial and Temporal Resolution}
\label{subsec:lsms}

The spatial grid for our calculation was constructed using the one-degree spatial grid of the World Ocean Atlas 2013 \cite[Chapter 3.4]{Garcia2014}, which is based on the global relief model ETOPO2v2 (\cite{NGDC2006}).

The spatial grid of the World Ocean Atlas 2013  has a resolution of one degree and 137 vertical layers with increasing thickness. The annual data provided by the World Ocean Atlas 2013  are available up to a depth of 5500 meters which represents 102 layers. The seasonal and monthly data are available up to a depth of 500 meters which represents 37 layers.

\begin{figure}[H]
	\centering
	\begin{subfigure}[b]{0.49\linewidth}
        \centering
        \includegraphics[width=1.0\linewidth]{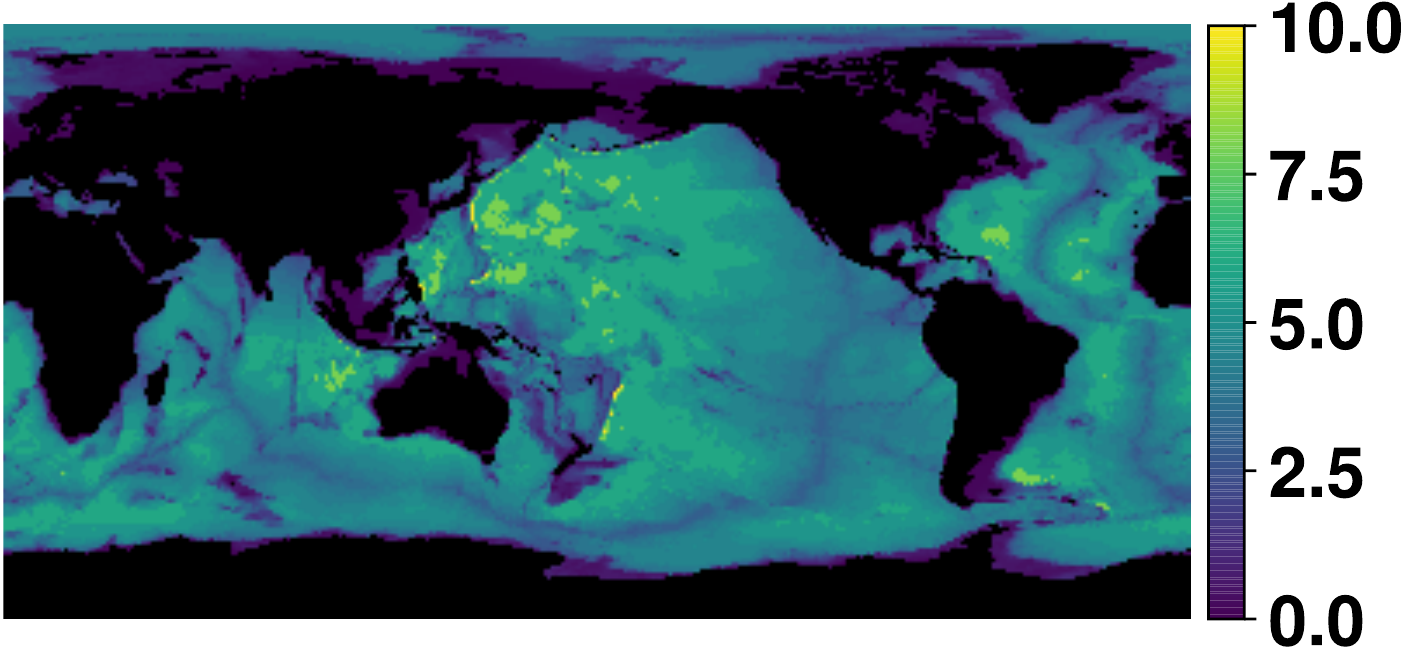}
        \caption{Depths in kilometer.} 
        \label{fig:data_lsm:depth}
	\end{subfigure}
	\hfill
	\begin{subfigure}[b]{0.49\linewidth}
        \centering
\small
        \begin{tabular}{l|l|l|l|l}
            0	    & 290       & 1080      & 2495	    & 4510	\\
            25      & 360	    & 1250      & 2740	    & 4855	\\
            50	    & 455       & 1420      & 3010		& 5200	\\
            85	    & 550	    & 1615      & 3280		& 6000	\\
            120	    & 670       & 1810      & 3575      & 8000	\\
            170	    & 790       & 2030      & 3870      & 10000	\\
            220	    & 935       & 2250		& 4190      &	\\
        \end{tabular}
        \caption{Depths possible for grid boxes in meters.}
\label{tab:data_lsm:depth_level}
	\end{subfigure}
    \caption{Depths in the spatial grid used for this statistical analysis.} 
\end{figure}

We decided to reduce the vertical resolution to 33 layers in order to increase the number of data in each grid box. The new depth in each grid box was chosen as the highest depth in Table \ref{tab:data_lsm:depth_level} which is less or equal to the depth in the World Ocean Atlas 2013. The resulting depths are plotted in Figure \ref{fig:data_lsm:depth}.

For the temporal resolution one month was used at all layers, allowing to cover time-dependent changes in all layers.

\section{Interpolation}
\label{subsec:interpolation}

The data in our calculations were linearly interpolated by triangulating the input data with the method of Qhull (\cite{Barber1996}) and performing linear barycentric interpolation on each triangle. Values for points outside the convex hull of the data points were interpolated using the value of the nearest data point. For this purpose, a kd-tree (\cite{Maneewongvatana1999}) was used to rapidly look up the nearest neighbor of each point. We used a Python (\cite{Python-3.7}) implementation of both algorithms, which is part of the SciPy library (\cite{Jones2019b} and \cite{Virtanen2019}).

The annual periodicity and the periodicity with respect to the longitude were included in the interpolation by assuming the same values for data points plus/minus the period.

Instead of the depth, the number of the corresponding depths level described in Table \ref{tab:data_lsm:depth_level} was used for the interpolation. Thus, a vertical distance deep down in the ocean is weighted less than near the surface. This takes into account that the changes of the values at great depth are smaller than those closer to the surface.

The points were scaled so that the distance between two consecutive depth levels corresponds to a distance of one degree, and the length of one year corresponds to the circumference of the earth.

\section{Software}
\label{subsec:software}

The results in Section \ref{sec: result} have been calculated and visualized using the measurements software package (\cite{measurements-0.3}) which is based on Python (\cite{Python-3.7}), NumPy (\cite{NumPy-1.17.3}), SciPy (\cite{Jones2019b} and \cite{Virtanen2019}), Matplotlib (\cite{matplotlib-3.1.1} and \cite{Hunter2007}), utillib (\cite{utillib-0.3}) and the matrix-decomposition library (\cite{matrix-decomposition-1.2}).

The measurements software package contains functions for all methods described in Section \ref{sec: method} as well as for visualizing corresponding result. It is especially suited for data from the World Ocean Database because it provides special functions for processing these data. However, it is not limited to these data.  \noappendix

\bibliographystyle{copernicus}
\bibliography{article.bib}

\end{document}